\documentclass[10pt]{article}
\usepackage{graphicx}
\usepackage{heppennames2}
\usepackage{lhchiggs}
\usepackage{cernunits}
\allowdisplaybreaks
\newcommand\pubdate{\today}

\def\csumb{Dipartimento di Fisica Teorica, Universit\`a di Torino, Italy\\
           INFN, Sezione di Torino, Italy}
\def\support{\footnote{Work supported by MIUR under contract
    2001023713$\_$006 and by UniTo - Compagnia di San Paolo under contract ORTO11TPXK.}}
\def\Title#1{\begin{center} {\Large\bf #1 } \end{center}}

\def\Author#1{\begin{center}{ \sc #1} \end{center}}
\def\Address#1{\begin{center}{ \it #1} \end{center}}

\newcommand\pubblock{\rightline{\begin{tabular}{l} \\
         \pubdate\\  \end{tabular}}}
\newenvironment{Abstract}{\begin{quotation}  }{\end{quotation}}

\def\Acknowledgments{\bigskip  \bigskip \begin{center}
          \large\bf Acknowledgments\end{center}}
\def\email#1{\footnote{#1}}
\makeatletter
\def\section{\@startsection{section}{0}{\z@}{5.5ex plus .5ex minus
 1.5ex}{2.3ex plus .2ex}{\large\bf}}
\def\subsection{\@startsection{subsection}{1}{\z@}{3.5ex plus .5ex minus
 1.5ex}{1.3ex plus .2ex}{\normalsize\bf}}
\def\subsubsection{\@startsection{subsubsection}{2}{\z@}{-3.5ex plus
-1ex minus  -.2ex}{2.3ex plus .2ex}{\normalsize\sl}}

\renewcommand{\@makecaption}[2]{%
   \vskip 10pt
   \setbox\@tempboxa\hbox{\small #1: #2}
   \ifdim \wd\@tempboxa >\hsize     
       \small #1: #2\par          
     \else                        
       \hbox to\hsize{\hfil\box\@tempboxa\hfil}
   \fi}
 \def\citenum#1{{\def\@cite##1##2{##1}\cite{#1}}}
\def\citea#1{\@cite{#1}{}}
\newcount\@tempcntc
\def\@citex[#1]#2{\if@filesw\immediate\write\@auxout{\string\citation{#2}}\fi
  \@tempcnta\z@\@tempcntb\m@ne\def\@citea{}\@cite{\@for\@citeb:=#2\do
    {\@ifundefined
       {b@\@citeb}{\@citeo\@tempcntb\m@ne\@citea\def\@citea{,}{\bf }\@warning
       {Citation `\@citeb' on page \thepage \space undefined}}%
    {\setbox\z@\hbox{\global\@tempcntc0\csname b@\@citeb\endcsname\relax}%
     \ifnum\@tempcntc=\z@ \@citeo\@tempcntb\m@ne
       \@citea\def\@citea{,}\hbox{\csname b@\@citeb\endcsname}%
     \else
      \advance\@tempcntb\@ne
      \ifnum\@tempcntb=\@tempcntc
      \else\advance\@tempcntb\m@ne\@citeo
      \@tempcnta\@tempcntc\@tempcntb\@tempcntc\fi\fi}}\@citeo}{#1}}
\def\@citeo{\ifnum\@tempcnta>\@tempcntb\else\@citea\def\@citea{,}%
  \ifnum\@tempcnta=\@tempcntb\the\@tempcnta\else
  {\advance\@tempcnta\@ne\ifnum\@tempcnta=\@tempcntb \else\def\@citea{--}\fi
    \advance\@tempcnta\m@ne\the\@tempcnta\@citea\the\@tempcntb}\fi\fi}
\makeatother
\DeclareRobustCommand{\PA}{\HepParticle{A}{}{}\Xspace}
\DeclareRobustCommand{\PV}{\HepParticle{V}{}{}\Xspace}
\DeclareRobustCommand{\PX}{\HepParticle{X}{}{}\Xspace}
\DeclareRobustCommand{\Pf}{\HepParticle{f}{}{}\Xspace}
\DeclareRobustCommand{\PAf}{\HepAntiParticle{\Pf}{}{}\Xspace}
\DeclareRobustCommand{\PF}{\HepParticle{F}{}{}\Xspace}
\DeclareRobustCommand{\PI}{\HepParticle{I}{}{}\Xspace}
\DeclareRobustCommand{\Phad}{\HepParticle{h}{}{}\Xspace}

\newcommand{\OS}{\mathrm{OS}}
\newcommand{\CPP}{\mathrm{CPP}}

\newcommand{\mBW}{\mathrm{BW}}
\newcommand{\OFS}{\mathrm{OFS}}
\newcommand{\intf}{\mathrm{int}}

\newcommand{\ssB}{{\mathrm{B}}}
\newcommand{\ssC}{{\mathrm{C}}}

\newcommand{\ssF}{{\mathrm{F}}}
\newcommand{\ssR}{{\mathrm{R}}}

\newcommand{\ssL}{{\mathrm{L}}}
\newcommand{\ssP}{{\mathrm{P}}}
\newcommand{\ssS}{{\mathrm{S}}}

\newcommand{\ssV}{{\mathrm{V}}}
\newcommand{\ssM}{{\mathrm{M}}}
\newcommand{\ssN}{{\mathrm{N}}}
\newcommand{\sH}{\mathrm{H}}
\newcommand{\HH}{\mathrm{HH}}
\newcommand{\VV}{\mathrm{VV}}

\newcommand{\bqas}{\begin{eqnarray*}}
\newcommand{\eqas}{\end{eqnarray*}}
\newcommand{\nl}{\nonumber\\}

\newcommand{\lpar}{\left(}                            
\newcommand{\rpar}{\right)}

\newcommand{\bq}{\begin{equation}}                    
\newcommand{\eq}{\end{equation}}
\newcommand{\bqa}{\arraycolsep 0.14em\begin{eqnarray}}
\newcommand{\eqa}{\end{eqnarray}}
\newcommand{\ba}[1]{\begin{array}{#1}}
\newcommand{\ea}{\end{array}}
\newcommand{\ben}{\begin{enumerate}}
\newcommand{\een}{\end{enumerate}}
\newcommand{\bei}{\begin{itemize}}
\newcommand{\eei}{\end{itemize}}
\newcommand{\eqn}[1]{Eq.(\ref{#1})}
\newcommand{\eqns}[2]{Eqs.(\ref{#1})--(\ref{#2})}

\newcommand{\eqnsc}[2]{Eqs.(\ref{#1}) and (\ref{#2})}

\newcommand{\hatp}{{\hat p}}

\newcommand{\bmid}{\Bigr|}

\newcommand{\ord}[1]{{\cal O}\lpar#1\rpar}

\newcommand{\Bref}[1]{Ref.~\cite{#1}}
\newcommand{\Brefs}[1]{Refs.~\cite{#1}}

\newcommand{\eg}{e.g.\xspace}
\newcommand{\ie}{i.e.\xspace}
\newcommand{\etc}{etc.\@\xspace}

\newcommand{\mv}{\mathswitch {M_{\PV}}}
\newcommand{\mh}{\mathswitch {M_{\PH}}}
\newcommand{\mw}{\mathswitch {M_{\PW}}}

\newcommand{\mz}{\mathswitch {M_{\PZ}}}
\newcommand{\mhs}{\mathswitch {M^2_{\PH}}}
\newcommand{\mBh}{\mathswitch {{\overline M}_{\PH}}}
\newcommand{\mBhs}{\mathswitch {{\overline M}^2_{\PH}}}
\newcommand{\mhq}{\mathswitch {M^4_{\PH}}}
\newcommand{\mws}{\mathswitch {M^2_{\PW}}}

\newcommand{\mt}{\mathswitch {M_{\PQt}}}
\newcommand{\cph}{\mathswitch {s_{\PH}}}
\newcommand{\cpw}{\mathswitch {s_{\PW}}}
\newcommand{\cpwz}{\mathswitch {s_{\PW}, s_{\PZ}}}
\newcommand{\cpz}{\mathswitch {s_{\PZ}}}
\newcommand{\cpt}{\mathswitch {s_{\PQt}}}
\newcommand{\muh}{\mathswitch {\mu_{\PH}}}
\newcommand{\muhs}{\mathswitch {\mu^2_{\PH}}}
\newcommand{\gh}{\mathswitch {\gamma_{\PH}}}
\newcommand{\GOS}{\mathswitch {\Gamma_{\OS}}}
\newcommand{\GOL}{\mathswitch {\overline\Gamma}}
\newcommand{\muol}{\mathswitch {\overline\mu}}
\newcommand{\GOSu}{\mathswitch {\Gamma^{\OS}}}

\newcommand{\muR}{\mathswitch {\mu_{\ssR}}}
\newcommand{\muF}{\mathswitch {\mu_{\ssF}}}
\newcommand{\muRs}{\mathswitch {\mu^2_{\ssR}}}
\newcommand{\muFs}{\mathswitch {\mu^2_{\ssF}}}
\newcommand{\tot}{{\mbox{\scriptsize tot}}}
\newcommand{\myprod}{{\mbox{\scriptsize prod}}}
\newcommand{\prop}{{\mbox{\scriptsize prop}}}
\newcommand{\dec}{{\mbox{\scriptsize dec}}}
\newcommand{\kcut}{{\mbox{\scriptsize cut}}}
\newcommand{\ch}{{\mbox{\scriptsize ch}}}
\newcommand{\inv}{{\mbox{\scriptsize inv}}}
\newcommand{\bos}{{\mbox{\scriptsize bos}}}
\newcommand{\peak}{{\mbox{\scriptsize peak}}}

\newcommand{\bin}{{\mbox{\scriptsize in}}}
\newcommand{\bout}{{\mbox{\scriptsize out}}}
\newcommand{\ac}{{\mbox{\scriptsize all}}}
\newcommand{\HO}{{\mathrm{HO}}}
\newcommand{\intfxy}[2]{\int_{\scriptstyle 0}^{\scriptstyle 1}\,d#1\,
                        \int_{\scriptstyle 0}^{\scriptstyle #1}\,d#2}
\newcommand{\li}[2]{\mathrm{Li}_{#1}\lpar\displaystyle{#2}\rpar} 

\newcommand{\ep}{\mathswitch \varepsilon}

\newcommand{\spro}[2]{{#1}\cdot{#2}}
\providecommand\HTO{{\sc HTO}}
\providecommand\POWHEG{{\sc POWHEG}}
\providecommand\iHixs{{\sc iHixs}}
\begin{document}
\begin{titlepage}
\pubblock
%
\vfill
\def\thefootnote{\fnsymbol{footnote}}
\Title{The Higgs Boson Lineshape\support
}
\vfill
\Author{Stefano Goria\email{goria@to.infn.it},               
Giampiero Passarino\email{giampiero@to.infn.it},
Dario Rosco\email{rosco@to.infn.it}}               
\Address{\csumb}
\vfill
\vfill
\begin{Abstract}
\noindent 
The heavy Higgs searches at LHC are carried out in $\Pg\Pg \to \PH \to \PW\PW \to \Pl\PGn\Pl\PGn$, 
$\Pg\Pg \to \PH \to \PZ\PZ \to \Pl\Pl\Pl\Pl,\Pl\Pl\PGn\PGn,\Pl\Pl\PQq\PQq$ channels. 
The current searches for a heavy Higgs boson assume on-shell (stable) Higgs-boson production. 
The Higgs-boson production cross section is then sampled with a Breit-Wigner distribution
(with fixed-width or running-width) and implemented in MonteCarlo simulations. Therefore the 
question remains of what is the limitation of the narrow Higgs-width approximation.
The main focus of this work is on the description of the Standard Model Higgs--boson-lineshape 
in the heavy Higgs region, typically $\MH$ above $600\UGeV$. The framework discussed in this paper 
is general enough and can be used for all processes and for all kinematical regions.
Numerical results are shown for the gluon-fusion process. Issues of gauge invariance and
residual theoretical uncertainties are also discussed. Limitations due to a breakdown of
the perturbative expansion are comprehensively discussed, including a discussion of th
equivalence theorem for (off-shell) virtual vector-bosons. Analytic continuation in a theory 
with unstable particles is thoroughly discussed.
\end{Abstract}
\vfill
\begin{center}
\end{center}
\end{titlepage}
\def\thefootnote{\arabic{footnote}}
\setcounter{footnote}{0}
\small
\thispagestyle{empty}
\tableofcontents
\normalsize
\clearpage
\setcounter{page}{1}
\section{Introduction \label{intro}}
At the beginning of $2011$ the status of the inclusive cross-section for Higgs-boson production 
in gluon fusion was summarized~\cite{Dittmaier:2011ti}. Corrections arising from 
higher-order QCD, electroweak effects, as well as contributions beyond the commonly-used 
effective theory approximation were analyzed. Uncertainties arising from missing terms in the 
perturbative expansion, as well as imprecise knowledge of parton distribution functions, 
were estimated to range from approximately $15{-}20\%$, with the precise value depending on 
the Higgs boson mass. For an updated study we we refer to \Bref{Dittmaier:2012vm}.

In this work we consider the problem of a consistent definition of the Higgs--boson-lineshape 
at LHC and address the question of optimal presentation of LHC data.

Some preliminary observations are needed: the Higgs system of the Standard Model has almost 
a non-perturbative behavior, even at a low scale. Let us consider the traditional on-shell 
approach and the values for the on-shell decay width of the Higgs boson~\cite{Dittmaier:2011ti}: 
for $\MH = 140\UGeV$ the total width is $8.12\times 10^{-3}\UGeV$; the process
$\PH \to \PAQb \PQb$ has a partial with of $2.55\times 10^{-3}\UGeV$, other two-body
decays are almost negligible while $\PH \to 4\,\Pf$ has $4.64\times 10^{-3}\UGeV$ well below
the $\PW\PW\,$-threshold. Therefore the four-body decay, even below threshold, is more 
important than the two-body ones. 
What are the corresponding implications? Since decay widths are related to the imaginary parts
of loop diagrams the above statement is equivalent to say that, in terms of imaginary parts 
of the $\PH$ self-energy, three-loop diagrams are as important as one-loop diagrams, or more.

We realize that most of the people just want to use some well-defined recipe
without having to dig any deeper; however, there is no alternative to a complete description
of LHC processes which has to include the complete matrix elements for all relevant
processes; splitting the whole $S\,$-matrix element into components is just conventional 
wisdom. However, the precise tone and degree of formality must be dictated by gauge invariance.

The framework discussed in this paper is general enough and can be used for all
processes and for all kinematical regions; however, the main focus of this work will be on the
description of the Higgs--boson-lineshape in the heavy Higgs region, typically above
$600\UGeV$. The general argument is well documented in the literature and, for a complete 
description of all technical details, we refer to \Bref{Actis:2008uh,Passarino:2010qk}.
Part of the results discussed in this paper and concerning a correct implementation 
of the Higgs--boson-lineshape have been summarized in a talk at the BNL meeting of the 
HXSWG\footnote{http://www.bnl.gov/hcs/}.
It is important to stress the exact content of this work: we deal with a gauge invariant definition
of the signal, which assumes that the non-resonant background and the signal/background
interference have been subtracted from the data.

One might wonder why considering a Standard Model (SM) Higgs boson in such a high-mass range. 
There are classic constraints on the Higgs boson mass coming from unitarity, triviality and
vacuum stability, precision electroweak data and absence of fine-tuning~\cite{Ellis:2009tp}. 
However, the search for a SM Higgs boson over a mass range from $80\UGeV$ to $1\UTeV$ is clearly 
indicated as a priority in many experimental papers, \eg 
{\it ATLAS: letter of intent for a general-purpose pp experiment at the large hadron collider 
at CERN}\footnote{http://atlas.web.cern.ch/Atlas/internal/tdr.html}. 
For recent results, see \Bref{Aad:2011ec,Collaboration:2012sm,Collaboration:2012sk,Collaboration:2012si,Graziano:2012yc,Chatrchyan:2012ft,Chatrchyan:2012hr}.

The situation is different if we consider extensions of the SM: in the two Higgs doublet (THD)
model, even if the SM-like Higgs boson is found to be light ($< 140\UGeV$), there is a possible 
range of mass splitting in the heavy Higgs boson. In general, for a given Higgs boson mass, 
the magnitude of the mass splittings among different heavy scalar bosons can be determined to 
satisfy the electroweak precision data, see \Bref{Kanemura:2011sj}.

In addition, the work of \Bref{Bai:2011aa} studies scenarios where a heavy Higgs boson can be 
made consistent with both the indirect constraints and the direct null searches by adding only 
one new particle beyond the Standard Model. 
Heavy Higgs effects in a theory with a fourth Standard-Model-like fermion generation have
been examined in \Brefs{Passarino:2011kv,Denner:2011vt}.

The work of \Bref{Peskin:2001rw} critically reviews models that allow a heavy Higgs boson 
consistent with the precision electroweak constraints. All have unusual features, and all can 
be distinguished from the Minimal SM either by improved precision measurements or by other 
signatures accessible to next-generation colliders.

Coming back to the framework that we are introducing, there is another important issue:
when working in the on-shell scheme one finds that the two-loop corrections to the on-shell 
Higgs width exceed the one-loop corrections if the on-shell Higgs mass is larger
than $900\UGeV$, as discussed in \Bref{Ghinculov:1996py}. This fact simply tells you
that perturbation theory diverges badly, starting from approximately $1\UTeV$.
In this work we will also illustrate the corresponding impact on the Higgs boson lineshape
(previous work can be found in \Brefs{Ghinculov:1995zs,Ghinculov:1995bz}).

Recently the problem of going beyond the zero-width approximation has received 
new boost from the work of \Brefs{Anastasiou:2011pi,Anastasiou:2012hx}: the program \iHixs{} allow
the study of the Higgs--boson-lineshape for a finite width of the Higgs boson and computes the
cross-section sampling over a Breit-Wigner distribution.
There is, however, a point that has been ignored in all calculations performed so far: the 
Higgs boson is an unstable particle and should be removed from the in/out bases in the
Hilbert space, without destroying the unitarity of the theory. Therefore,
concepts as the {\em production} of an unstable particle or its {\em partial decay 
widths} do not have a precise meaning and should be replaced by a conventionalized definition 
which respects first principles of Quantum Field Theory (QFT).

This paper is organized as follows. In \refS{Sect_prop} we introduce and discuss complex
poles for unstable particles.
In \refS{Sect_prodec} we analyze production and decay of a Higgs boson at LHC.
A discussion on gauge invariance is presented in \refS{Sect_gi}.
In \refS{Sect_err} we present a short discussion on the QCD scale error.
In \refS{Sect_num} we present numerical results while in \refS{Sect_THU} we discuss the
residual theoretical uncertainty.
Finally, technical details are discussed in Appendices, in particular in \refA{appB} we discuss 
how to apply the equivalence theorem for virtual vector-bosons and in \refA{appCa} we discuss
analytic continuation in a theory with unstable particles.
\section{Propagation\label{Sect_prop}}
To start our discussion we consider the process $i j \to \PH ( \to \PF )  + \PX$ where 
$i,j\,\in\,$partons and $\PF$ is a generic final state (\eg $\PF= \PGg \PGg, 4\,\Pf$, \etc). For 
the sake of simplicity we neglect, for a moment, folding the partonic process with parton
distribution functions (PDFs). Since the Higgs boson is a scalar resonance we can split the 
whole process into three parts, {\em production}, {\em propagation} and {\em decay}.
In QFT all amplitudes are made out of propagators and vertices and the (Dyson-resummed) 
propagator for the Higgs boson reads as follows:
\bq
\Delta_{\PH}(s) = \Bigl[ s - \mhs + S_{\HH}\lpar s,\mt^2,\mh^2,\mw^2,\mz^2\rpar \Bigr]^{-1},
\label{DRHP}
\eq
where $M_i$ is a renormalized mass and $S_{\HH}$ is the renormalized Higgs 
self-energy (to all orders but with one-particle-irreducible diagrams). The first 
argument of the self-energy in \eqn{DRHP} is the external momentum squared, the 
remaining ones are (renormalized) masses in the loops. 
We define complex poles for unstable particles as the (complex) solutions of the following 
system:
\bqa
\cph - \mhs + S_{\HH}\lpar \cph,\mt^2,\mh^2,\mw^2,\mz^2\rpar &=& 0,
\nl
\cpw - \mws + S_{\WW}\lpar \cpw,\mt^2,\mh^2,\mw^2,\mz^2\rpar &=& 0,
\eqa
\etc To lowest order accuracy the Higgs propagator can be rewritten as
\bq
\Delta^{-1}_{\PH} =  s - \cph.
\label{CMSprop}
\eq
The complex pole describing an unstable particle is conventionally parametrized as
\bq
s_i = \mu^2_i - i\,\mu_i\,\gamma_i,
\label{CPpar}
\eq
where $\mu_i$ is an input parameter (similar to the on-shell mass) while $\gamma_i$ can be 
computed (as the on-shell total width), say within the Standard Model. 
There are other, equivalent, parametrizations~\cite{Ghinculov:1996py}, 
\eg $\sqrt{\cph}= \muh - i/2\,\gh$.
Note that the the pole of $\Delta$ fully embodies the propagation properties of a particle. 
We know that even in ordinary Quantum Mechanics the resonances are described as the 
complex energy poles in the scattering amplitude. 
The general formalism for describing unstable particles in QFT was developed long ago,
see \Brefs{Veltman:1963th,Jacob:1961zz,Valent:1974bd,Lukierski:1978ke,Bollini:1993yp};
for an implementation in gauge theories we refer to the work of
\Brefs{Grassi:2001bz,Grassi:2000dz,Kniehl:1998vy,Kniehl:1998fn}, for complex
poles in general to \Brefs{Stuart:1991xk,Argyres:1995ym,Beenakker:1996kn} and
\Brefs{Jegerlehner:2002em,Jegerlehner:2001fb}.
For alternative approaches we refer to the work of \Brefs{Beneke:2004km,Beneke:2003xh}.

We can summarize by saying that unstable particles are described by irreducible, 
nonunitary representations of the Poincare group, corresponding to the complex 
eigenvalues of the four-momentum  $p^{\mu}$ satisfying the condition 
$p^2= - \mu^2 + i\,\mu\gamma$. For a discussion on violation of the spectral condition
we refer to \Bref{Lukierski:1972tx}.

Consider the complex mass scheme (CMS) introduced in \Bref{Denner:2005fg,Denner:2006ic}
(see also \Bref{Passarino:2000mt}) and extended at two-loop level in \Bref{Actis:2006rc}: here, 
at lowest level of accuracy, we use
\bq
\cph - \mhs + S_{\HH}\lpar \cph,\cpt,\cph,\cpw,\cpz\rpar = 0,
\label{exact}
\eq
where now $S_{\HH}$ is computed at one loop level and $\cpw$ \etc are the experimental 
complex poles, derived from the corresponding (experimental) on-shell masses and widths. For 
the $\PW$ and $\PZ$ bosons the input parameter set (IPS) is defined in terms of 
pseudo-observables; at first, on-shell pseudo-quantities are derived by fitting the 
experimental lineshapes with
\bq
\Sigma_{\VV}(s) = \frac{N}{(s - M^2_{\OS})^2 + 
s^2\,\Gamma^2_{\OS}/M^2_{\OS}}, \qquad V = \PW, \PZ,
\eq
where $N$ is an irrelevant (for our purposes) normalization constant.
Secondly, we define pseudo-observables 
\bq
M_{\ssP} = M_{\OS}\,\cos\psi, \qquad  
\Gamma_{\ssP} = \GOS\,\sin\psi, \qquad
\psi = \arctan \frac{\GOS}{M_{\OS}},
\eq
which are inserted in the IPS.
Renormalization with complex poles should not be confused with a simple recipe for the 
replacement of running widths with constant widths; there are many more ingredients in the scheme.
It is worth noting that perturbation theory based on $\cph$ instead of the on-shell mass has
much better convergence properties; indeed, as shown in \Bref{Ghinculov:1996py}, the two-loop 
corrections to the imaginary part of $\cph$ become as large as the one-loop ones {\em only}
when $\muh = 1.74\UTeV$. This suggests that the complex pole scheme is preferable also
from the point of view of describing the heavy Higgs-boson production at LHC. 

There is a substantial difference between $\PW,\PZ$ complex poles and the Higgs 
complex pole. In the first case $\PW,\PZ$ bosons decay predominantly into two (massless) 
fermions while for the Higgs boson below the $\PW\PW\,$-threshold the decay into 
four fermions is even larger than the decay into a $\PAQb\PQb$ pair, while the other
fermion pairs are heavily suppressed by vanishing Yukawa couplings. 
Therefore we cannot use for the Higgs boson the well-known result, 
\bq
\Im S_{\VV}(s) \approx \frac{\GOSu_{\ssV}}{M^{\OS}_{\ssV}}\,s.
\eq
which is valid for $\PW,\PZ$. As a consequence of this fact we have
\bq
\mu^2_{\ssV} = M^2_{\ssV\,\OS} - \Gamma^2_{\ssV\,\OS} + \HO,
\quad
\gamma_{\ssV} = \GOSu_{\ssV}\,\Bigl[ 1 - \frac{1}{2}\,
\Bigl(\frac{\GOSu_{\ssV}}{M^{\OS}_{\ssV}}\Bigr)^2\Bigr] + \HO,
\label{cpV}
\eq
where the perturbative expansion is well under control since 
$\GOSu_{\ssV}/M^{\OS}_{\ssV} \muchless 1$. For the Higgs boson we have a different expansion,
\bqa
\gh &=& \Gamma_{\PH}\,\bigl[ 1 + \sum_n a_n\,\Gamma^n_{\PH}\bigr] +
\sum_{\ssV=\PW,\PZ}\,\frac{\mv}{\mh}
\sum_{n,l} b^{\ssV}_{nl}\,\Gamma^n_{\ssV}\,\Gamma^l_{\PH},
\nl
a_0 &=& - \bigl[ \Im\,S^h_{\HH\,,\,\OS}\bigr]^2,
\quad
a_1 = \frac{1}{2}\,\Bigl[ 
\frac{1}{\mh}\,\Im\,S^h_{\HH\,,\,\OS}
-\,\mh\,\Im\,S^{hh}_{\HH\,,\,\OS} \Bigr],
\nl
b^{\ssV}_{10} &=& -\,\Re\,S^v_{\HH\,,\,\OS}
+ \Re\,S^v_{\HH\,,\,\OS}\,\Re\,S^h_{\HH\,,\,\OS}
- \Im\,S^v_{\HH\,,\,\OS}\,\Im\,S^h_{\HH\,,\,\OS}
\nl
b^{\ssV}_{11} &=& \frac{1}{2\,\mh}\,\Im\,S^v_{\HH\,,\,\OS} -
\mh\,\Im\,S^{hv}_{\HH\,,\,\OS},
\quad
b^{\ssV}_{20} = - \frac{1}{2}\,\frac{\mv}{\mh}\,\Im\,S^{vv}_{\HH\,,\,\OS}
\label{HO}
\eqa
Here we have $M_i = M^{\OS}_i$, $\Gamma_i = \GOSu_i$ (on-shell quantities) and
\bq
S^{a_1\cdots a_n}_{\HH\,,\,\OS} = 
\frac{\partial^n}{\partial M^2_{a_1}\,\cdots\, \partial M^2_{a_n}}\,
S_{\HH}\lpar \mhs,\mt^2,\mh^2,\mw^2,\mz^2\rpar,
\eq
where only the first few terms in the perturbative expansion are explicitly given.
In \eqn{HO} we used the following definition of the on-shell width
\bq
\GOSu_{\PH}\,M^{\OS}_{\PH} = \Im\,S_{\HH\,,\,\OS} + \HO
\eq
Only the complex pole is gauge-parameter independent to all orders of perturbation 
theory while on-shell quantities are ill-defined beyond lowest order. Indeed, in the 
$R_{\xi}$ gauge, at lowest order, one has the following expression for the 
bosonic part of the Higgs self-energy:
\bq
\Im\,S_{\HH\,,\,\bos}(s) = \frac{g^2}{4\,\mws}\,s^2\,\Bigl[
\lpar \frac{\mhq}{s^2} - 1\rpar\,\lpar 1 - 4\,\xi_{\PW}\,
\frac{\mws}{s}\rpar^{1/2}\,\theta\lpar s - 4\,\xi_{\PW} \mws\rpar
+ \frac{1}{2}\,\lpar \PW \to \PZ\rpar \Bigr],
\label{Imbos}
\eq
where $\xi_{\PV}$ ($\PV = \PW,\PZ$) are gauge parameters. Note that \eqn{HO} (the expansion) 
involves derivatives. 

Starting from \eqn{Imbos} we understand the problem with the on-shell definition of the mass: 
once again, let $\MH$ be the renormalized Higgs mass and $\cph= \muhs - i\,\muh\gh$ the 
parametrization of the complex pole; it follows
\bqa
\MH^2\mid_{\OS} = M^2_{\PH} + \Re\,S_{\HH}(\MH^2),
&\qquad&
\muh^2 = M^2_{\PH} + \Re\,S_{\HH}(\cph),
\nl
\MH\mid_{\OS} - \muh &=& -\,\frac{1}{2}\,\gh\,\Im\,S'_{\HH}(\muh^2),
\eqa
showing that at order $g^4$ the on-shell mass is ill-defined since $\cph$ is $\xi\,$-independent.

A technical remark: suppose that we use $\muh$ as a {\em free} input parameter and derive 
$\gh$ in the SM from the equation
\bq
\muh\,\gh =  \Im\,S_{\HH}\lpar \cph,\cpt,\cph,\cpw,\cpz\rpar.
\label{geq}
\eq
Since the bare $\PW,\PZ$ masses are replaced with complex poles also in couplings (to 
preserve gauge invariance) it follows that $\gh$ (solution of \eqn{geq}) is 
renormalization scale dependent; while counterterms can be introduced to make the
self-energy ultraviolet finite, the $\muR$ dependence drops only in subtracted 
quantities, \ie after finite renormalization, something that would require knowledge of
the experimental Higgs complex pole~\cite{Actis:2006rb,Actis:2006rc}. Following our intuition 
we fix the scale at $\muR = \muh$.

After evaluating the coefficients of \eqn{HO} (\eg in the $\xi = 1$ gauge) it is easily 
seen that the expanded solution of \eqn{exact} is not a good approximation to exact one, 
especially for high values of $\muh$. For instance, for $\muh = 500\UGeV$ we have an exact 
solution $\gh= 58.70\UGeV$, an expanded one of $62.87\UGeV$ with an on-shell with of $68.0\UGeV$.
Therefore, we will use an exact, numerical, solution for $\gh$; details on the
structure of the Higgs self-energy to all orders of perturbation theory can be found
in \Bref{Actis:2006rb,Actis:2006rc}.

It is important to remark that an evaluation of $\gh$ at the same level of accuracy to 
which $\GOSu_{\PH}$ is known (see \Bref{Dittmaier:2011ti}) would require, at least, a 
three-loop calculation (the first instance where we have a four-fermion cut of the $\PH$ 
self-energy). 

Complex poles for unstable $\PW, \PZ, \PH$ and $\PQt$ also tell us that it is 
very difficult for an heavy Higgs boson to come out right within the SM;
these complex poles are solutions of a (coupled) system of equations
but for $\PW, \PZ$ and (partially) $\PQt$ we can compare with the corresponding
experimental quantities. Results are shown in \refT{tab:HTO_1} and clearly indicate
mismatches between predicted and experimental $\PW, \PQt$ complex poles.
Indeed, as soon as $\muh$ increases, it becomes more and more difficult to find complex
$\PW, \PQt$ and $\PZ$ poles with an imaginary part compatible with measurements.

\begin{table}
\begin{center}
\caption[]{\label{tab:HTO_1}{The Higgs boson complex pole at fixed values of
the $\PW, \PQt$ complex poles compared with the complete solution for $\cph, \cpw$ and
$\cpt$}}
\vspace{0.2cm}
\begin{tabular}{rrrr}
\hline 
$\muh\,$[GeV] & $\gamma_{\PW}\,$ [GeV] fixed & $\gamma_{\PQt}\,$ [GeV] fixed &
$\gamma_{\PH}\,$ [GeV] derived \\ 
$200$ & $2.088$ & $1.481$ & $1.355$ \\
$250$ &         &         & $3.865$ \\
$300$ &         &         & $8.137$ \\ 
$350$ &         &         & $14.886$ \\ 
$400$ &         &         & $26.598$ \\ 
\hline
$\muh\,$[GeV] & $\gamma_{\PW}\,$ [GeV] derived & $\gamma_{\PQt}\,$ [GeV] derived &
$\gamma_{\PH}\,$ [GeV] derived \\ 
$200$ & $2.130$ & $1.085$ & $1.356$ \\
$250$ & $2.119$ & $0.962$ & $3.823$ \\
$300$ & $2.193$ & $0.836$ & $8.139$ \\
$350$ & $2.607$ & $0.711$ & $14.653$ \\
$400$ & $3.922$ & $0.566$ & $25.498$ \\
\hline
\end{tabular}
\end{center}
\end{table}

This simple fact raises the following question: what is the physical meaning of 
an heavy Higgs boson search? We have the usual and well-known 
considerations~\cite{Ellis:2009tp}:
a Higgs boson above $600\UGeV$ requires new physics at $1\UTeV$, argument
based on partial-wave unitarity~\cite{Passarino:1990hk,Passarino:1985ax} (which should not be 
taken quantitatively or too literally); violation of unitarity bound possibly implies the presence 
of $J=0,1$ resonances but there is no way to predict their masses, simply scaling the 
$\PGp{-}\PGp$ system gives resonances in the $1\,\UTeV$ range. 
Generally speaking, it would be a good idea to address this search as {\em search} for 
$J=0,1$ heavy resonances decaying into $\PV\PV \to 4\,\Pf$. 
In a model independent approach both $\muh$ and $\gh$ should be kept free in order to perform 
a $2\,$dim scan of the Higgs-boson lineshape. For the high-mass region this remains our
recommended strategy. Once the fits are performed it will be left to theorists to struggle 
with a model-dependent interpretation of the results.

To summarize, we have addressed the following question: what is the common sense 
definition of mass and width of the Higgs boson? We have several options,
\bq
\cph = \muhs - i\,\muh\,\gh,
\quad
\cph = \lpar \mu'_{\PH} - \frac{i}{2}\,\gamma'_{\PH}\rpar^2, 
\quad
\cph = \frac{\mBhs - i\,{\GOL}_{\PH}\,\mBh}
{1 + {\GOL}^2_{\PH}/\mBhs}.
\eq
We may ask which one is correct, approximate or closer to the experimental peak. Here
we have to distinguish: for $\gh \muchless \muh$ $\mBh$ is a good approximation 
to the on-shell mass and it is closer to the experimental peak; for instance, for the $\PZ$ 
boson ${\overline M}_{\PZ}$ is equivalent to the mass measured at Lep.
However, in the high-mass scenario, where $\gh \sim \muh$, the situation 
changes an ${\overline O}_{\PH} \not= O^{\OS}_{\PH}$ for any observable (or pseudo-observable). 
Therefore, the message is: do not use the on-shell width to estimate $\mBh$ in the high-mass
region. 

Missing a calculation of the three-loop Higgs self-energy we can analyze the low-mass region
by using an approximation to the exact Higgs complex-pole which is based on the expansion
of $\gh$ around the best-known on-shell calculation, $\Gamma_p$ from 
{\sc Prophecy4f}~\cite{Prophecy4f}. Therefore, below $\muh = 200\,\UGeV$ we will use
\bq
\frac{\gh}{\Gamma_p} \approx 1 + \frac{1}{2}\,X_{\PW}\,\lpar d_1 - \muhs\,d_2\rpar\,
\frac{\Gamma_p}{\muh} - X^2_{\PW}\,d^2_1,
\label{expI}
\eq
\bq
X_{\PW} = 4\,\sqrt{2}\,\frac{G_{\ssF}\,\mws}{\pi^2}, \qquad
d_n = \frac{\partial^n}{\partial \muh^n}\,S_{\HH}.
\eq
In the expansion all masses but the Higgs mass are kept real and the higher-order effects are
simulated by the expansion parameter, $\Gamma_p/\muh$.
\section{Production and decay\label{Sect_prodec}}
Before describing production and decay of an Higgs boson we underline the general structure
of any process containing a Higgs boson intermediate state. The corresponding amplitude is 
schematically given by
\bq
A(s) = \frac{f(s)}{s - \cph} + N(s),
\eq
where $N(s)$ denotes the part of the amplitude which is non-Higgs-resonant.
Signal ($S$) and background ($B$) are defined as follows:
\bq
A(s) = S(s) + B(s),
\qquad
S(s)= \frac{f(\cph)}{s - \cph}, \quad
B(s)= \frac{f(s) - f(\cph)}{s - \cph} + N(s).
\label{split}
\eq
As a first step we will show how to write $f(s)$ in a way such that pseudo-observables 
make their appearance. Consider the process $i j \to \PH \to \PF$ where $i,j\,\in\,$partons and
$\PF$ is a generic final state; the complete cross-section will be written as follows:
\bq
\sigma_{i j \to \PH \to \PF}(s) = 
\frac{1}{2\,s}\,\int\,d\Phi_{i j \to \PF}\,
\Bigl[ \sum_{s,c} \bmid A_{i j \to \PH} \bmid^2 \Bigr]\,
\frac{1}{\bmid s - \cph\bmid^2}\,
\Bigl[ \sum_{s,c} \bmid A_{\PH \to \PF} \bmid^2 \Bigr],
\eq
where $\sum_{s,c}$ is over spin and colors (averaging on the initial state). 
Note that the background (\eg $\Pg\Pg \to 4\,\Pf$, see 
\Bref{Binoth:2006mf,Campbell:2011cu,Kauer:2012ma}) 
has not been included and, strictly speaking and for reasons of gauge invariance, one should 
consider only the residue of the Higgs-resonant amplitude at the complex pole, as described in 
\eqn{split}. For the moment we will argue that the dominant corrections are the QCD ones 
where we have no problem of gauge parameter dependence. If we decide to keep the Higgs 
boson off-shell also in the resonant part of the amplitude (interference signal/background 
remains unaddressed) then we can write
\bq
\int\,d\Phi_{i j \to \PH}\,\sum_{s,c} \bmid A_{i j \to \PH} \bmid^2 = s\,{\overline A}_{ij}(s).
\eq
For instance, we have
\bq
{\overline A}_{\Pg\Pg}(s) = \frac{\alphas^2}{\pi^2}\,\frac{G_{\ssF}\,s}{288\,\sqrt{2}}\,
\bmid \sum_q\,f\lpar \tau_q\rpar \bmid^2\,\lpar 1 + \delta_{\QCD}\rpar,
\eq
where $\tau_q = 4\,m^2_q/s$, $f(\tau_q)$ is defined in Eq.(3) of \Bref{Spira:1995rr} and where 
$\delta_{\QCD}$ gives the QCD corrections to $\Pg\Pg \to \PH$ up to 
next-to-next-to-leading-order (NNLO) + next-to-leading logarithms (NLL) resummation. 
Furthermore, we define
\bq
\Gamma_{\PH \to \PF}(s) =
\frac{1}{2\,\sqrt{s}}\,\int\,d\Phi_{\PH \to \PF}\,\sum_{s,c} \bmid A_{\PH \to \PF} \bmid^2, 
\label{offw}
\eq
which gives the partial decay width of a Higgs boson of virtuality $s$ into a final 
state $\PF$.
\bq
\sigma_{i j \to \PH} = \frac{{\overline A}_{ij}(s)}{s},
\eq
which gives the production cross-section of a Higgs boson of virtuality $s$.
We can write the final result in terms of pseudo-observables
\bq
\sigma_{i j \to \PH \to \PF}(s) = \frac{1}{\pi}\,
\sigma_{i j \to \PH}\,\frac{s^2}{\bmid s - \cph\bmid^2}\,
\frac{\Gamma_{\PH \to \PF}}{\sqrt{s}}.
\eq
It is also convenient to rewrite the result as 
\bq
\sigma_{i j \to \PH \to \PF}(s) = \frac{1}{\pi}\,
\sigma_{i j \to \PH}\,\frac{s^2}{\bmid s - \cph\bmid^2}\,
\frac{\Gamma^{\tot}_{\PH}}{\sqrt{s}}\,\hbox{BR}\lpar \PH \to \PF\rpar,
\label{QFT}
\eq
where we have introduced a sum over all final states,
\bq 
\Gamma^{\tot}_{\PH} = \sum_{f\in\PF}\,\Gamma_{\PH \to \PF}.
\label{Gtot}
\eq
Note that we have written the phase-space integral for $i(p_1) + j(p_2) \to \PF$ as
\bq
\int\,d\Phi_{i j \to \PF} =
\int\,d^4k\,\delta^4( k - p_1 - p_2)\,
\int\,\prod_f\,d^4p_f\,\delta^+( p^2_f)\,\delta^4( k - \sum_f p_f),
\eq
where we assume that all initial and final states (\eg $\PGg \PGg$, $4\,\Pf$, \etc) 
are massless. 

It is worth noting that the introduction of complex poles does not 
imply complex kinematics. According to \eqn{split} only the residue of the propagator 
at the complex pole becomes complex, not any element of the phase-space integral. 
Details of the procedure are discussed in \refA{appC}.

On a more formal bases one should say that unstable states lie in a natural
extension of the usual Hilbert space that corresponds to the second sheet of
the $S\,$-matrix; these states have zero norm and, therefore, escape the
usual prohibition of having an hermitian Hamiltonian with complex 
energy~\cite{Weldon:1975gu}.
On a more pragmatic level we use the guiding principle that Green's functions 
involving unstable particles should smoothly approach the value for stable 
ones (the usual Feynman $-\,i\,0$ prescription) when the couplings of the theory 
tend to zero. 

If we insist that $|\PH>$ is an asymptotic state in the Hilbert space 
then the observable to consider will be $< \PF\,{\rm out}\,|\,\PH\,{\rm in} >$, 
otherwise one should realize that for stable particles the proof of the LSZ 
reduction formulas~\cite{Lehmann:1957zz} depends on the existence of asymptotic states
\bq
|\,p\,{\bin}\, > = \lim_{t \to -\,\infty}\int\,d^3 x\,H(x)\,i\,
{\partial_t}\!\!\!\!\!^{^{\leftrightarrow}} \,
e^{i\,\spro{p}{x}}\,|\,0\,>,
\eq
(in the weak operator sense). For unstable particles the energy is complex so that this 
limit either diverges or vanishes. Although a modification of the LSZ reduction formulas has 
been proposed long ago for unstable particles, see \Bref{Weldon:1975gu}, we prefer an
alternative approach where one considers extracting information on the Higgs 
boson directly from
\bq
<\,\PF\;{\bout}\,|\,\PH\,>\,<\,\PH\,|\,\PI\;{\bin}\,> +
\sum_{n\,\not=\,\PH}\,
<\,\PF\;{\bout}\,|\,n\,>\,<\,n\,|\,\PI\;{\bin}\,>,
\eq
for some initial state $\PI$ and some final state $\PF$ and where 
$\{n\}\,\oplus\,\PH$ is a complete set of states (not as in the in/out bases). 
As we have seen, the price to be paid is the necessity of moving into the complex plane.
Why do we need pseudo-observables? Ideally experimenters (should) extract so-called 
{\em realistic observables} from raw data, \eg 
$\sigma \lpar \Pp\Pp \to \PGg\PGg + \PX\rpar$ and (should) present results in a form that 
can be useful for comparing them with theoretical predictions, i.e. the results should be 
transformed into pseudo-observables; during the deconvolution procedure one should also account 
for the interference background -- signal;
theorists (should) compute pseudo-observables using the best available technology and satisfying 
a list of demands from the self-consistency of the underlying theory.
\begin{itemize}
\item {\underline{Definition}}
We define an off-shell production cross-section (for all channels) as follows:
\end{itemize}
\bq
\sigma^\prop_{i j \to \ac} =
\frac{1}{\pi}\,\sigma_{i j \to \PH}\,\frac{s^2}{\bmid s - \cph\bmid^2}\,
\frac{\Gamma^{\tot}_{\PH}}{\sqrt{s}}.
\label{sigmaPR}
\eq
When the cross-section $i j \to \PH$ refers to an off-shell Higgs boson
the choice of the QCD scales should be made according to the virtuality 
and not to a fixed value. Therefore, always referring to \refF{HLS:complete},
for the PDFs and $\sigma_{i j \to \PH  + \PX}$ one should select 
$\muFs = \muRs = z\,s/4$ ($z\,s$ being the invariant mass of the
detectable final state). Indeed, beyond lowest order (LO) one must not choose the invariant 
mass of the incoming partons for the renormalization and factorization scales, with the factor 
$1/2$ motivated by an improved convergency of fixed order expansion, but an infrared safe 
quantity fixed from the detectable final state, see \Bref{Collins:1989gx}. The argument is based 
on minimization of the universal logarithms (DGLAP) and not the process-dependent ones\footnote{We 
gratefully acknowledge S.~Forte and M.~Spira for an important discussion on this point.}.

The off-shell Higgs-boson production is currently computed according to the replacement
\bq
\sigma_{\OS}(\muhs)\,\delta( \zeta - \muhs)
\Longrightarrow
\sigma_{\OFS}(\zeta)\,\hbox{BW}(\zeta),
\label{sigmaBW}
\eq
(\eg see \Bref{Alioli:2008tz}) at least at lowest QCD order, where the so-called modified 
Breit--Wigner distribution is defined by
\bq 
\hbox{BW}(s) = \frac{1}{\pi}\,\frac{s\,\GOSu_{\PH}/\muh}
{\lpar s - \muhs\rpar^2 + \lpar s\,\GOSu_{\PH}/\muh\rpar^2},
\label{BWdef}
\eq
where now $\muh = M^{\OS}_{\PH}$ and $\zeta= z\,s$.
This ad-hoc Breit--Wigner cannot be derived from QFT and also is not normalizable
in $[0\,,\,+\infty]$. 
Note that this Breit--Wigner for a running width comes from the 
substitution of $\Gamma \to \Gamma(s) = \Gamma\,s/M^2$ in the Breit--Wigner for a 
fixed width $\Gamma$. This substitution is not justifiable. Its practical purpose is to 
enforce a {\em physical} behavior for low virtualities of the Higgs boson but the usage 
cannot be justified or recommended.
For instance, if one considers $\PV\PV$ scattering and uses this distribution in the 
$s\,$-channel Higgs exchange, the behavior for large values of $s$ spoils unitarity 
cancellation with the contact diagram. 
It is worth noting that the alternative replacement
\bq
\sigma_{\OS}(\muhs)\,\delta( \zeta - \muhs)
\Longrightarrow
\sigma_{\OFS}(\zeta)\,\hbox{BW}(\zeta),
\quad
\hbox{BW}(\zeta) = 
\frac{1}{\pi}\,\frac{\muh\,\GOSu_{\PH}}
{\lpar \zeta - \muhs\rpar^2 + \lpar \muh\,\GOSu_{\PH}\rpar^2}, 
\label{BWFW}
\eq
has additional problems at the low energy tail of the resonance due to the $\Pg\Pg\,$-luminosity,
creating an artificial increase of the lineshape at low virtualities\footnote{We 
gratefully acknowledge S.~Frixione for an important discussion on this point.}.

Another important issue is that $\gh$ which appears in the imaginary part of the inverse 
Dyson-resummed propagator is not the on-shell width since they differ by higher-order 
terms and their relations becomes non-perturbative when the on-shell width becomes of the
same order of the on-shell mass (typically, for on-shell masses above $800\,$GeV).
For nonlinear parameterizations of the scalar sector of the standard model ande Dyson summation 
of the Higgs self energy we refer to the work of \Bref{Schwinn:2004bk}. 

The complex-mass scheme can be translated into a more familiar language by introducing the
Bar-scheme. Using \eqn{CMSprop} with the parametrization of \eqn{CPpar}
we perform the well-known transformation
\bq
\mBhs = \muhs + \gamma^2_{\PH} 
\qquad
\mu_{\sH}\,{\GOL}_{\PH} = \mBh\,\gh.
\label{Bars}
\eq
A remarkable identity follows (defining the Bar-scheme):
\bq
\frac{1}{s - \cph} =
\Bigl( 1 + i\,\frac{{\GOL}_{\PH}}{\mBh}\Bigr)\,
\Bigr( s - \mBhs + 
i\,\frac{{\GOL}_{\PH}}{\mBh}\,s \Bigr)^{-1},
\label{barid}
\eq
showing that the Bar-scheme is equivalent to introducing a running width in the propagator
with parameters that are not the on-shell ones. Special attention goes to the
numerator in \eqn{barid} which is essential in providing the right asymptotic
behavior when $s \to \infty$, as needed for cancellations with contact terms in
$\PV\PV\,$scattering. 
If we compare the result of \eqn{barid} with Eq.(4) of \Bref{Seymour:1995qg} and interpret
$m_{\PH}$ and $\Gamma_{\PH}$ of that equation as the corresponding Bar - scheme quantities
$\mBh$ and $\GOL_{\PH}$ of \eqn{Bars}, we see that the so-called Seymour - scheme of
\Brefs{Anastasiou:2011pi,Anastasiou:2012hx} is exactly giving the Higgs propagator with a
complex pole. There is a second variant for the Higgs propagator in \Bref{Seymour:1995qg},
\ie (in the notation of that paper)
\bq
\Delta_{\PH}(s) = \frac{m^2_{\PH}}{s}\,\Bigl[  
s - m^2_{\PH} + i\,\frac{\Gamma_{\PH}}{m_{\PH}}\,s\Bigr]^{-1}
\eq
which has been used afterwards with the motivation that the prescription accounts for 
signal-background interference effects. 
If we denote by $A_{\PH}$ the Higgs-resonant amplitude and by $A_{\ssB}$ the box
contribution in $\Pg\Pg \to \PZ\PZ$ then the interference is
\bq
A_{\intf} = 2\,\Re \lpar \Delta_{\PH}\,I^{\dagger}\rpar, 
\quad
I = {\overline A}_{\PH}\,A^{\dagger}_{\ssB},
\quad
{\overline A}_{\PH} = (s -\cph)\,A_{\PH}.
\label{interf}
\eq
Note that for large values of the Higgs mass the term in \eqn{interf} proportional to
$\Im\,\Delta_{\PH}$ is not negligible.
The main issue in \Bref{Seymour:1995qg} is on unitarity cancellations at high energy. 
Of course, the behavior of both amplitudes for $s \to \infty$ is known and simple
and any correct treatment of perturbation theory (no mixing of different orders) will 
respect the unitarity cancellations; from this point of view, the numerator in
\eqn{barid} is essential. Since the Higgs boson decays almost completely into 
longitudinal $\PZ$s, for $s \to \infty$ we have~\cite{Glover:1988rg} (for a single quark $q$)
\bq
A_{\PH} \sim \frac{s m^2_q}{2 M^2_{\PZ}}\,\Delta_{\PH}\,\ln^2\frac{s}{m^2_q}
\qquad
A_{\ssB} \sim -\,\frac{m^2_q}{2 M^2_{\PZ}}\,\ln^2\frac{s}{m^2_q}
\eq
showing cancellation in the limit ($s\,\Delta_{\PH} \to 1$). However, the behavior for 
$s \to \infty$ (unitarity) should not/cannot be used to simulate the interference for 
$s < M^2_{\PH}$.
The only relevant message to be derived here is that unitarity requires the interference
to be destructive at large values of $s$, see also \Bref{Valencia:1992ix}.

A sample of numerical results is shown in \refT{tab:HTO_2} where we compare the Higgs
boson complex pole to the corresponding quantities in the Bar-scheme.
\begin{table}
\begin{center}
\caption[]{\label{tab:HTO_2}{Higgs boson complex pole; 
$\GOSu_{\PH}$ is the on-shell width, $\gh$ is defined in \eqn{CPpar} and the Bar-scheme
in \eqn{Bars}.}}
\vspace{0.2cm}
\begin{tabular}{rrrrr}
\hline 
$\muh$[GeV]                                      & 
$\GOSu_{\PH}$[GeV]  (\Bref{Dittmaier:2011ti})    &
$\gh\,$[GeV]                                     &
${\overline M}_{\PH}$[GeV]                       &
$\GOL_{\PH}$[GeV]  \\
\hline 
$200$  & $1.43$   & $1.35$   & $200$     & $1.35$   \\
$400$  & $29.2$   & $25.60$  & $400.9$   & $26.66$  \\
$600$  & $123$    & $103.93$ & $608.9$   & $105.48$ \\
$700$  & $199$    & $162.97$ & $718.7$   & $167.33$ \\
$800$  & $304$    & $235.57$ & $834.0$   & $245.57$ \\
$900$  & $449$    & $320.55$ & $955.4$   & $340.28$ \\
$1000$ & $647$    & $416.12$ & $1083.1$  & $450.71$ \\
\hline
\end{tabular}
\end{center}
\end{table}

In conclusion, the use of the complex pole is recommended even if the accuracy at which
its imaginary part can be computed is not of the same quality as the 
next-to-leading-order (NLO) accuracy of the on-shell width. Nevertheless, the use of a solid 
prediction (from a  theoretical point of view) should always be preferred to the introduction 
of ill-defined quantities (lack of gauge invariance).
\subsection{Schemes\label{schemes}}
\begin{figure}
\vspace{-0.5cm}
\centering
\begin{minipage}{.9\textwidth}
\centerline{\includegraphics[bb=  0 0 595 842,width=14.cm]{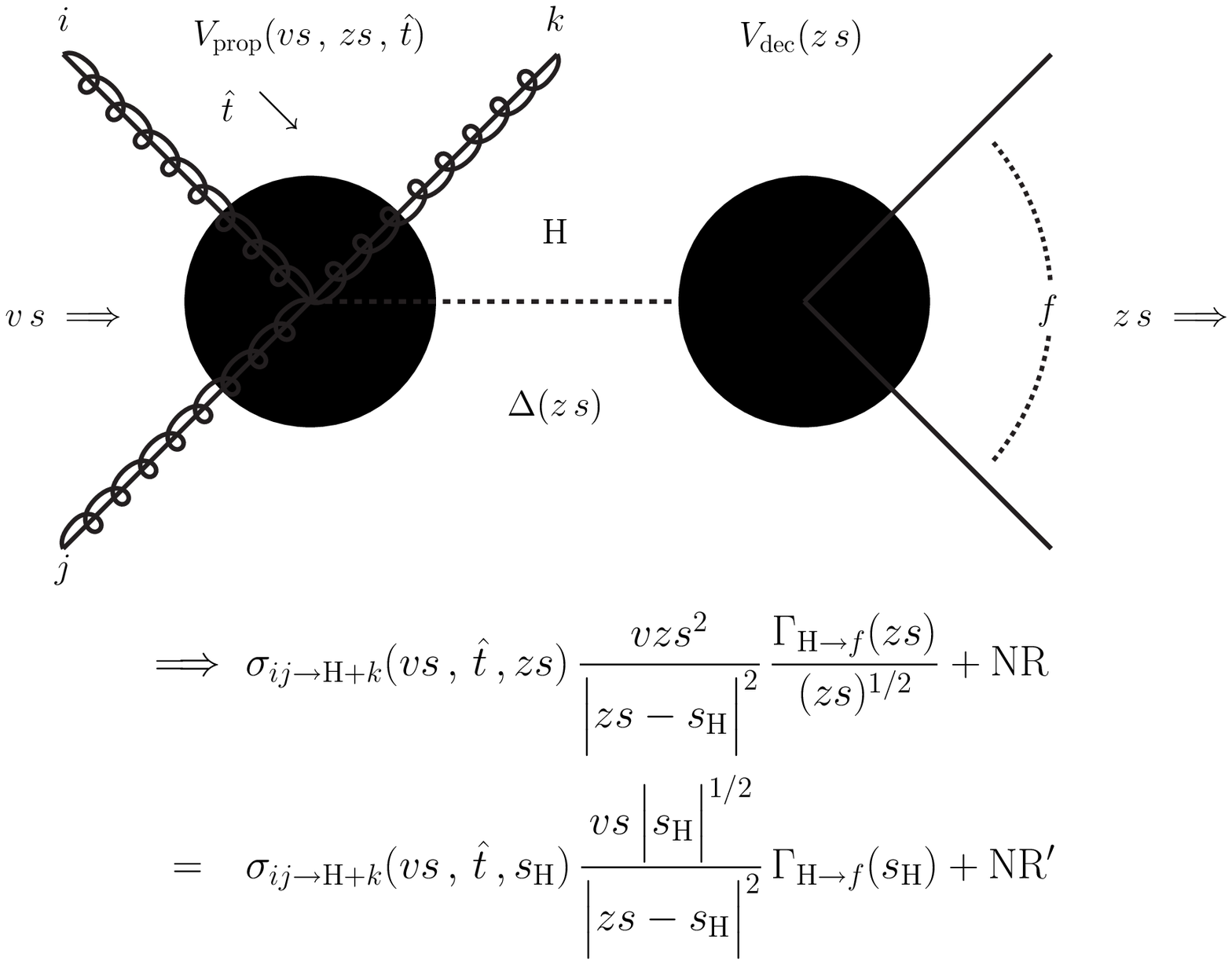}}
\vspace{-8.4cm}
\caption[]{The resonant part of the process $i j \to \PH + k$ where $i,j$ and $k$
are partons ($\Pg$ or $\PQq$). We distinguish between production, propagation 
and decay, indicating the explicit dependence on Mandelstam invariants. If $s$ denotes 
the $\Pp\Pp$ invariant mass we distinguish between $v s$, the $i j$ invariant mass, $z s$, 
the Higgs boson virtuality and ${\hat t}$ which is related to the Higgs boson transverse 
momentum. Within NR we include the Non-Resonant contributions as well as their interference 
with the signal. NR' includes those terms that are of higher order in the Laurent expansion 
of the signal around the complex pole. At NNLO QCD the parton $k$ is a pair $k_1{-}k_2$ and
more invariants are needed to characterize the production mechanism.}
\label{HLS:complete}
\end{minipage}
\end{figure}

We are now in a position to give a more detailed description of the strategy behind
\eqn{split}. Consider the complete amplitude for a given process, \eg the one in
\refF{HLS:complete}; let $\lpar \zeta= z s,\,\dots\rpar$ the the full list of
Mandelstam invariants characterizing the process, then
\bq
A\lpar \zeta,\,\dots\rpar = V_{\myprod}\lpar \zeta,\,\dots\rpar\,
\Delta_{\prop}(\zeta)\,V_{\dec}(\zeta) + N\lpar \zeta,\,\dots\rpar.
\eq
Here $V_{\myprod}$ denotes the amplitude for production, \eg $\Pg\Pg \to \PH(\zeta)  + \PX$,
$\Delta_{\prop}(\zeta)$ is the propagation function, $V_{\dec}(\zeta)$ is the
amplitude for decay, \eg $\PH(\zeta) \to 4\,\Pf$.
If no attempt is made to split $A(s)$ no ambiguity arises but, usually, the two components 
are known at different orders. Ho to define the signal? The following schemes are available:
\begin{description}
\item[ONBW]
\bq
S\lpar \zeta,\,\dots\rpar = 
V_{\myprod}\lpar \muhs,\,\dots\rpar\,\Delta_{\prop}(\zeta)\,V_{\dec}(\muhs)
\qquad
\Delta_{\prop}(\zeta) = \hbox{Breit--Wigner}.
\eq
in general violates gauge invariance, neglects the Higgs off-shellness and introduces the
ad hoc Breit--Wigner of \eqn{BWdef}.
\item[OFFBW]
\bq
S\lpar \zeta,\,\dots\rpar = 
V_{\myprod}\lpar \zeta,\,\dots\rpar\,\Delta_{\prop}(\zeta)\,V_{\dec}(\zeta),
\qquad
\Delta_{\prop}(\zeta) = \hbox{Breit--Wigner},
\label{offbw}
\eq
in general violates gauge invariance, and introduces an ad hoc Breit--Wigner.
\item[ONP]
\bq
S\lpar \zeta,\,\dots\rpar = 
V_{\myprod}\lpar \muhs,\,\dots\rpar\,\Delta_{\prop}(\zeta)\,V_{\dec}(\muhs),
\qquad
\Delta_{\prop}(\zeta) = \hbox{propagator}
\eq
in general violates gauge invariance and neglects the Higgs off-shellness.
\item[OFFP]
\bq
S\lpar \zeta,\,\dots\rpar = 
V_{\myprod}\lpar \zeta,\,\dots\rpar\,\Delta_{\prop}(\zeta)\,V_{\dec}(\zeta),
\qquad
\Delta_{\prop}(\zeta) = \hbox{propagator},
\label{offp}
\eq
in general violates gauge invariance
\item[CPP]
\bq
S\lpar \zeta,\,\dots\rpar = 
V_{\myprod}\lpar \cph,\,\dots\rpar\,\Delta_{\prop}(\zeta)\,V_{\dec}(\cph),
\qquad
\Delta_{\prop}(\zeta) = \hbox{propagator}.
\label{cpp}
\eq
This scheme respects all requirements: only the pole, the residue and the reminder of 
$A(\zeta)$ are gauge invariant. Furthermore the CPP-scheme allows to identify POs like 
the production cross-section and any partial decay width by putting in one-to-one correspondence 
robust theoretical quantities and experimental data.
\end{description}
From the list above it follows that only CPP-scheme should be used, once the
signal/background interference becomes available. However, the largest part of the available 
calculations is not yet equipped with Feynman integrals on the second Riemann sheet 
(complex poles lie there); furthermore, the largest corrections in the production are from 
QCD and we could argue that gauge invariance is not an issue over there, so that the OFFP-scheme 
remains, at the moment, the most pragmatic alternative. Additional considerations with more 
detailed descriptions are however postponed until \refS{Sect_gi}.

Implementation of the CPP-scheme requires some care in the presence of four-leg
processes (or more). The natural choice is to perform analytical continuation
from real kinematics to complex invariants; consider, for instance, the process
$\Pg(p_1) + \Pg(p_2) \to \Pg(p_3) + \PH(p_4) $ where the external $\PH$ leg is continued from 
real on-shell mass to $\cph$. For the three invariants, 
\bq
s = -\,\lpar p_1 + p_2\rpar^2, \qquad
t = -\,\lpar p_1 - p_4\rpar^2, \qquad
u = -\,\lpar p_1 - p_3\rpar^2, 
\eq
we use the following continuation: 
\bq
t= -\,\frac{s}{2}\,\lpar 1 - \frac{\cph}{s}\rpar\,\lpar 1 - \cos\theta\rpar,
\qquad
u= -\,\frac{s}{2}\,\lpar 1 - \frac{\cph}{s}\rpar\,\lpar 1 + \cos\theta\rpar,
\label{mACone}
\eq
where $\theta$ is the (real) scattering angle (between $p_2$ and $p_3$) and $s = 4\,E^2$, where 
$2\,E$ is the (real) energy of center-of-mass system.
For the process $\Pg(p_1) + \Pg(p_2) \to \PH(p_3) + \PH(p_4) $ where both external $\PH$ legs 
are continued from real on-shell mass to $\cph$ we obtain
\bq
t = -\,\frac{s}{2}\,\Bigl[ 1 - \frac{\cph}{s} - 
\bigl( 1 - 4\,\frac{\cph}{s} \bigr)^{1/2}\,\cos\theta \Bigr]
\quad
u = -\,\frac{s}{2}\,\Bigl[ 1 - \frac{\cph}{s} + 
\bigl( 1 - 4\,\frac{\cph}{s} \bigr)^{1/2}\,\cos\theta \Bigr].
\label{mACtwo}
\eq
For a more detailed discussion of the analytic continuation we refer to \refA{appCa}.
Here we observe that when the problem is with extracting pseudo-observables, \eg 
$\Gamma_{\PH \to 4\,\Pf}$, analytic continuation is performed only after integration over 
all variables but the Higgs virtuality. The role of pseudo-observables is then very similar to
what one had for Lep physics, \eg the hadronic and leptonic peak cross-sections
\bq
\sigma^0_{\Phad} = 12\,\pi\,\frac{\Gamma_{\Pe}\,\Gamma_{\Phad}}{M^2_{\PZ}\,\Gamma^2_{\PZ}},
\qquad
\sigma^0_{\Pl} = 12\,\pi\,\frac{\Gamma_{\Pe}\,\Gamma_{\Pl}}{M^2_{\PZ}\,\Gamma^2_{\PZ}},
\eq
where the POs are fully inclusive. Already at that time it was clear that the expression
for realistic observables (RO) at arbitrary $s$ requires a careful examination because
of non triviality in off-shell gauge invariance, see \Bref{Grunewald:2000ju}.

To summarize, our proposal aims to support the use of \eqn{QFT} for producing
accurate predictions for the Higgs lineshape in the SM; \eqn{QFT} can be adapted easily for 
a large class of extensions of the SM. 
Similarly \eqn{sigmaPR} should be used to define the Higgs-boson production cross-section
instead of $\sigma_{\OS}$ or $\sigma_{\OFS}$ of \eqn{sigmaBW}.

To repeat an obvious argument the zero-width approximation for the Higgs boson is usually
reported in comparing experimental studies and theoretical predictions. The approximation
has very low quality in the high-mass region where the Higgs boson has a non negligible
width. An integration over some distribution is a more accurate estimate of the signal
cross-section and we claim that this distribution should be given by the complex propagator
and not by some ad hoc Breit--Wigner.

As we have already mentioned most of this paper is devoted to studying the Higgs-boson
lineshape in the high-mass region. However, nothing in the formalism is peculiar to that
application and the formalism itself forms the basis for extracting pseudo-observables from 
experimental data. This is a timely contribution to the relevant literature.
\subsection{Production cross-section}
Before presenting detailed numerical results and comparisons we give the complete definition of
the production cross-section; let us define $\zeta= z\,s$, $\kappa= v\,s$, and write
\bq
\sigma^{\myprod} = 
\sum_{i,j}\,\int \hbox{PDF}\,\otimes\,\sigma^{\myprod}_{i j \to \ac} = 
\sum_{i,j}\,\int_{z_0}^1 dz \int_z^1 \frac{dv}{v}\,{\cal L}_{ij}(v)
\sigma^\prop_{ij \to \ac}(\zeta, \kappa, \muR, \muF),
\label{PDFprod_1}
\eq
where $z_0$ is a lower bound on the invariant mass of the $\PH$ decay products,
the luminosity is defined by
\bq
{\cal L}_{ij}(v) = \int_v^1 \frac{dx}{x}\,
f_i\lpar x,\muF\rpar\,f_j\lpar \frac{v}{x},\muF\rpar,
\label{PDFprod_2}
\eq
where $f_i$ is a parton distribution function and
\bq
\sigma^\prop_{ij \to \ac}(\zeta, \kappa, \muR, \muF) = \frac{1}{\pi}\, 
\sigma_{ij \to \PH  + \PX}(\zeta, \kappa, \muR, \muF)\,\frac{ \zeta\,\kappa}{\bmid \zeta - \cph\bmid^2}\,
\frac{\Gamma^{\tot}_{\PH}(\zeta)}{\sqrt{\zeta}}.
\label{PDFprod_3}
\eq
Therefore, $\sigma_{ij \to \PH  + \PX}(\zeta, \kappa, \muR)$ is the cross section for two
partons of invariant mass $\kappa$ ($z \le v \le 1$) to produce a final state containing 
a $\PH$ of virtuality $\zeta= z\,s$ plus jets (X); it is made of several 
terms (see \Bref{Spira:1995rr} for a definition of $\Delta\sigma$),
\bq
\sum_{ij}\,\sigma_{ij \to \PH  + \PX}(\zeta, \kappa, \muR, \muF) = 
\sigma_{\Pg\Pg \to \PH}\,\delta\lpar 1 - \frac{z}{v}\rpar + \frac{s}{\kappa}\,\lpar
\Delta\sigma_{\Pg\Pg \to \PH\Pg} + \Delta\sigma_{\PQq\Pg \to \PH\PQq} + 
\Delta\sigma_{\PAQq\PQq \to \PH\Pg} + \mbox{NNLO}\rpar.  
\label{PDFprod_4}
\eq
As a technical remark the complete phase-space integral for the process
$\hatp_i + \hatp_j \to p_k + \{f\}$ ($\hatp_i = x_i\,p_i$ \etc) is written as
\bqa
\int\,d\Phi_{ij \to f} &=& 
\int\,d\Phi_{\myprod}\,\int\,d\Phi_{\dec} =
\int d^4 p_k\,\delta^+(p^2_k)\,\prod_{l=1,n} d^4 q_l\,\delta^+(q^2_l)\,
\delta^4\lpar \hatp_i + \hatp_j - p_k - \sum_l q_l\rpar 
\nl
{}&=&
\int d^4k d^4Q\,\delta^+(p^2_k)\,\delta^4 \lpar \hatp_i + \hatp_j - p_k - Q\rpar\, 
\int \prod_{l=1,n} d^4q_l\,\delta^+(q^2_l)\,\delta^4 \lpar Q - \sum_l q_l\rpar
\eqa
where $\int\,d\Phi_{\dec}$ is the phase-space for the process $Q \to \{f\}$ and
\bqa
\int\,d\Phi_{\myprod} &=& s\,\int dz\,\int d^4p_k d^4Q\,\delta^+(p^2_k)\,
\delta\lpar Q^2 - \zeta\rpar\,\theta(Q_0)\,
\delta^4 \lpar \hatp_i + \hatp_j - p_k - Q\rpar
\nl
{}&=&
s^2\,\int dz dv d{\hat t}\,\int d^4p_k d^4Q\,\delta^+(p^2_k)\,
\delta\lpar Q^2 - \zeta\rpar\,\theta(Q_0)\,
\delta^4 \lpar \hatp_i + \hatp_j - p_k - Q\rpar
\nl
{}&\times&
\delta\lpar (\hatp_i + \hatp_j)^2 - \kappa\rpar\,
\delta\lpar (\hatp_i + Q)^2 - {\hat t}\rpar.
\eqa
\eqnsc{PDFprod_1}{PDFprod_3} follow after folding with PDFs of argument $x_i$ and $x_j$, after 
using $x_i = x$, $x_j = v/x$ and after integration over ${\hat t}$. At NNLO there is an 
additional parton in the final state and five invariants are need to describe the partonic 
process, plus the $\PH$ virtuality. However, one should remember that at NNLO use is made 
of the effective theory approximation where the Higgs-gluon interaction is described by a 
local operator. Our variables $z, v$ are related to \POWHEG\, 
parametrization~\cite{Alioli:2008tz}, $Y, \xi$, by $Y= \ln(x/\sqrt{z})$ and $v= z/(1-\xi)$. 

As already mentioned perturbation theory breaks down in for the complex pole when $\muh = 
1.74\UTeV$ (around $930\UGeV$ for on-shell mass). When we consider the lineshape for 
$\muh = 900\UGeV$ ($\GOS = 647\UGeV$) in a window of, at least, $\pm 2\,\GOS$ around the peak 
we go above $2\UTeV$ in the Higgs virtuality; it is worth noting that for $\muh= 1.5(2)\UTeV$ 
the on-shell width is $3.38(15.8)\UTeV$; these results follows from using 
{\sc Prophecy4f}~\cite{Prophecy4f} which includes two-loop leading corrections and it is worth
observing that the LO result at $\muh= 2\UTeV$ is $3.7\UTeV$ with a $58\%$ one-loop
corrections and an estimated $270\%$ two-loop contribution/uncertainty.    

These considerations give additional support to the CPP - scheme as the only viable option
for defining the Higgs signal in the very high mass region; indeed the $\Gamma^{\tot}_{\PH}(s)$
in \eqn{QFT} is replaced by a $\Gamma^{\tot}_{\PH}(\cph)$ in the CPP - scheme.
The OFFP-scheme remains the most pragmatic alternative but the tail of an heavy Higgs boson 
lineshape extending above $1.3{-}1.5\UTeV$ should be taken with due caution.
\section{The issue of gauge invariance\label{Sect_gi}}
As discussed in \refS{Sect_prop} only the CPP-scheme of \eqn{cpp} should be used and the
background should be included in the calculation. 
Let us consider the {\em pragmatic} OFFP-scheme of \eqn{offp} in more detail: here
we use the Higgs propagator with its complex pole but production and decay are computed at 
arbitrary Higgs virtuality and not at the complex pole.
As far as LO production is concerned, \eg the $\Pg\Pg\PH$ one-loop fermion triangle, there
is never an issue of gauge-parameter dependence in going off-shell; in this respect higher
order QCD corrections are not a problem.

Consider now the decay, \ie $\Gamma^{\tot}_{\PH}$ in \eqn{sigmaPR}: the
amplitude $A(\PH \to \PF)$, for each final state $|\,\PF\,>$ and as long as we include the 
complete set of diagrams at one-loop order, is gauge-parameter independent if the Higgs boson is
on its mass-shell. However, as soon as we put an external leg off-shell, the amplitude
must be coupled to the corresponding physical source $\PI$, \ie $A(\PI \to \PH \to \PF)$,
and only the complete process $\PI \to \PF$ is gauge-parameter independent. The latter does not 
exclude the existence of subsets of diagrams that satisfy the requirement but this can only
be examined case-by-case. To rephrase it, if the Higgs boson is off shell, in LO and NLO QCD 
in most cases the matrix element still respects gauge invariance, but in NLO EW gauge invariance 
is lost, unless the CPP-scheme is used.

How to deal with the OFFP-scheme? Technically speaking, we have a matrix element 
\bq
\Gamma\lpar \PH \to \PF \rpar = f\lpar s\,,\,\muhs\rpar,
\eq
where $s$ is the virtuality of the external Higgs boson, $\muh$ is the mass of internal 
Higgs lines and Higgs wave-function renormalization has been included.  
The following happens: $f(\cph\,,\,\cph)$ is gauge-parameter independent (CPP-scheme)
to all orders while $f(\muhs\,,\,\muhs)$ is gauge-parameter independent at one-loop but not
beyond, $f(s\,,\,\muhs)$ is not. In order to account for the off-shellness of
the Higgs boson we defined the OFFP-scheme by choosing (at one loop level) $f(s\,,\,s)$, \ie
we intuitively replace the on-shell decay of the Higgs boson of mass $\muh$ with the
{\em on-shell} decay of an Higgs boson of mass $\sqrt{s}$ and not with the off-shell decay
of an Higgs boson of mass $\muh$. The same applies for the NLO EW correction to production.

The proof that CPP-scheme satisfies gauge-parameter independence can be sketched as follows:
\bq
S(s) = \frac{A_{\myprod}(\cph)\,A_{\dec}(\cph)}{\bigl[ 1 + S'_{\HH}(\cph)\bigr]\,
(s - \cph)},
\qquad
\frac{\partial}{\partial\,\xi}\,A_{\myprod,\dec}(\cph)\,
\bigl[1 + S'_{\HH}(\cph)\bigr]^{-1/2} = 0,
\label{NI}
\eq
where $\xi$ is an arbitrary gauge parameter, $S'_{\HH}(\cph)$ is the derivative of
$S_{\HH}(s)$ computed at $s = \cph$ and $A_{\myprod,\dec}$ is the amplitude for
production/decay including vertices and wave-function renormalization constants for the
external lines. Note that \eqn{NI} follows from the use of Nielsen 
identities, see \Bref{Gambino:1999ai}. An example is discussed in \refA{appA}.

At one loop the renormalized Higgs self-energy, computed in the $R_{\xi}\,$-gauge, can 
be written as
\bq
S^{(1)}_{\HH}\lpar s\,,\,\MH\,,\,\xi\rpar = 
S^{(1)}_{\HH}\lpar s\,,\,\MH\,,\,\xi=1\rpar +
(s - \MH^2)\,\Delta S^{(1)}_{\HH}\lpar s\,,\,\MH\,,\,\xi\rpar.
\eq 
Finally, to clarify the statement that $f(\muhs\,,\,\muhs)$ may violate gauge invariance
beyond one loop we consider the $\PH \to \PGg\PGg$ decay. The LO amplitudes depends on the 
bare Higgs mass and standard renormalization, introducing the on-shell mass through the real 
part of the one-loop self-energy, leads to a violation of the Ward-Slavnov-Taylor identities 
(WSTI), see \Brefs{Veltman:1970nh,Taylor:1971ff,Slavnov:1972fg}; this will happen at two-loop 
above the $\PW\PW$ threshold. Therefore, the real part of the Higgs self-energy 
stemming from mass renormalization must be traded for the complex expression, even if the 
external Higgs boson is assumed to be an on-shell particle.

\eqn{NI} gives an additional argument against Breit-Wigner distributions: only the propagator
\bq
\frac{1}{s - \MH^2 + S_{\HH}(s)} =
Z_{\HH}^{-1/2}\,\frac{1}{s -\cph}\,Z_{\HH}^{-1/2},
\qquad
Z_{\HH} = 1 + \frac{S_{\HH}(s) - S_{\HH}(\cph)}{s - \cph},
\eq
reproduces the correct factors to be combined with production and decay vertices.

To conclude the discussion we summarize few well-known points which, however, are not always taken 
into proper account. Consider the partial decay width $\PH \to \PW\PW$, hiding the
fact that also the $\PW$s are unstable objects; one way of constructing it is to consider
the amplitude for the process and derive at tree-level the following expression:
\bqa
\Gamma^{\OS}_{\PH \to \PW\PW} &=& \frac{1}{\MH}\,\int\,d\Phi_{\PH \to \PW\PW}\,
\bmid A_{\PH \to \PW\PW} \bmid^2
\nl
{} &=& \frac{G_{\ssF} \MH^3}{8 \sqrt{2} \pi}\,
\Bigl( 1 - 4\,\frac{\MW^2}{\MH^2} + 12\,\frac{\MW^4}{\MH^4} \Bigr)\,
\beta_{\PW}(\MH^2),
\quad 
\beta^2_{\PW}(\MH^2) = 1 - 4\,\frac{\MW^2}{\MH^2}.
\label{defone}
\eqa
Next we compute the imaginary part of the one-loop $\PH$ self-energy due to internal
charged lines. We obtain:
\bqa
\Im\,S^{\ch}_{\HH}(s) &=& s^{1/2}\,\Gamma^{\ch,\inv}_{\PH}(s)\,\theta\lpar s - 4\,\MW^2\rpar +
\Delta^{\ch}_{\PH}(s)\,\theta\lpar s - 4\,\xi\,\MW^2\rpar,
\nl
\Gamma^{\ch,\inv}(s) &=& \frac{G_{\ssF} s^{3/2}}{8 \sqrt{2} \pi}\,
\Bigl( 1 - 4\,\frac{\MW^2}{s} + 12\,\frac{\MW^4}{s^2} \Bigr)\,
\beta_{\PW}(s),
\nl
\Delta^{\ch}_{\PH}(s) &=& -\,\frac{G_{\ssF} s^2}{8 \sqrt{2} \pi}\,   
\bigl( 1 - \frac{\MH^4}{s^2}\bigr)\,\bigl( 1 - 4\,\xi\,\frac{\MW^2}{s}\bigr)^{1/2}.
\label{deftwo}
\eqa
Unitarity of the theory requires proportionality between $\Gamma^{\OS}_{\PH \to \PW\PW}$ and
$\Im\,S^{\ch}_{\HH}$. This is never a problem for the fermion final states but we
see already the paradox in the bosonic sector: from one side we compute $\PH \to \PW\PW$,
from the other all charged bosonic lines (physical or not) circulate in the loop.
It is trivially true that
\bq
\Gamma^{\OS}_{\PH \to \PW\PW} = \Gamma^{\ch,\inv}_{\PH}(\MH^2) =
\frac{1}{\MH}\,\Im\,S^{\ch}_{\HH}(\MH^2),
\eq
a result that satisfies unitarity and gauge invariance; however, going off-shell is different.
There are matters of principle that do not matter in practice: if we consider a `light' Higgs,
say below $200\UGeV$, the difference between the on-shell width computed in the
$R_{\xi}\,$-gauge and $\gamma_{\PH}$ is tiny and, aside from the neighborhoods of unphysical 
thresholds ($\xi = \MH^2/(4 \MW^2),\xi = \MH^2/(4 \MZ^2)$), negligible over a wide range of 
values of the gauge parameter $\xi$~\cite{Kniehl:1998fn}. However, consider a typical
implementation of the $\PH$ production cross-section:
\bq
\int ds \frac{s^{1/2}\,\Gamma_{\PH}(s)}{\pi}\,
\frac{\sigma_{ij \to \PH}(s)}{(s - \MH^2)^2 + \MH^2\,\Gamma_{\PH}(\MH^2)}.
\label{someimp}
\eq
What it is meant by $\Gamma_{\PH}(\MH^2)$ in the denominator of \eqn{someimp}?
Since we are dealing with the denominator of the propagation function for a Higgs boson
of virtuality $s$ and since we would like to respect unitarity, it has to be interpreted
as the imaginary part of the self-energy. At one loop, for $s = \MH^2$ gauge invariance
is not violated. But what about the $\Gamma_{\PH}(s)$ that appears in the numerator? 
Again, because of unitarity, it should be interpreted as the imaginary part of the off-shell
self-energy, \ie the self-energy for a $\PH$ of mass $\MH$ and virtuality $s$; however, this 
quantity has a $\xi\,$-dependent part which is a function of the Higgs virtuality. Therefore, 
no matter how small is the on-shell width for a light Higgs boson, there will be an
$s$ and $\xi\,$-dependent distortion of the lineshape.
Of course there will never be a problem for a calculation assembling the whole process in
one stroke; however, defining the signal is another story. 

We conclude this Section by showing explicitly the delicate point in extracting the
pseudo-observable {\em decay width} from the complete amplitude. We have
\bqa
S(s) &=& 
\Bigl[ 1 + \frac{S_{\HH}(s) - S_{\HH}(\cph)}{s - \cph} \Bigr]^{-1}\,A_{\myprod}(s)\,
\,\frac{1}{s - \cph}\,A_{\dec}(s)
\nl
{}&=& A_{\myprod}(s)\,
\Bigl[ 1 + \frac{S_{\HH}(s) - S_{\HH}(\cph)}{s - \cph} \Bigr]^{-1/2}
\,\frac{1}{s - \cph}\,
\Bigl[ 1 + \frac{S_{\HH}(s) - S_{\HH}(\cph)}{s - \cph} \Bigr]^{-1/2}
\nl
{}&\times& \Bigl\{ A_{\dec}(\cph) + \Bigl[ A_{\dec}(s) - A_{\dec}(\cph)\Bigr]\Bigr\}
\nl
{}&=& S_{\myprod}(s)\,\frac{1}{s - \cph}\,S_{\dec}(s),
\eqa
where the matrix element for the {\em decay} can be decomposed into
\bqa
S_{\dec}(s) &=& \Bigl[ 1 + \frac{1}{2}\,X_{\HH}(s) \Bigr]\,Z^{-1/2}_{\PH}\,A_{\dec}(\cph) +
\Bigl[ 1 - \frac{1}{2}\,Y_{\HH}(s) \Bigr]\,\Bigl[ A_{\dec}(s) - A_{\dec}(\cph)\Bigr],
\nl
X_{\HH}(s) &=& S'_{\HH}(\cph) - \frac{S_{\HH}(s) - S_{\HH}(\cph)}{s - \cph},
\qquad
Y_{\HH}(s) = \frac{S_{\HH}(s) - S_{\HH}(\cph)}{s - \cph}.
\eqa
The decay width is defined in terms of $Z^{-1/2}_{\PH}\,A_{\dec}(\cph)$ while the rest
is $\ord{s - \cph}$.
\section{Numerical results\label{Sect_num}}
In the following we will present numerical results obtained with the program 
\HTO{} (G.~Passarino, unpublished) that allows for the study of the Higgs--boson-lineshape, in
gluon-gluon fusion (ggF), using complex poles. \HTO{} is a FORTRAN $95$ program that contains a 
translation of the subroutine {\tt HIGGSNNLO} written by M.~Grazzini for computing the total 
(on-shell) cross-section for Higgs-boson production (in ggF) at NLO and 
NNLO~\cite{deFlorian:2009hc,Grazzini:2008tf,Catani:2001ic} (\HTO{} has a library for
one-loop integrals that extends the one of \Bref{Montagna:1998kp}). 
It is worth noting that the ggF production mechanism dominates up to $\muh = 700\UGeV$ but
for higher masses the vector-boson fusion (VBF) mechanism starts do compete, with a ratio
ggF/VBF of $1.11$ at $1\UTeV$.
The following acronyms will be used:
\begin{description}
\item[FW] Breit--Wigner Fixed Width (\eqn{BWFW})
\item[RW] Breit--Wigner Running Width (\eqn{BWdef})
\item[OS] parameters in the On-Shell scheme
\item[Bar] parameters in Bar-scheme (\eqn{Bars})
\item[FS] QCD renormalization (factorization) scales fixed
\item[RS] QCD renormalization (factorization) scales running 
\end{description}
All results in this paper (but those in \refT{tab:HTO_12}) refer to $\sqrt{s} = 7\UTeV$ and 
are based on the MSTW2008 PDF sets~\cite{Martin:2009iq}. For complex $\PW,\PZ$ poles we use 
\eqn{cpV} with HXSWG standard input. For the $\PQt$ complex pole we use~\cite{Kniehl:2008cj} 
\bq
\sqrt{\cpt}= M^{\OS}_{\PQt} - \frac{i}{2}\,\Gamma^{\OS}_{\PQt},
\eq
with a LO on-shell width
\bq
\Gamma^{\OS}_{\PQt} =
\frac{G_{\ssF}}{8\sqrt{2}\pi}\,\lpar M_{\PQt}^2-\MW^2\rpar^2\,
\frac{M_{\PQt}^2+2\,\MW^2}{M_{\PQt}^3},
\eq
with an on-shell mass $M_{\PQt} = M^{\OS}_{\PQt} = 172.5\UGeV$.

The integrand in \eqn{PDFprod_1} has several peaks; the most 
evident is in the propagator and in \HTO{} a change of variable is performed,
\bqa
\zeta &=& \frac{1}{1+(\muh/\gh)^2}\,\Bigl[ \muhs + \muh\,\gh\,
\tan\lpar \muh \gh\,z'\rpar \Bigr],
\nl
z' &=& \frac{s}{\muh \gh}\,\Bigl[ \lpar a_m + a_{\ssM}\rpar\,\zeta - a_m \Bigr],
\quad a_m = \arctan\frac{\muhs-z_0 s}{\muh \gh},
\quad a_{\ssM} = \arctan\frac{s-\muhs}{\muh \gh}.
\eqa
Similarly, another change is performed
\bq
v = \exp\{ \lpar 1 - u^2 \rpar\,\ln z\}, 
\qquad
x = \exp\{ \lpar 1 - y^2 \rpar\,\ln v\}. 
\eq
In comparing the OFFBW-scheme with the OFFP-scheme one should realize that there are
two sources of difference, the functional form of the distributions and the different
numerical values of the parameters in the distributions. To understand the impact of
the functional form we have performed a comparison where (unrealistically)
$\gh = \GOSu_{\PH}$ is used in the Higgs propagator; results are shown 
in \refT{tab:HTO_3}.
\begin{table}
\begin{center}
\caption[]{\label{tab:HTO_3}
{The re-weighting factor $w$ for the (total) integrated Higgs lineshape,
$\sigma^{\prop}/\sigma^{\mBW}$ using $\gh = \GOSu_{\PH}$
in \eqn{sigmaPR}.}}
\vspace{0.2cm}
\begin{tabular}{crllll}
\hline 
$\muh\;$[GeV]               & 
$\GOSu_{\PH}\;$[GeV]        &
$\sigma^{\OS}\;$[pb]        &
$\sigma^{\mBW}\;$[pb]       &
$\sigma^{\prop}\;$[pb]      &
$w$                        \\
\hline
$ 200 $ & $ 1.43  $ & $ 5.249  $ & $ 5.674  $ & $ 5.459  $ & $ 0.962 $    \\
$ 300 $ & $ 8.43  $ & $ 2.418  $ & $ 2.724  $ & $ 2.585  $ & $ 0.949 $    \\
$ 400 $ & $ 29.2  $ & $ 2.035  $ & $ 1.998  $ & $ 1.927  $ & $ 0.964 $    \\
$ 500 $ & $ 68.0  $ & $ 0.8497 $ & $ 0.8108 $ & $ 0.7827 $ & $ 0.965 $    \\
$ 600 $ & $ 123.0 $ & $ 0.3275 $ & $ 0.3231 $ & $ 0.3073 $ & $ 0.951 $    \\
\hline
\end{tabular}
\end{center}
\end{table}

In \refT{tab:HTO_4} we compare the on-shell production cross-section as given
in \Bref{Dittmaier:2011ti} with the off-shell cross-section in the OFFBW-scheme
(\eqn{offbw}) and in the OFFP-scheme (\eqn{offp}).
The total Higgs boson width, present in \eqn{Gtot}, is computed by interpolating 
the tables of \Bref{Dittmaier:2011ti}.

An important question regarding the numerical impact of a calculation where the
on-shell Higgs boson is left off-shell and convoluted with some distribution is
how much of the effect survives the inclusion of theoretical uncertainties. In the
OFFBW-scheme we can compare with the results of \Bref{Anastasiou:2011pi}, at least
for values of the Higgs boson mass where their results are available, \ie below $300\UGeV$. 
We compare with Tables~$2{-}3$ of \Bref{Dittmaier:2011ti} at $300\,$GeV and obtain the 
results shown in \refT{tab:HTO_5} where only the QCD scale uncertainty is included. At these 
values of the Higgs boson mass both on-shell production and off-shell production, sampled over 
a Breit--Wigner, are compatible within the errors. The OFFP-scheme gives a slightly higher 
central value of $2.86\Upb$.
The comparison shows drastically different results for high values of the mass
where the difference in central values is much bigger than the theoretical
uncertainty.
\begin{table}
\begin{center}
\caption[]{\label{tab:HTO_4}{Comparison of the on-shell production cross-section as given
in \Bref{Dittmaier:2011ti} with the off-shell cross-section in the OFFBW-scheme
(\eqn{offbw}) and in the OFFP-scheme (\eqn{offp}).}}
\vspace{0.2cm}
\begin{tabular}{crllll}
\hline 
$\muh$[GeV]               & 
$\GOSu_{\PH}$[GeV]        &
$\gh$[GeV]                &
$\sigma^{\OS}$[pb]        &
$\sigma^{\mBW}$[pb]       &
$\sigma^{\prop}$[pb]      \\
\hline
$ 500 $ & $ 68.0  $ & $ 60.2 $ & $ 0.8497  $ & $ 0.8239  $ & $ 0.9367 $    \\
$ 550 $ & $ 93.1  $ & $ 82.8 $ & $ 0.5259  $ & $ 0.5161  $ & $ 0.5912 $    \\
$ 600 $ & $ 123   $ & $ 109  $ & $ 0.3275  $ & $ 0.3287  $ & $ 0.3784 $    \\
$ 650 $ & $ 158   $ & $ 139  $ & $ 0.2064  $ & $ 0.2154  $ & $ 0.2482 $    \\
$ 700 $ & $ 199   $ & $ 174  $ & $ 0.1320  $ & $ 0.1456  $ & $ 0.1677 $    \\
$ 750 $ & $ 248   $ & $ 205  $ & $ 0.0859  $ & $ 0.1013  $ & $ 0.1171 $    \\
$ 800 $ & $ 304   $ & $ 245  $ & $ 0.0567  $ & $ 0.0733  $ & $ 0.0850 $    \\
$ 850 $ & $ 371   $ & $ 277  $ & $ 0.0379  $ & $ 0.0545  $ & $ 0.0643 $    \\
$ 900 $ & $ 449   $ & $ 331  $ & $ 0.0256  $ & $ 0.0417  $ & $ 0.0509 $    \\
\hline
\end{tabular}
\end{center}
\end{table}
\begin{table}
\begin{center}
\caption[]{\label{tab:HTO_5}{The production cross-section in $\Upb$ at $\muh= 300\UGeV$.
Result from \HTO{} is computed with running QCD scales.}}
\vspace{0.2cm}
\begin{tabular}{cccc}
\hline 
Tab.~2 of \Bref{Dittmaier:2011ti}  &
Tab.~3 of \Bref{Dittmaier:2011ti}  &
Tab.~5 of \Bref{Anastasiou:2011pi} &
\HTO{} RS-option\\
&&&\\
$2.42^{+0.14}_{-0.15}$ &
$2.45^{+0.16}_{-0.22}$ &
$2.57^{+0.15}_{-0.22}$ &
$2.81^{+0.25}_{-0.23}$ \\
\hline
\end{tabular}
\end{center}
\end{table}

In \refT{tab:HTO_5b} we show a more detailed comparison of the production cross-section in the
OFFBW-scheme between \iHixs{} of \Bref{Anastasiou:2011pi} and our calculation.
$\Delta$ is the percentage error due to QCD scales and $\delta$ is the percentage ratio
\HTO{}/\iHixs{}.

\begin{table}
\begin{center}
\caption[]{\label{tab:HTO_5b}{Comparison of the production cross-section in the
OFFBW-scheme between \iHixs{} (Table~5 of \Bref{Anastasiou:2011pi}) and our calculation.
$\Delta$ is the percentage error due to QCD scales and $\delta$ is the percentage ratio
\HTO{}/\iHixs{}.}}
\vspace{0.2cm}
\begin{tabular}{cccccc}
\hline 
$\muh$[GeV]              &
$\sigma_{\mathrm{iHixs}}$[pb] &
$\Delta_{\mathrm{iHixs}}[\%]$   &
$\sigma_{\mathrm{HTO}}$[pb]   &
$\Delta_{\mathrm{HTO}}[\%]$     &
$\delta[\%]$ \\
\hline 
$200 $&$  5.57 $&$   {+}7.19 \; {-}9.06 $&$ 5.63 $&$   {+}9.12 \; {-}9.30 $&$ 1.08$ \\
$220 $&$  4.54 $&$   {+}6.92 \; {-}8.99 $&$ 4.63 $&$   {+}8.93 \; {-}8.85 $&$ 1.98$ \\
$240 $&$  3.80 $&$   {+}6.68 \; {-}8.91 $&$ 3.91 $&$   {+}8.76 \; {-}8.51 $&$ 2.89$ \\
$260 $&$  3.25 $&$   {+}6.44 \; {-}8.84 $&$ 3.37 $&$   {+}8.61 \; {-}8.22 $&$ 3.69$ \\
$280 $&$  2.85 $&$   {+}6.18 \; {-}8.74 $&$ 2.97 $&$   {+}8.49 \; {-}7.98 $&$ 4.21$ \\
$300 $&$  2.57 $&$   {+}5.89 \; {-}8.58 $&$ 2.69 $&$   {+}8.36 \; {-}7.75 $&$ 4.67$ \\
\hline
\end{tabular}
\end{center}
\end{table}

Another quantity that is useful in describing the lineshape is obtained by introducing
the peak mass,
\bq
M^2_{\peak} = \bigl( M^{\OS}_{\PH}\bigr)^2 + \bigl( \Gamma^{\OS}_{\PH}\bigr)^2,
\label{peak}
\eq
an by considering the following windows in the invariant mass, 
\bq
\Bigl[ M_{\peak} + \frac{n}{2}\,\Gamma^{\OS}_{\PH}\,,\,
       M_{\peak} + \frac{n+1}{2}\,\Gamma^{\OS}_{\PH} \Bigr],
\qquad n= 0,\pm 1, \dots.
\label{windows}
\eq
In \refT{tab:HTO_6} we present the ratio OFFP/OFFBW for the invariant mass distribution 
in the windows of \eqn{windows}.
\begin{table}
\begin{center}
\caption[]{\label{tab:HTO_6}{Scaling factor OFFP/OFFBW schemes (\eqnsc{offp}{offbw})
for the invariant mass windows of \eqn{windows}.}}
\vspace{0.2cm}
\begin{tabular}{ccccccccc}
\hline 
$\muh$[GeV] & $n=-4$ & $n=-3$ & $n=-2$ & $n=-1$ & $n=0$ & $n=1$ & $n=2$ & $n=3$ \\
$500$ & $1.24$ & $1.13$ & $1.08$ & $1.23$ & $1.32$ & $1.11$ & $0.97$ & $0.88$ \\
$600$ & $1.39$ & $1.20$ & $1.08$ & $1.27$ & $1.40$ & $1.12$ & $0.94$ & $0.84$ \\
$650$ & $1.50$ & $1.24$ & $1.08$ & $1.30$ & $1.44$ & $1.13$ & $0.93$ & $0.81$ \\
$700$ & $1.62$ & $1.28$ & $1.09$ & $1.34$ & $1.49$ & $1.13$ & $0.92$ & $0.79$ \\
$750$ & $1.78$ & $1.34$ & $1.10$ & $1.41$ & $1.57$ & $1.14$ & $0.90$ & $0.77$ \\
$800$ & $1.99$ & $1.41$ & $1.12$ & $1.49$ & $1.64$ & $1.14$ & $0.89$ & $0.75$ \\
\hline
\end{tabular}
\end{center}
\end{table}

In \refT{tab:HTO_7} we include uncertainties in the comparison; both OFFBW and OFFP are
computed with the RS option, \ie running QCD scales instead of a fixed one.  
Note that for $\sigma(\mu)$ we define a central value $\sigma_c = \sigma(\muol)$ 
and a scale error as $[ \sigma^-\,,\,\sigma^+ ]$ where
\bq
\sigma^- = \min_{\mu \in [\muol/2,2 \muol]}\,\sigma(\mu),
\qquad 
\sigma^+ = \max_{\mu \in [\muol/2,2 \muol]}\,\sigma(\mu),
\label{error}
\eq
and where $\mu = \muR = \muF$ and ${\muol}$ is the reference scale, static or
dynamic.
\begin{table}
\begin{center}
\caption[]{\label{tab:HTO_7}{Production cross section with errors due to QCD scale
variation and PDF uncertainty. First entry is the on-shell cross-section of
Table~2 of \Bref{Dittmaier:2011ti}, second entry is OFFBW (\eqn{offbw}), last entry 
is OFFP \eqn{offp}.}}
\vspace{0.2cm}
\begin{tabular}{cccc}
\hline 
$\muh$[GeV]               & 
$\sigma$[pb]              &
Scale\,[\%]               &
PDF\,[\%]                 \\
\hline
$600$ & $0.336$  & $+\,6.1\;-\,5.2$   & $+\,6.2\;-\,5.3$ \\ 
      & $0.329$  & $+\,11.2\;-\,11.3$ & $+\,4.7\;-\,4.8$ \\ 
      & $0.378$  & $+\,9.7\;-\,10.0$  & $+\,5.0\;-\,3.9$ \\ 
$650$ & $0.212$  & $+\,6.2\;-\,5.2$   & $+\,6.5\;-\,5.5$ \\ 
      & $0.215$  & $+\,12.4\;-\,12.2$ & $+\,5.1\;-\,5.3$ \\ 
      & $0.248$  & $+\,10.2\;-\,10.0$ & $+\,5.3\;-\,4.2$ \\ 
$700$ & $0.136$  & $+\,6.3\;-\,5.3$   & $+\,6.9\;-\,5.8$ \\ 
      & $0.146$  & $+\,13.9\;-\,13.3$ & $+\,5.6\;-\,5.8$ \\ 
      & $0.168$  & $+\,11.0\;-\,11.1$ & $+\,5.5\;-\,4.6$ \\ 
$750$ & $0.0889$ & $+\,6.4\;-\,5.4$   & $+\,7.2\;-\,6.1$ \\ 
      & $0.101$  & $+\,15.8\;-\,14.5$ & $+\,6.1\;-\,6.3$ \\ 
      & $0.117$  & $+\,12.2\;-\,11.9$ & $+\,5.8\;-\,5.1$ \\ 
$800$ & $0.0588$ & $+\,6.5\;-\,5.4$   & $+\,7.6\;-\,6.3$ \\ 
      & $0.0733$ & $+\,18.0\;-\,16.0$ & $+\,6.7\;-\,6.9$ \\ 
      & $0.0850$ & $+\,13.7\;-\,13.0$ & $+\,6.1\;-\,5.5$ \\ 
$850$ & $0.0394$ & $+\,6.5\;-\,5.5$   & $+\,8.0\;-\,6.6$ \\ 
      & $0.0545$ & $+\,20.4\;-\,17.6$ & $+\,7.2\;-\,7.5$ \\ 
      & $0.0643$ & $+\,15.7\;-\,14.3$ & $+\,6.5\;-\,6.0$ \\ 
$900$ & $0.0267$ & $+\,6.7\;-\,5.6$   & $+\,8.3\;-\,6.9$ \\ 
      & $0.0417$ & $+\,22.8\;-\,19.1$ & $+\,7.7\;-\,8.0$ \\ 
      & $0.0509$ & $+\,18.1\;-\,16.0$ & $+\,7.0\;-\,6.6$ \\ 
\hline
\end{tabular}
\end{center}
\end{table}

In \refT{tab:HTO_8} we compare results from Table~5 of \Bref{Anastasiou:2011pi}
with our OFFBW results with two options: 1) $\muRs, \muFs$ are fixed, 2) they run with 
$\zeta/4$. For three values of $\muh$ reported we find differences of $2.0\%, 3.7\%$ and
$4.7\%$, compatible with the scale uncertainty. Furthermore, we use 
\Bref{Dittmaier:2011ti} for input parameters which differ slightly from the one used
in \Bref{Anastasiou:2011pi}. Note that also the functional form ot the Breit--Wigner
is different since their default value is not the one in \eqn{BWdef} but
\bq 
\hbox{BW}(s) = \frac{1}{\pi}\,\frac{\sqrt{s}\,\Gamma_{\PH}(\sqrt{s})}
{\lpar  s - \muhs\rpar^2 + \muhs\,\Gamma^2_{\PH}(\muh)},
\label{XBWdef}
\eq
where $\Gamma_{\PH}(\sqrt{s})$ id the decay width of a Higgs boson at rest with mass
$\sqrt{s}$.
\begin{table}
\begin{center}
\caption[]{\label{tab:HTO_8}{Comparison of results from Table~5 of \Bref{Anastasiou:2011pi}
with our OFFBW results (\eqn{offbw}) with two options: 1) $\muRs, \muFs$ are fixed, 2) they run 
with $\zeta/4$.}}
\vspace{0.2cm}
\begin{tabular}{cccccccc}
\hline 
&  \iHixs{} & & \HTO{} &&&& \\
$\muh$[GeV] &  $\sigma$[pb]   & Scale\,[\%] &
                 $\sigma_1$[pb] & Scale\,[\%] &
                 $\sigma_2$[pb] & Scale\,[\%] \\
\hline
$220$ & $4.54$ & $+\,6.9\;-\,9.0$ & $4.63$ & $+\,8.9\,-\,8.9$ & $4.74$ & $+\,7.3\;-\,8.7$ \\
$260$ & $3.25$ & $+\,6.4\;-\,8.8$ & $3.37$ & $+\,8.6\,-\,8.2$ & $3.49$ & $+\,7.9\;-\,8.6$ \\
$300$ & $2.57$ & $+\,5.9\;-\,8.6$ & $2.69$ & $+\,8.4\,-\,7.8$ & $2.81$ & $+\,8.4\;-\,8.5$ \\
\hline
\end{tabular}
\end{center}
\end{table}

In \refT{tab:HTO_9} we use the OFFP-scheme of \eqn{offp} and look for the effect of
running QCD scales in the $\PH$ invariant mass distribution. Differences are of the order of
$2{-}3\%$ apart from the high-mass side of the distribution for very high $\PH$ masses. 
\begin{table}
\begin{center}
\caption[]{\label{tab:HTO_9}{Scaling factor in the OFFP-scheme (\eqn{offp})
running/fixed QCD scales for the invariant mass distribution.
Here $[x,y] = [M_{\peak} + x\,\Gamma^{\OS}_{\PH}\,,\,
M_{\peak} + y\,\Gamma^{\OS}_{\PH}]$.}}
\vspace{0.2cm}
\begin{tabular}{ccccccc}
\hline 
$\muh$[GeV] & 
$[-1\,,\,-\frac{1}{2}]$             &
$[-\frac{1}{2}\,,\,-\frac{1}{4}]$   &
$[-\frac{1}{4}\,,\,0]$              &
$[0\,,\,\frac{1}{4}]$               &
$[\frac{1}{4}\,,\,\frac{1}{2}]$     &
$[\frac{1}{2}\,,\,1]$               \\
$600$ & $1.023$ & $1.021$ & $1.019$ & $1.018$ & $1.017$ & $1.016$  \\
$700$ & $1.027$ & $1.023$ & $1.021$ & $1.019$ & $1.018$ & $1.019$  \\
$800$ & $1.030$ & $1.024$ & $1.021$ & $1.021$ & $1.021$ & $1.136$  \\
\hline
\end{tabular}
\end{center}
\end{table}
As we have mentioned already, the only scheme respecting gauge invariance that allows us for
a proper definition of pseudo-observables is the CPP-scheme of \eqn{cpp}. It requires
analytical continuation of the Feynman integrals into the second Riemann sheet.
Once again, the definition of POs is conventional but should put in one-to-one
correspondence well-defined theoretical predictions with derived experimental data.

In \refT{tab:HTO_10} we give a simple example by considering the process $\Pg\Pg \to \PH$ 
at lowest order and compare the traditional on-shell production cross-section, see
the l.h.s. of \eqn{sigmaBW}, with the production cross-section as defined in the CPP-scheme. 
Therefore we only consider $\Pg\Pg \to H$ at LO (\ie $v = z$) and put $\PH$ on its real 
mass-shell or on the complex one.
Below $300\UGeV$ there is no visible difference, at $300\UGeV$ the two results start to
differ with an increasing gap up to $600\UGeV$ after which there is a plateau of about $25\%$ 
for higher values of $\muh$.
\begin{table}[t]
\begin{center}
\caption[]{\label{tab:HTO_10}{$\gh$ and $\GOSu_{\PH}$ as a function of $\muh$. We also report
the ratio between the LO cross-sections for $\Pg\Pg \to \PH$ in the CPP-scheme and in the
OS-scheme.}} 
\vspace{0.2cm}
\begin{tabular}{rrrr}
\hline 
$\muh$[GeV]  & $\gh$[GeV]  & $\GOSu_{\PH}$[GeV] &
$\sigma_{\CPP}(\cph)\,/\,\sigma_{\OS}(M^{\OS}_{\PH})$ \\
\hline
$   300 $&$    7.58  $&$     8.43 $&$      0.999 $ \\
$   350 $&$    14.93 $&$    15.20 $&$      1.086 $ \\
$   400 $&$    26.66 $&$    29.20 $&$      1.155 $ \\
$   450 $&$    41.18 $&$    46.90 $&$      1.183 $ \\
$   500 $&$    58.85 $&$    68.00 $&$      1.204 $ \\
$   550 $&$    79.80 $&$    93.10 $&$      1.221 $ \\
$   600 $&$   104.16 $&$   123.00 $&$      1.236 $ \\
$   650 $&$   131.98 $&$   158.00 $&$      1.248 $ \\
$   700 $&$   163.26 $&$   199.00 $&$      1.256 $ \\
$   750 $&$   197.96 $&$   248.00 $&$      1.262 $ \\
$   800 $&$   235.93 $&$   304.00 $&$      1.264 $ \\
$   850 $&$   277.00 $&$   371.00 $&$      1.263 $ \\
$   900 $&$   320.96 $&$   449.00 $&$      1.258 $ \\
$   950 $&$   367.57 $&$   540.00 $&$      1.252 $ \\
$  1000 $&$   416.57 $&$   647.00 $&$      1.242 $ \\
\hline
\end{tabular}
\end{center}
\end{table}

The production cross-section in \eqn{PDFprod_4} is made of one term corresponding to
$\Pg\Pg \to H$ which we denote by $\sigma^0$ and terms with at least one additional jet,
$\sigma^j$. The invariant mass of the two partons in the initial state is $\kappa= v\,s$ with
$z \le v \le 1$, $\zeta$ being the Higgs virtuality. In \refT{tab:HTO_11} we introduce a cut
such that
\bq
z \le v \le \min\{ v_c\,z\,,\,1\}
\eq
and study the dependence of the result on $v_c$ by considering $\sigma^t = \sigma^0 + \sigma^j$ 
and $\delta= \sigma^t_c/\sigma^t$. The results show that a cut $z \le v \le 1.01\,z$ gives
already $85\%$ of the total answer.
\begin{table}
\begin{center}
\caption[]{\label{tab:HTO_11}{The effect of introducing a cut on the invariant mass
of the initial-state partons.}} 
\vspace{0.2cm}
\begin{tabular}{cccccc}
\hline 
$\muh$[GeV]        & 
$v_c$              &
$\sigma^0$[pb]     &
$\sigma^j$[pb]     &
$\sigma^t$[pb]     &
$\delta[\%]$         \\
\hline
$600$ & no cut  & $0.2677$ & $0.1107$ &  $0.3784$  &      \\ 
      & $1.001$ &          & $0.0127$ &  $0.2804$  & $74$ \\
      & $1.01$  &          & $0.0473$ &  $0.3149$  & $83$ \\ 
      & $1.05$  &          & $0.0918$ &  $0.3595$  & $95$ \\
$700$ & no cut  & $0.1208$ & $0.0469$ &  $0.1677$  &      \\    
      & $1.001$ &          & $0.0054$ &  $0.1262$  & $75$ \\
      & $1.01$  &          & $0.0201$ &  $0.1409$  & $84$ \\ 
      & $1.05$  &          & $0.0391$ &  $0.1599$  & $95$ \\
$800$ & no cut  & $0.0632$ & $0.0218$ &  $0.0850$  &      \\
      & $1.001$ &          & $0.0025$ &  $0.0657$  & $77$ \\ 
      & $1.01$  &          & $0.0094$ &  $0.0726$  & $85$ \\
      & $1.05$  &          & $0.0182$ &  $0.0814$  & $96$ \\ 
\hline
\end{tabular}
\end{center}
\end{table}

LHC is now running with a center of mass energy of $8\UTeV$; in \refT{tab:HTO_12} we have shown 
a comparison for the production cross-section at $7\UTeV$ and $8\UTeV$.
\begin{table}[t]
\begin{center}
\caption[]{\label{tab:HTO_12}{Comparison of the production cross-sections
in the OFFP-scheme (\eqn{offp}) at $7\UTeV$ and $8\UTeV$.}} 
\vspace{0.2cm}
\begin{tabular}{cccc}
\hline 
$\muh$[GeV]                 & 
$\sigma$[pb] $7\UTeV$       &
$\sigma$[pb] $8\UTeV$       &
ratio                         \\
\hline
$600$  & $0.378$  & $0.587$  & $1.55$ \\
$650$  & $0.248$  & $0.392$  & $1.58$ \\
$700$  & $0.168$  & $0.271$  & $1.67$ \\
$750$  & $0.117$  & $0.194$  & $1.67$ \\
$800$  & $0.0850$ & $0.145$  & $1.77$ \\
$850$  & $0.0643$ & $0.113$  & $1.77$ \\
$900$  & $0.0509$ & $0.0922$ & $1.87$ \\
\hline
\end{tabular}
\end{center}
\end{table}

\begin{figure}
\begin{minipage}{.9\textwidth}
  \includegraphics[width=0.5\textwidth, bb = 0 0 595 842]{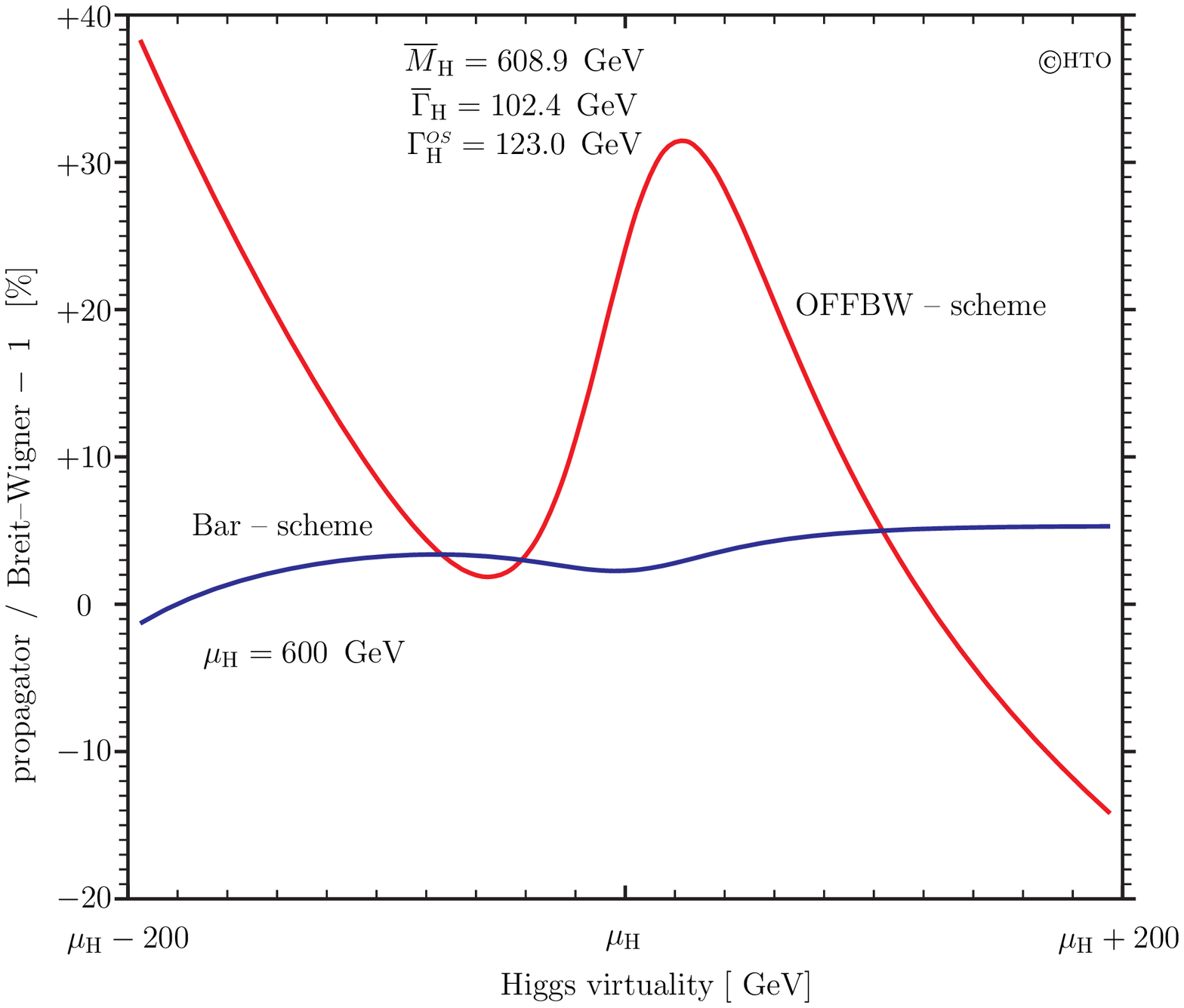}
  \includegraphics[width=0.5\textwidth, bb = 0 0 595 842]{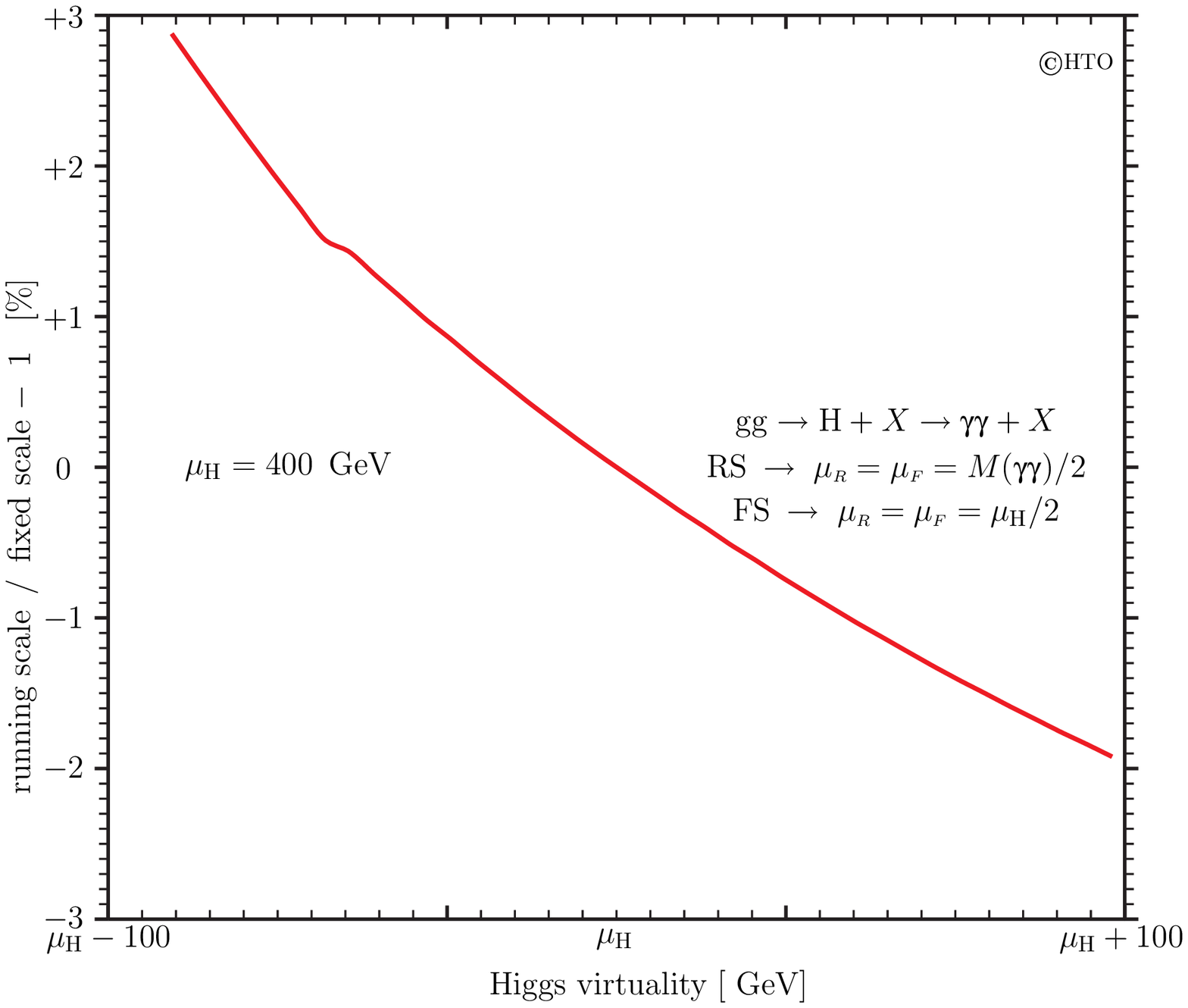}
  \vspace{-2.cm}
  \caption{
In the left figure we show a comparison of production cross-section as computed with the 
OFFP-scheme of \eqn{offp} or with the OFFBW-scheme of \eqn{offbw}. The red curve
gives Breit--Wigner parameters in the OS-scheme and the blue one in the Bar-scheme of
\eqn{Bars}.
In the right figure we show the effect of using dynamical QCD scales for the production 
cross-section of \eqns{PDFprod_1}{PDFprod_3}.} 
\label{fig:HTO_12}
\end{minipage}


\begin{minipage}{.9\textwidth}
  \includegraphics[width=0.5\textwidth, bb = 0 0 595 842]{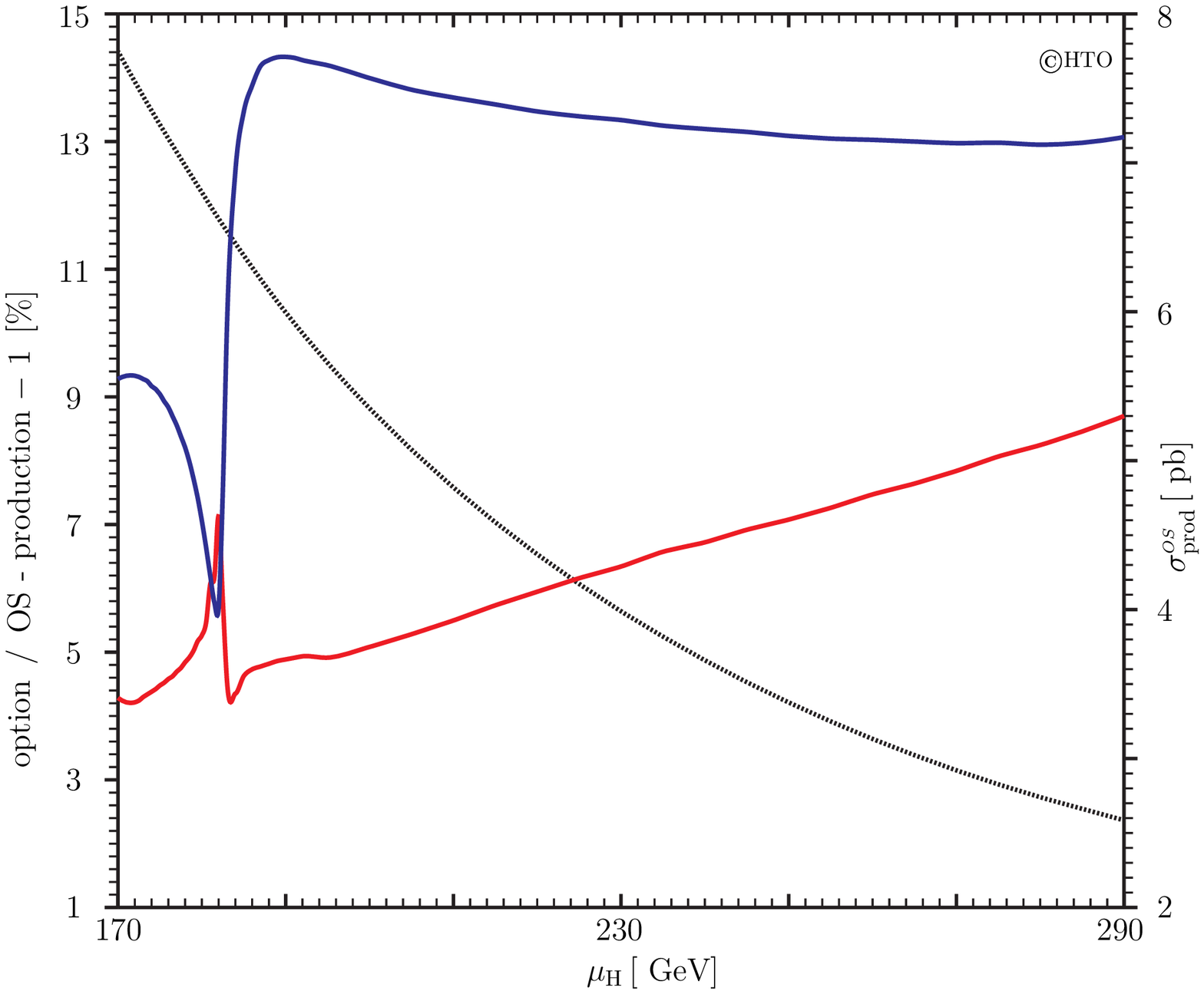}
  \includegraphics[width=0.5\textwidth, bb = 0 0 595 842]{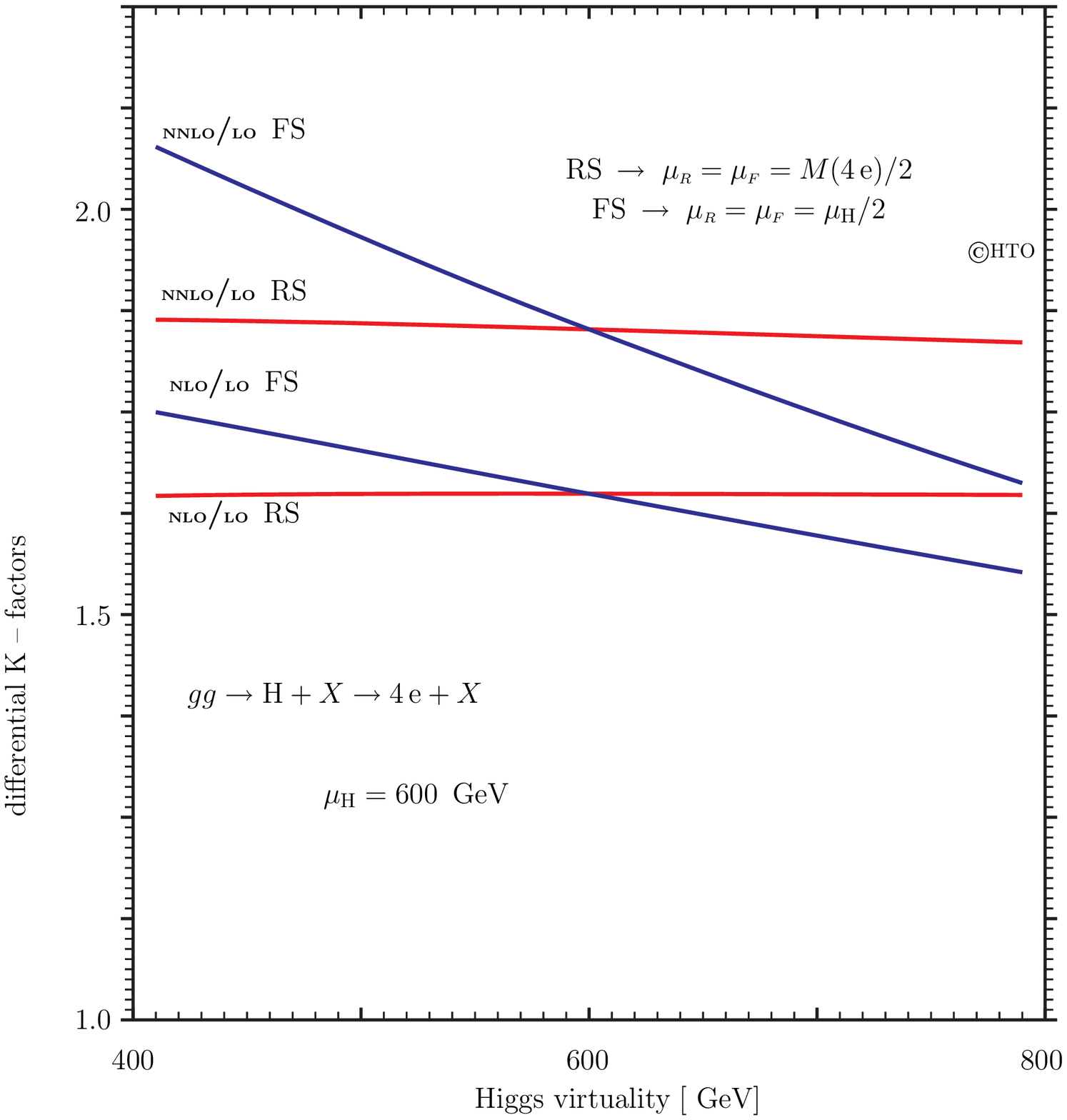}
  \vspace{-2.cm}
  \caption{
In the left figure the blue curve gives the off-shell production cross-section sampled over 
the (complex) Higgs propagator while the red curves is sampled over a Breit--Wigner distribution
The black curve gives the on-shell production cross-section.
The right figure shows differential $K\,$factor for the process $\Pp\Pp \to (\PH \to 4\,\Pe)  
+ \PX$, comparing the fixed scale option, $\muR = \muF = \MH/2$, and the running scale option,
$\muR = \muF = M(4\,\Pe)/2$.}
\label{fig:HTO_35}
\end{minipage}
\end{figure}

On the left-hand side of \refF{fig:HTO_12} we compare the production cross-section as 
computed with the OFFP-scheme of \eqn{offp} or with the OFFBW-scheme of \eqn{offbw}
for $\muh = 600\UGeV$. For the OFFBW-scheme we use Breit--Wigner parameters in the OS-scheme 
(red curve) and in the Bar-scheme of \eqn{Bars} (blue curve). Deviations from the OFFP-scheme are 
maximal in the OS-scheme and much less pronounced in the Bar-scheme.

On the right-hand side of \refF{fig:HTO_12} we show the effect of using dynamical QCD 
scales (invariant mass of the di-photon system) for the 
$\Pg\Pg \to \PH + \PX \to \PX + \PGg\PGg$ cross-section at $\muh= 400\UGeV$. 

On the left-hand side of \refF{fig:HTO_35} we consider the on-shell production cross-section 
$\sigma^{\OS}( \Pp\Pp \to \PH )$ (black curve) which includes convolution with PDFs.
The blue curve gives the off-shell production cross-section sampled over the (complex) Higgs
propagator while the red curves is sampled over a Breit--Wigner distribution. The
observed effect is substantial even in the low mass region. 

On the right-hand side of \refF{fig:HTO_35} we show differential $K\,$factors
($\sigma_{\NLO}/\sigma_{\LO}$ \etc) for the process
$\Pp\Pp \to (\PH \to 4\,\Pe)  + \PX$, comparing the fixed QCD scale option,
$\muR = \muF = \MH/2$, and the running scale option, $\muR = \muF = M(4\,\Pe)/2$.
For running QCD scales the $K\,$factor is practically constant over a wide range of the Higgs
virtuality.

\begin{figure}
\begin{minipage}{.9\textwidth}
  \includegraphics[width=0.5\textwidth, bb = 0 0 595 842]{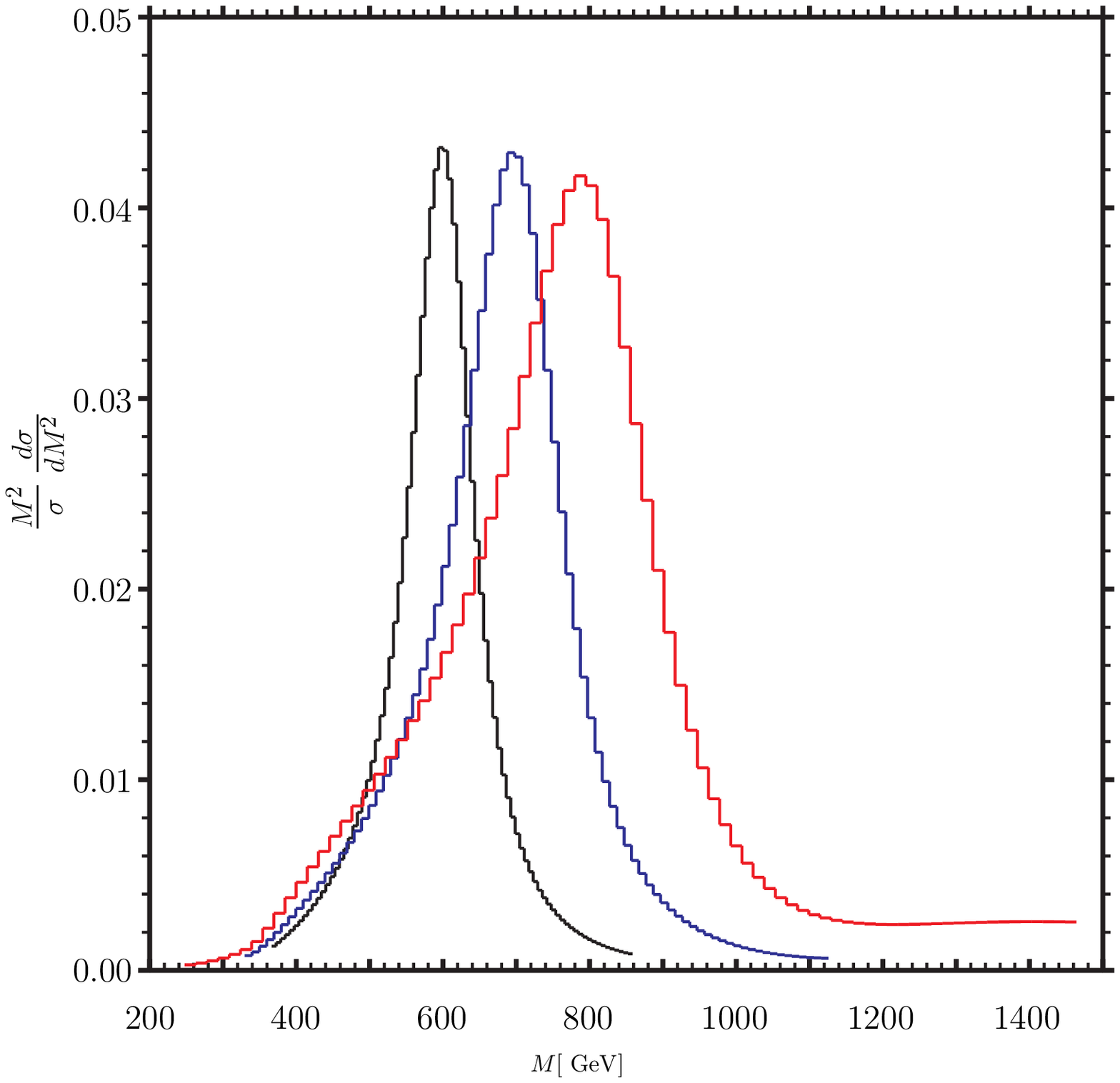}
  \includegraphics[width=0.5\textwidth, bb = 0 0 595 842]{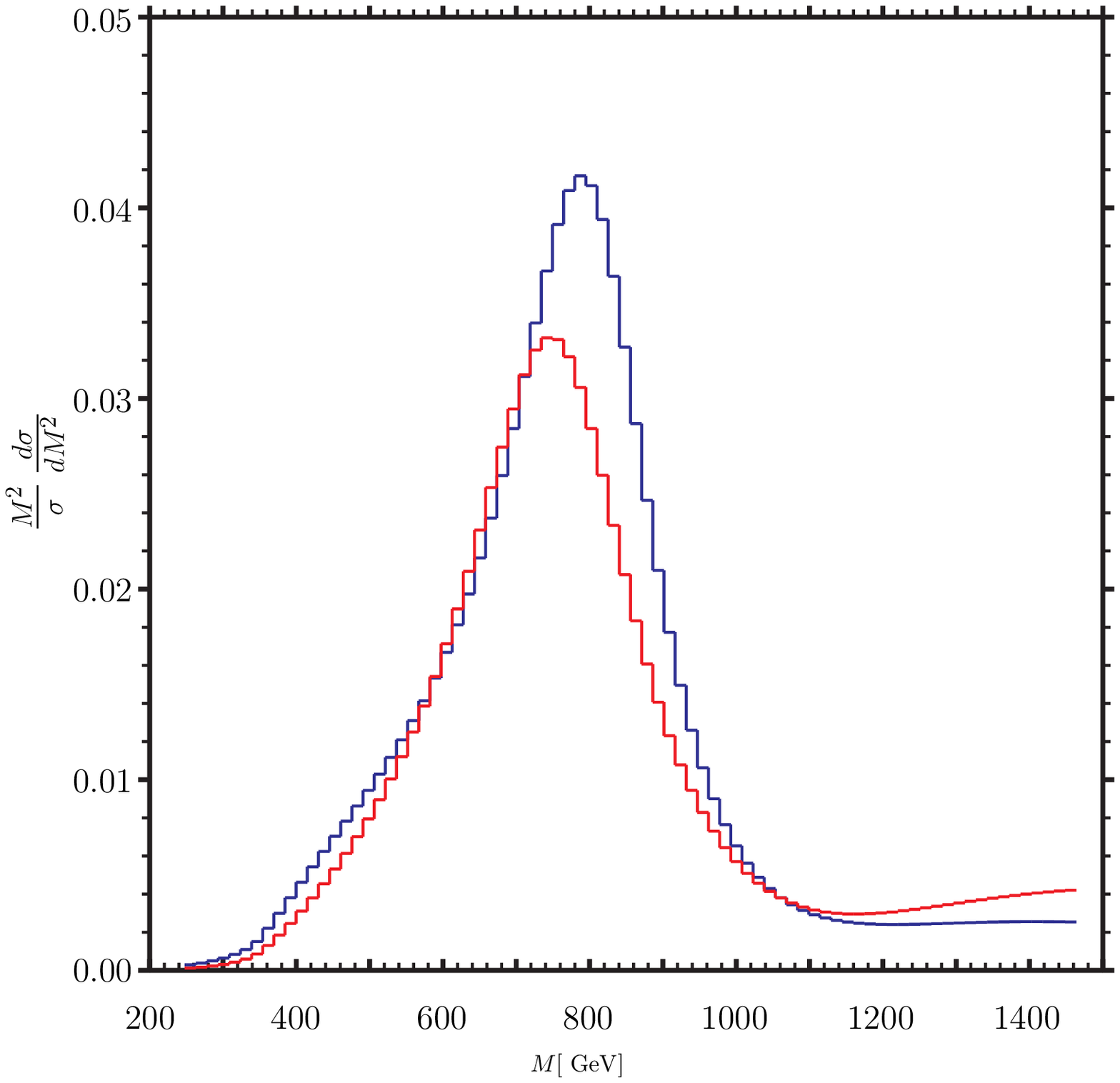}
  \vspace{-2.cm}
  \caption{
The normalized invariant mass distribution in the OFFP-scheme with
running QCD scales (left) for $600\UGeV$ (black), $700\UGeV$ (blue), $800\UGeV$ (red) in
the windows $M_{\peak} \pm 2\,\GOS$.
The normalized invariant mass distribution in the OFFP-scheme (blue) and OFFBW-scheme (red)
with running QCD scales (right) at $800\UGeV$ in the window $M_{\peak} \pm 2\,\GOS$.}
\label{fig:HTO_67}
\end{minipage}


\begin{minipage}{.9\textwidth}
  \includegraphics[width=0.5\textwidth, bb = 0 0 595 842]{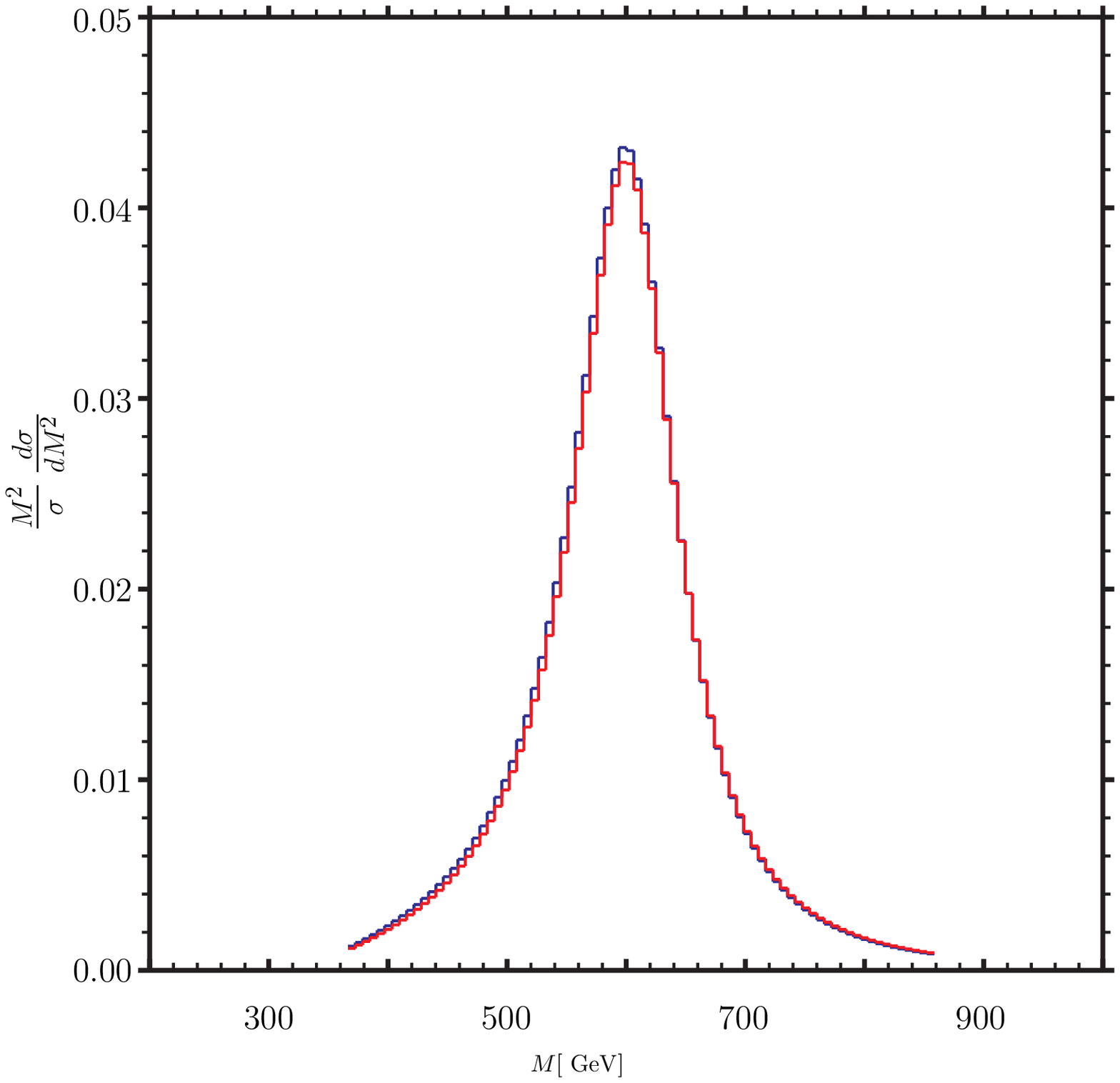}
  \includegraphics[width=0.5\textwidth, bb = 0 0 595 842]{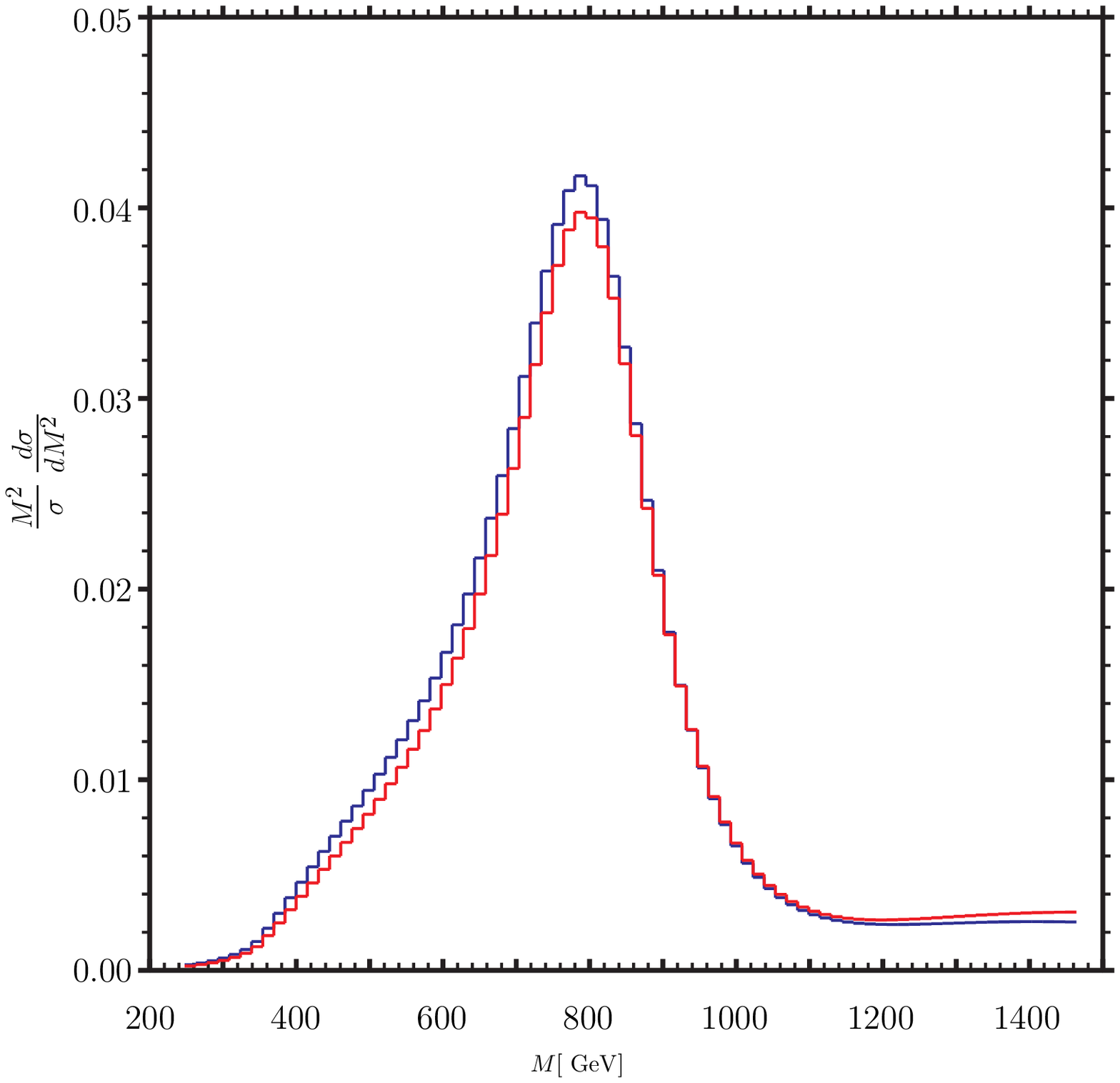}
  \vspace{-3.cm}
  \caption{
The normalized invariant mass distribution in the OFFP-scheme with
running QCD scales for $600\UGeV$ (left), $800\UGeV$ (right) in the windows 
$M_{\peak} \pm 2\,\GOS$. The blue line refers to $8\UTeV$, the red one to $7\UTeV$.}
\label{fig:HTO_89}
\end{minipage}
\end{figure}

On the left-hand side of \refF{fig:HTO_67} we show the normalized invariant mass distribution 
in the OFFP-scheme with running QCD scales for Higgs-boson masses of $600\UGeV$ (black), 
$700\UGeV$ (blue), $800\UGeV$ (red) in the windows $M_{\peak} \pm 2\,\GOS$ where $M_{\peak}$ 
is defined in \eqn{peak}.

On the right-hand side of \refF{fig:HTO_67} we show the normalized invariant mass 
distribution in the OFFP-scheme (blue) and OFFBW-scheme (red) with running QCD scales 
for a Higgs-boson mass of $800\UGeV$ in the window $M_{\peak} \pm 2\,\GOS$.

\begin{figure}
\vspace{-4cm}
\begin{minipage}{.9\textwidth}
\begin{center}
  \includegraphics[width=0.7\textwidth, bb = 0 0 595 842]{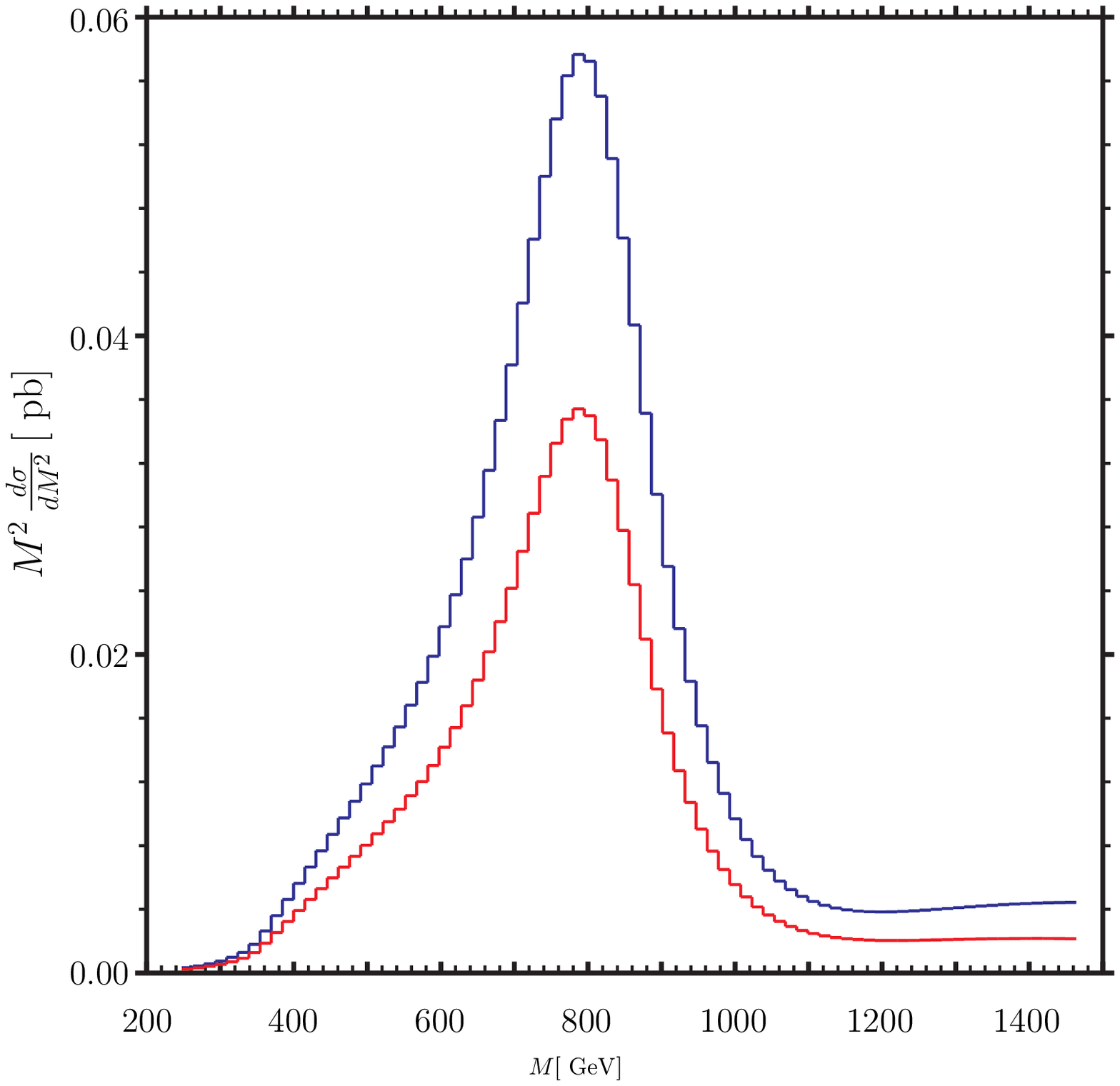}
  \vspace{-3.6cm}
  \caption{
The invariant mass distribution in the OFFP-scheme with
running QCD scales for $800\UGeV$ in the window $M_{\peak} \pm 2\,\GOS$. The blue line refers 
to $8\UTeV$, the red one to $7\UTeV$.}
\label{fig:HTO_10}
\end{center}
\end{minipage}
\end{figure}

Finally, in \refF{fig:HTO_89} we show the normalized invariant mass distribution in the 
OFFP-scheme with running QCD scales at $600{-}800\UGeV$ in the window $M_{\peak} \pm 2\,\GOS$
for $7\UTeV$ (blue) and $8\UTeV$ (red).
The invariant mass distribution in the OFFP-scheme with running QCD scales for 
a Higgs-boson mass of $800\UGeV$ (in the window $M_{\peak} \pm 2\,\GOS$) is shown in
\refF{fig:HTO_10}; the blue line refers to a center-of-mass energy of $8\UTeV$, the red one 
to $7\UTeV$.
\subsection{Results in CPP-scheme}
Here, we will briefly summarize what was covered in \refS{schemes} for the CPP-scheme.
referring to \eqn{PDFprod_3} we can say the following: the production amplitude, including
QCD corrections, is always $\xi\,$-independent and we can keep full off-shellness; the decay
amplitude, beyond LO, is not and should be evaluated at the Higgs complex pole in the
CPP-scheme. To give an example of the results we select the process $\Pg\Pg \to \PH \to
\PZ^c\PZ^c$, where the two $\PZ\,$-bosons in the final state are also taken at their
complex pole, thus giving a full gauge-invariant quantity.  
We will show results in the OFFP-scheme (\eqn{offp}) and in the CPP-scheme (\eqn{cpp}).
Caveat emptor: one should not interpret their difference as a source of theoretical
uncertainty since comparisons should only take place for the full S$\,\oplus\,$B matrix
element and what we are doing in the two schemes simply amounts to shifting some part
from the known S-amplitude to the unknown S/B-interference. Furthermore, only the
CPP-scheme is a fully consistent recipe to {\em define} a gauge-invariant signal; the
only reason why we presented most of our results in the OFFP-scheme is that it is much
simpler to implement in any preexisting calculation while the CPP-scheme requires
analytical continuation into the second Riemann sheet. 

\begin{figure}
\vspace{-10cm}
\begin{minipage}{.9\textwidth}
  \includegraphics[width=0.5\textwidth, bb = 0 0 595 842]{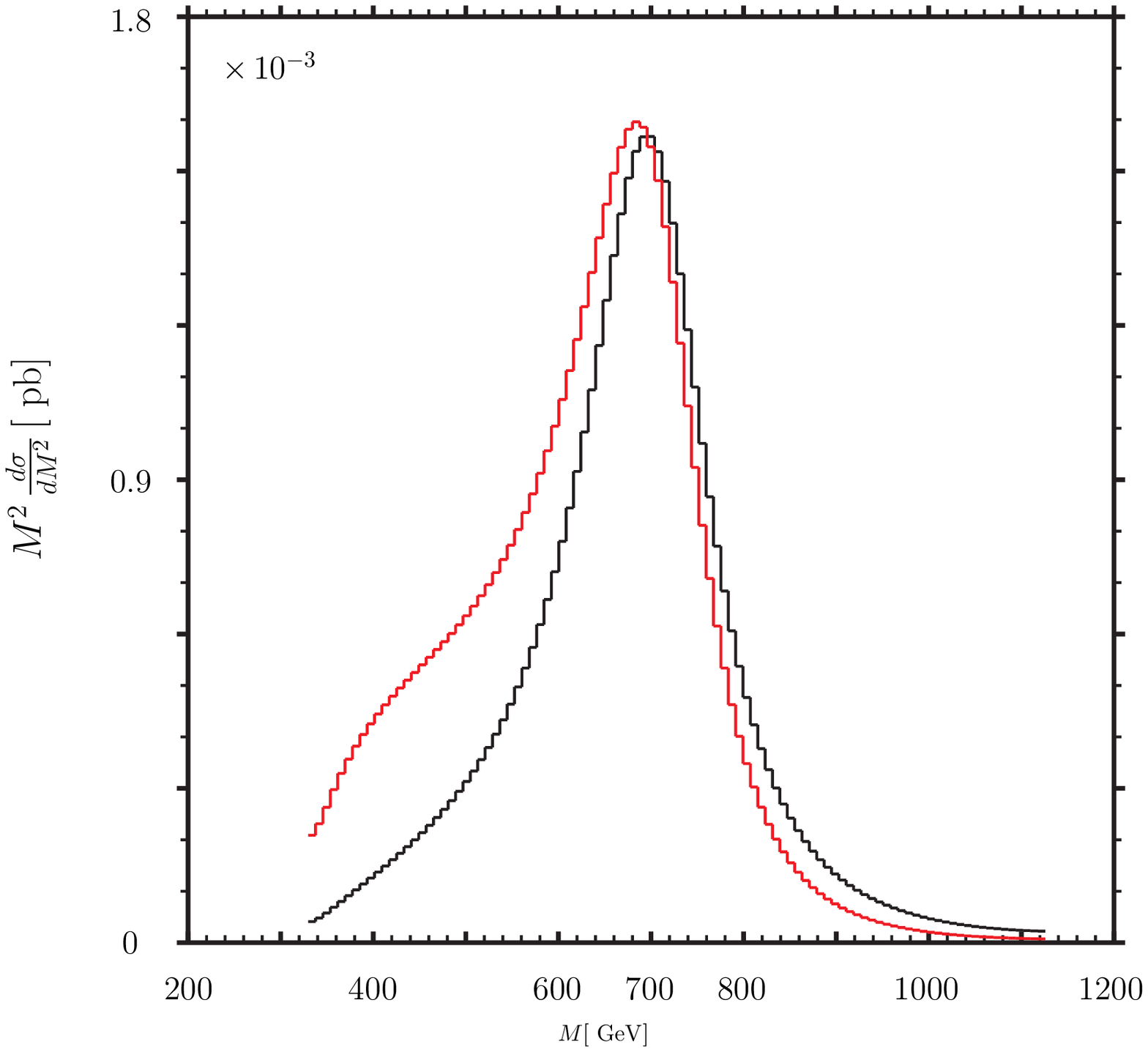}
  \includegraphics[width=0.5\textwidth, bb = 0 0 595 842]{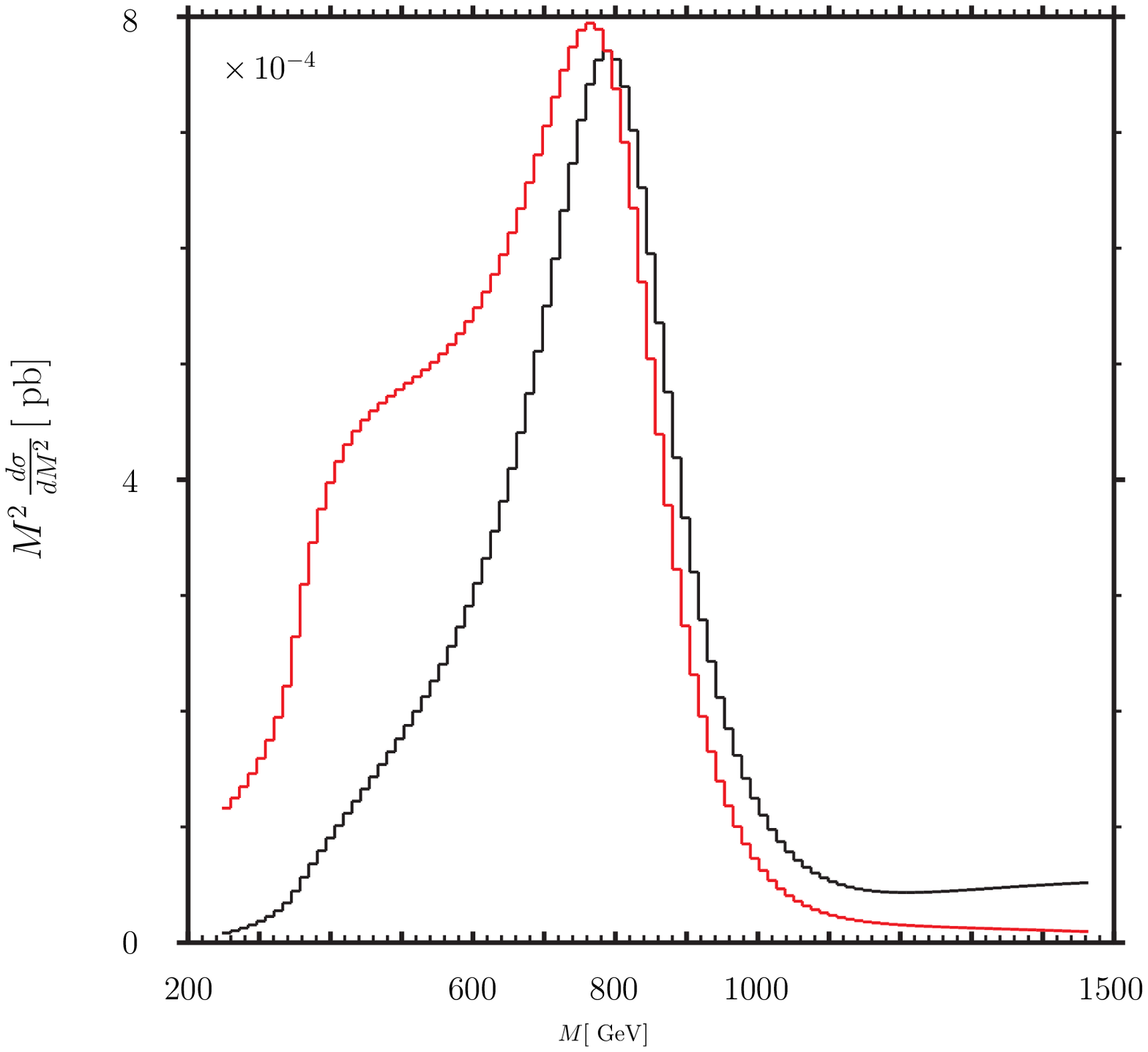}
  \vspace{-1.6cm}
  \caption{
The invariant mass distribution in the OFFP-scheme (black) and in the CPP-scheme (red)
for $\muh= 700\UGeV$ (left) and $\muh= 800\UGeV$ (right) in the  window 
$[M_{\peak} - 2\,\GOS\,{-}\,M_{\peak} + 2\,\GOS]$ for the process $\Pg\Pg \to \PH \to \PZ^c\PZ^c$.}
\label{fig:HTO_15}
\end{minipage}
\end{figure}

In \refF{fig:HTO_15} we show the invariant mass distribution in the OFFP-scheme (black) and 
in the CPP-scheme (red) for $\muh= 700\UGeV$ (left) and $\muh= 800\UGeV$ (right)
in the window $[M_{\peak}{-}M_{\peak} + 2\,\GOS]$ 
for the process $\Pg\Pg \to \PH \to \PZ^c\PZ^c$. The shoulder in the low mass side of the
resonance (in the CPP-scheme) is due to the fact that the $\PH\,$-width is very large,
the production cross-section is increasing for decreasing invariant mass and there is no
suppression from the decay which is taken at the Higgs complex pole. Thus, the decrease in
the distribution for small invariant masses is delayed by the large width.
The high invariant mass behavior reflects again a constant (at the complex pole) decay
amplitude.
\subsection{Results in the low Higgs mass region}
For lower values of the Higgs mass we compare the OFFBW-scheme and the OFFP one 
(based on \eqn{expI}) at $8\UTeV$ in \refT{tab:HTO_14}; here we also show the on-shell 
results of \Bref{8TeV} and the Breit-Wigner (running width scheme) approach of 
\Bref{Anastasiou:2012hx}.
The results of \refT{tab:HTO_14} show a good convergence of OFFBW/OFFP results to the
on-shell ones for low values of the Higgs mass.
\begin{table}[t]
\begin{center}
\caption[]{\label{tab:HTO_14}{Comparison of the production cross-sections at
$8\UTeV$: OFFBW-scheme, OFFP-scheme, on-shell results of \Bref{8TeV} and the Breit-Wigner 
approach of \Bref{Anastasiou:2012hx}.}} 
\vspace{0.2cm}
\begin{tabular}{ccccc}
\hline 
$\muh$[GeV]                              & 
$\sigma$[pb] OFFBW                       &
$\sigma$[pb] OFFP                        &
$\sigma$[pb] \Bref{8TeV}                 &
$\sigma$[pb] \Bref{Anastasiou:2012hx}    \\
\hline
$300$  & $3.74$  & $3.87$  & $3.32$ & $3.45$\\
$280$  & $4.10$  & $4.25$  & $3.67$ & $3.82$\\
$260$  & $4.60$  & $4.79$  & $4.16$ & $4.34$\\
$240$  & $5.29$  & $5.53$  & $4.82$ & $5.05$\\
$220$  & $6.21$  & $6.43$  & $5.71$ & $6.00$\\
$200$  & $7.48$  & $7.53$  & $6.95$ & $7.32$\\
$190$  & $8.31$  & $8.35$  & $7.78$ & $8.21$\\
$180$  & $9.42$  & $9.45$  & $8.85$ & $9.42$\\
$170$  & $10.59$ & $10.62$ & $10.11$ & $10.72$\\
$160$  & $12.24$ & $12.24$ & $11.76$ & $12.66$\\
$150$  & $13.88$ & $13.88$ & $13.57$ & $14.43$\\
$140$  & $15.78$ & $15.79$ & $15.59$ & $16.53$\\
$130$  & $18.05$ & $18.05$ & $18.04$ & $19.14$\\
$125$  & $19.41$ & $19.41$ & $19.49$ & $20.69$\\
\hline
\end{tabular}
\end{center}
\end{table}
\section{QCD scale error\label{Sect_err}}
The conventional theoretical uncertainty associated with QCD scale variation is defined in 
\eqn{error}. Let us consider the production cross-section at $800\UGeV$ in the OFFP-scheme with
running or fixed QCD scales, we obtain
\bq
\mbox{RS}\quad 0.08497\,{}^{+13.7\%}_{-13.0\%},
\qquad
\mbox{FS}\quad 0.07185\,{}^{+6.7\%}_{-8.5\%}.
\eq
The uncertainty with running scales is about twice the one where the scales are kept fixed.
One might wonder which is the major source and we have done the following; in estimating
the conventional uncertainty in the RS-option we keep $\muFs = \muol^2 = \zeta/4$ and 
vary $\muR$ between $\muol/2$ and $2\,\muol$. The corresponding result is
\bq
\mbox{RS}\quad 0.08497\,{}^{+3.9\%}_{-4.4\%}, \qquad\quad \muR\quad\mbox{only}.
\eq
Therefore, the main source of uncertainty comes from varying the factorization scale.
However, in the fixed-scale option we obtain
\bq
\mbox{FS}\quad 0.07185\,{}^{+3.6\%}_{-4.4\%}, \qquad\quad \muR\quad\mbox{only}
\eq
and $\muF\,$-variation is less dominant. In any case the question of which variation is
dominating depends on the mass-range; indeed, in the ONBW-scheme with fixed scales at
$140\,\UGeV$ we have
\bqa
\mbox{ONBW}\quad &12.27&\,{}^{+11.1\%}_{-10.1\%}, \qquad\quad \muR\quad\mbox{and}\quad \muF
\nl
\mbox{ONBW}\quad &12.27&\,{}^{+8.7\%}_{-9.6\%}, \qquad\quad \muR\quad\mbox{only}
\eqa
showing $\muR$ dominance.

In order to compare this conventional definition of uncertainty with the work of 
\Bref{Cacciari:2011ze} we recall their definition of $p\%\,$ credible interval; given a series
\bq
\sigma = \sum_{n=l}^k\,c_n\,\alphas^n,
\eq
the $p\%\,$ credible interval $\sigma \pm \Delta\sigma^p$ is defined as
\bqa
\Delta\sigma^p &=& \alpha_s^{k+1}\,\max\lpar\{c\}\rpar\,\frac{n_c+1}{n_c}\,p\%,
\qquad p\% \le \frac{n_c}{n_c+1},
\nl
\Delta\sigma^p &=& \alpha_s^{k+1}\,\max\lpar\{c\}\rpar\,
\Bigl[(n_c+1)\,(1-p\%)\Bigr]^{-1/n_c},
\qquad p\% \ge \frac{n_c}{n_c+1},
\eqa
with $n_c= k-l+1$. Using this definition we find that the $68\%(90\%)$ credible interval for 
$\muR$ uncertainty in the production cross-section at $800\UGeV$ is $3.1\%(5.3\%)$ in substantial 
agreement with the conventional method. For ONBW-scheme at $140\UGeV$ the $90\%$
credible interval for $\muR$ uncertainty is $8.0\%$. It could be interesting to extend the 
method of \Bref{Cacciari:2011ze} to cover $\muF$ uncertainty.
\section{Residual theoretical uncertainty\label{Sect_THU}}
\begin{figure}
\vspace{-10cm}
\begin{minipage}{.9\textwidth}
\begin{center}
  \includegraphics[width=0.7\textwidth, bb = 0 0 595 842]{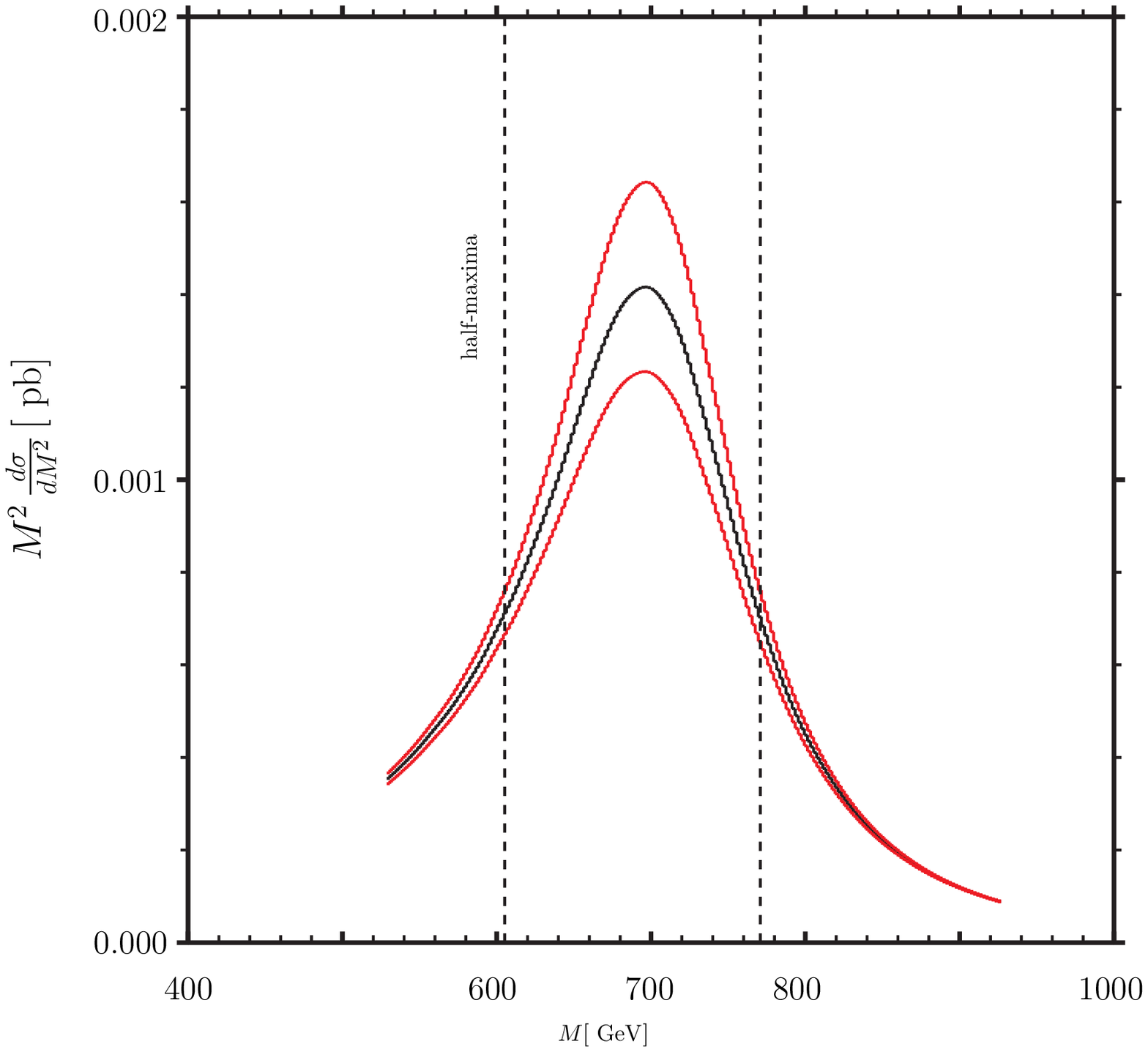}
  \vspace{-1.6cm}
  \caption{
The invariant mass distribution in the OFFP-scheme with
running QCD scales for $\muh= 700\UGeV$ in the window $M_{\peak} \pm \GOS$. The red lines give
the associated theoretical uncertainty as discussed is \refS{Sect_THU}.}
\label{fig:HTO_11}
\end{center}
\end{minipage}
\end{figure}

As we mentioned before the accuracy at which the imaginary part of $\cph$ can be computed is not 
of the same quality as the next-to-leading-order (NLO) accuracy of the on-shell width.
It would we desirable to include two- and three-loop contributions as well in $\gh$ and for some
of these contributions only on-shell results have been computed so far. 
Here we follow the analysis of \Bref{Ghinculov:1996py} where the authors try to improve the 
solution of the complex pole equation by using the available on-shell information. It is again a 
method based on some expansion and does not include the full SM, QCD corrections and complex 
poles for internal $\PW, \PZ$ lines; it amounts to consider only the Higgs-Goldstone Lagrangian 
of the SM. Nevertheless it is very useful to give a rough estimate of the missing orders. The 
authors of \Bref{Ghinculov:1996py} define a double expansion of the self-energy,
\bq
S_{\HH}(s) = A\,M^2_{\PH} + B\,(s - M^2_{\PH}) + \frac{C}{M^2_{\PH}}\,(s - M^2_{\PH})^2 + \dots
\quad
A = \sum_{n=0}^{\infty}\,a_n\,\Bigl( \frac{G_{\ssF}\,M^2_{\PH}}{2 \sqrt{2} \pi^2}\Bigr)^n,
\eq
etc. In this way we can write
\bq
\gh = \frac{G_{\ssF}\,M^3_{\PH}}{2 \sqrt{2} \pi^2}\,a_1\,\lpar
1 + \frac{G_{\ssF}\,M^2_{\PH}}{2 \sqrt{2} \pi^2}\,\frac{a_2}{a_1} + \dots\rpar.
\eq
The ratio $a_2/a_1$ has been computed and we can estimate that the first correction to
$\gh$ is roughly given by
\bq
\delta_{\PH} = 0.350119\,\frac{G_{\ssF}\,\muhs}{2 \sqrt{2} \pi^2}.
\eq
Changes in $\gh$ range from $2.3\%$ at $400\UGeV$ to $9.4\%$ at $750\UGeV$. In general,
from \Bref{Ghinculov:1996py} we do not see very large variations up to $1TeV$
with a breakdown of the perturbative expansion around when $1.74\UTeV$.
Let us elaborate a bit more: in the Higgs-Goldstone model one has
\bq
\frac{\gh}{\muh} =  1.1781\,g_{\PH} + 0.4125\,g^2_{\PH} + 1.1445\,g^3_{\PH}
\qquad g_{\PH} = \frac{G_{\ssF} \muhs}{2\sqrt{2} \pi^2},
\eq
giving $\gh = 168.84(\LO), 180.94(\NLO), 186.59(\NNLO)\UGeV$ for $\muh= 700\UGeV$.
Using the three known terms in the series we estimate a $68\%\,$ credible interval
of $\gh = 186.59 \pm 1.93\UGeV$. The difference $\NNLO - \LO$ is 17.8\UGeV and in the
full SM our estimate is $\gh = 163.26 \pm 11.75\UGeV$.
Therefore, using $\gh\,(1 \pm \delta_{\PH})$ we can give a rough but reasonable estimate of 
the remaining uncertainty on the lineshape. Of course, one could plot the lineshape with an 
error band (an example is shown in \refF{fig:HTO_11} for $\muh= 700\UGeV$) but it is better 
to quantify the uncertainty at the level of those quantities that characterize the resonance. 
Referring to \eqn{PDFprod_1} we introduce the following quantities:
\bqa
\sigma^{\myprod} \qquad &{}& \qquad \mbox{total production cross-section}
\nl
\Sigma = \frac{d\sigma^{\myprod}}{d\zeta} \qquad &{}& \qquad \mbox{differential distribution}
\nl
\{\zeta_{\max}\,,\,\Sigma_{\max}\} \qquad &{}& \qquad \mbox{the maximum of the lineshape}
\nl
\{\zeta_{\pm}\,,\,\frac{1}{2}\,\Sigma_{\max}\} \qquad &{}& \qquad 
\mbox{the half-maxima of the lineshape}
\nl
\PA = \int_{\zeta_-}^{\zeta_+}\,d\zeta\,\Sigma &{}& \qquad 
\mbox{the area of the resonance between half-maxima}
\eqa
where $\zeta$ is the Higgs virtuality. Changing $\gh$ to $1 \pm \delta_{\PH}$ gives
a variation on the production cross-section which is linear in $\delta_{\PH}$, as shown
in \refT{tab:HTO_THU}; similarly, the variation in the peak height goes as
$2\,\delta_{\PH}$ with no appreciable variation in its position, $\zeta_{\max}$.
In \refT{tab:HTO_THU} we also show the variation (in $\UGeV$) in the position of the
half-maxima and in the area of the resonance.
\begin{table}[t]
\begin{center}
\caption[]{\label{tab:HTO_THU}{Theoretical uncertainty on the production cross-section,
the height of the maximum, the position of the half-maxima and the area of the resonance.} }
\vspace{0.2cm}
\begin{tabular}{cccccc}
\hline 
$\muh$[GeV]                  & 
$\delta_{\PH}[\%]$           &
$\delta\sigma^{\myprod}[\%]$ &
$\delta\Sigma_{\max}[\%]$    &
$\Delta\zeta_-\,,\,\Delta\zeta_+$[GeV] &
$\delta\PA[\%]$              \\
\hline
$600$  & $5.3$  & ${-}6.0\;{}\;{+}6.3$ & ${-}10.8\;{}\;{+}11.4$ &
$({-}2.5\,,\,{+}2.5)\;{}\;({+}2.5\,,\,{-}2.5)$ &
${-}4.8\;{}\;{+}6.0$ \\
$700$  & $7.2$  & ${-}8.0\;{}\;{+}8.6$ & ${-}14.8\;{}\;{+}16.0$ &
$({-}9.0\,,\,{+}8.0)\;{}\;({+}4.0\,,\,{-}4.0)$ &
${-}7.0\;{}\;{+}11.8$ \\
$800$  & $9.4$  & ${-}9.7\;{}\;{+}10.6$ & ${-}19.3\;{}\;{+}21.5$ &
$({-}18.2\,,\,{+}18.2)\;{}\;({+}6.1\,,\,{-}6.1)$ &
${-}8.7\;{}\;{+}9.5$ \\
\hline
\end{tabular}
\end{center}
\end{table}

To summarize our estimate of the theoretical uncertainty associated to the signal: 
\bei
\item the remaining uncertainty on the production cross-section is typically well 
reproduced by $(\delta_{\PH} +1)[\%]$ 
\item $\Sigma_{\max}$ changes approximately with the naive expectation, $2\,\delta_{\PH}[\%]$ 
\item The shift induced in the position of the half-maxima are more complicated but a rough 
approximation is given by
\bq
\Delta\zeta_{\pm} \approx \pm \frac{1}{2}\,\frac{\muh}{M_{\pm}}\,\gh\,\delta_{\PH},
\quad M_{\pm} = \muh^{1/2}\,\lpar \muh \pm \gh\rpar^{1/2}.
\eq 
\eei

How to use these results depends on the specific analysis, 
$\PH \to \PW\PW \to \Pl\PGn\Pl\PGn$, $\Pl\PGn\PQq\PQq$ and 
$\PH \to \PZ\PZ \to \Pl\Pl\Pl\PL$ \etc
For example, the lineshape information is used in $\PH \to \PW\PW \to \Pl\PGn\PQq\PQq$ 
analysis~\cite{ATLAS:2011aa}.
The $\PH \to \PZ\PZ \to \Pl\Pl\PGn\PGn$ analysis use transverse mass spectrum for 
the $\Pl\Pl\PGn\PGn$ system.
If the analysis is a simple ``counting experiment'', therefore not taking the Higgs lineshape 
into account, the uncertainty on the area and the half-maxima can be neglected.
Therefore, informations on the shape of the resonance will become more relevant once more 
advanced analysis techniques (based on differential distributions as discriminant 
variable) will be used. 

The factor $\Gamma^{\tot}_{\PH}(\zeta)$ in \eqn{QFT} deserves a separate discussion.
As explained in \refS{Sect_gi} it represents the ``on-shell'' decay of an Higgs boson of 
mass $\sqrt{\zeta}$ and we have to quantify the corresponding uncertainty.
The staring point is $\Gamma^{\tot}$ computed by {\sc Prophecy4f}~\cite{Prophecy4f}
which includes two-loop leading corrections in $G_{\ssF} M^2_{\PH}$, where $M_{\PH}$ is now 
the on-shell mass. Next we consider the on-shell width in the Higgs-Goldstone model,
discussed in \Brefs{Ghinculov:1996py,Frink:1996sv}. We have
\bq
\frac{\Gamma_{\PH}}{\sqrt{\zeta}}\bmid_{HG} = \sum_{n=1}^3\,a_n\,\lambda^n = X_{HG},
\qquad \lambda = \frac{G_{\ssF} \zeta}{2 \sqrt{2} \pi^2}.
\label{HGG}
\eq
Let $\Gamma_p= X_p\,\sqrt{\zeta}$ the width computed by {\sc Prophecy4f}, we redefine
the total width as
\bq
\frac{\Gamma_{\tot}}{\sqrt{\zeta}} = \lpar X_p - X_{HG} \rpar + X_{HG} =
\sum_{n=0}^3\,a_n\,\lambda^n,
\eq
where now $a_0 = X_p - X_{HG}$. As long as $\lambda$ is not too large we can define
a $p\% < 80\%$ credible interval as (following from $a_{2,3} < a_1$)
\bq
\Gamma_{\tot}(\zeta) = \Gamma_p(\zeta) \pm \Delta\Gamma, \qquad
\Delta\Gamma = \frac{5}{4}\,\max\{\mid a_0\mid,a_1\}\,p\%\,\lambda^4\,\sqrt{\zeta}.
\eq
It is easily seen that for $\sqrt{\zeta} = 929\UGeV$ the two-loop corrections are of the
same size of the one-loop corrections and for $\sqrt{\zeta} = 2.6\UTeV$ one-loop and Born
become of the same size. Not that the expansion parameter in \eqn{HGG} is one for 
$\sqrt{\zeta} = 1.57\UTeV$. In \refT{tab:HTO_THUW} we present both the $68\%$ and the 
$95\%$ credible intervals. This should be compared with the corresponding estimate in
\Bref{Dittmaier:2012vm} of $17\%(M^4_{\PH}/1\UTeV)$ for $M_{\PH} > 500\UGeV$.
Due to the large effect induced by radiative corrections for large values of $M_{\PH}$
and to the fact that the two-loop correction is only known in the Higgs-Goldstone
Lagrangian we have devoted \refA{appB} to a detailed discussion of the validity of the
whole approach.
\begin{table}[t]
\begin{center}
\caption[]{\label{tab:HTO_THUW}{Theoretical uncertainty on the total decay width,
$\Gamma^{\tot}_{\PH}$ in \eqn{QFT}. $\Gamma_p$ is the total width computed by 
{\sc Prophecy4f}{}, $\Delta\Gamma$ gives the credible intervals and $\delta\Gamma[\%]$ is the ratio
$\Delta\Gamma/Gamma_p$.}}
\vspace{0.2cm}
\begin{tabular}{rrcc}
\hline 
$\sqrt{\zeta}$[GeV]          & 
$\Gamma_p$[GeV]              &
$\delta\Gamma[68\%]$         &
$\delta\Gamma[95\%]$        \\
\hline
$600$   & $123$   &  $0.25$  & $0.42$ \\
$700$   & $199$   &  $0.62$  & $1.03$ \\
$800$   & $304$   &  $1.35$  & $2.24$ \\ 
$900$   & $449$   &  $2.63$  & $4.38$ \\
$1000$  & $647$   &  $4.72$  & $7.85$ \\
$1200$  & $1205$  &  $13.1$  & $21.7$ \\
$1500$  & $3380$  &  $34.7$  & $57.8$ \\
$2000$  & $15800$ &  $98.9$  & $165$ \\
\hline
\end{tabular}
\end{center}
\end{table}

It is clear that it does not make much sense to have an error estimate beyond $1.3\UTeV$
and, therefore, all results for the Higgs lineshape that have a sizable fraction of
events in this high-mass region should not be taken too seriously. Here, once again, the
only viable alternative to define the Higgs signal is the CPP-scheme.

\begin{figure}
\vspace{-10cm}
\begin{minipage}{.9\textwidth}
\begin{center}
  \includegraphics[width=0.8\textwidth, bb = 0 0 595 842]{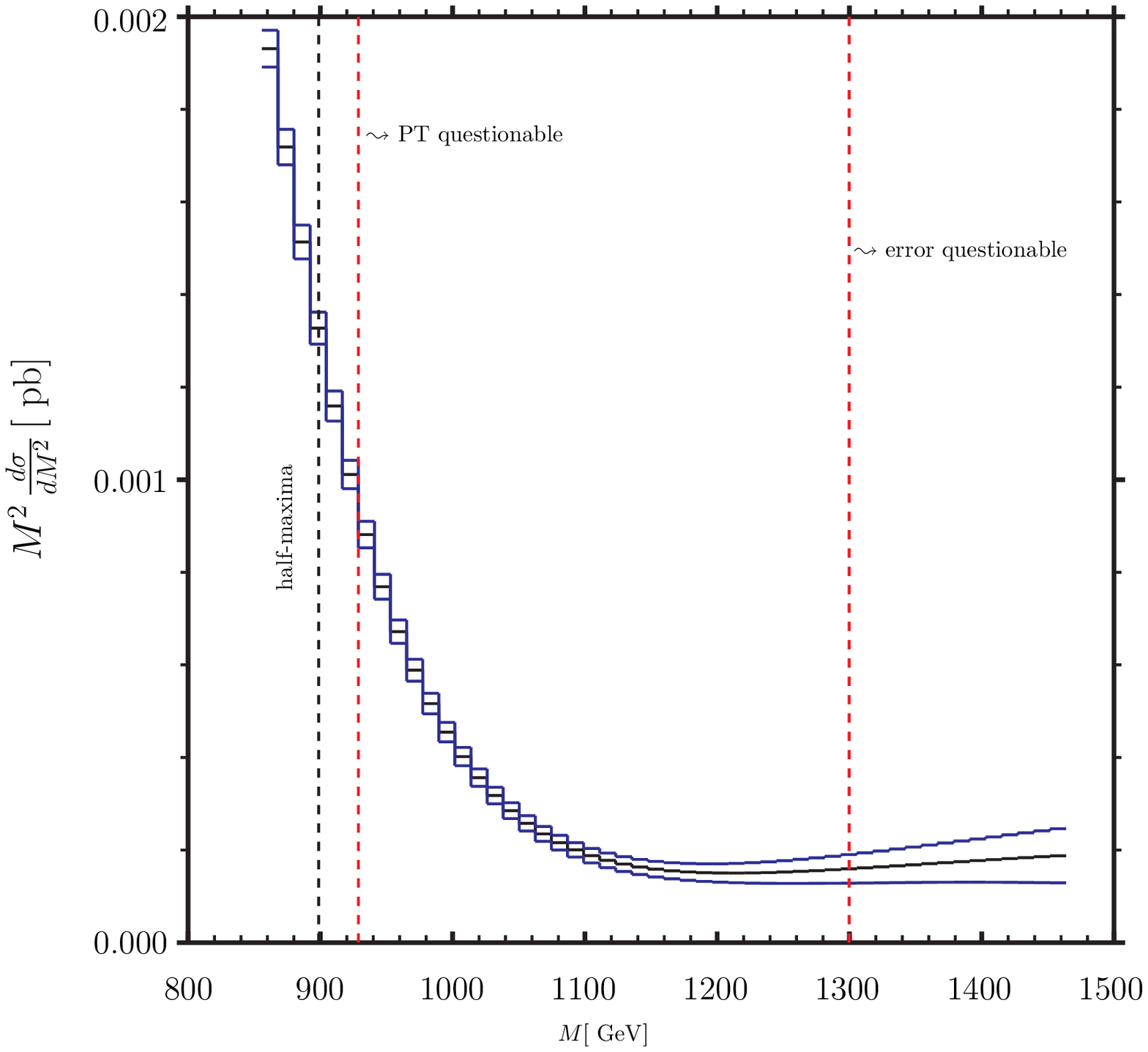}
  \vspace{-1.6cm}
  \caption{
The invariant mass distribution in the OFFP-scheme for $\muh= 800\UGeV$ in the 
window $[M_{\peak}{-}M_{\peak} + 2\,\GOS]$ with the error band due to the
theoretical uncertainty introduced by $\Gamma^{\tot}_{\PH}(\zeta)$
discussed is \refS{Sect_THU}.}
\label{fig:HTO_14}
\end{center}
\end{minipage}
\end{figure}

To summarize: above $0.93\UTeV$ perturbation theory becomes questionable since the two-loop 
corrections start be become larger than the one-loop ones; above $1.3\UTeV$ the error
estimate also becomes questionable since the expansion parameter is $\lambda = 0.7$ and
the $95\%$ credible interval (after inclusion of the leading two-loop effects) is 
$32.2\%$. The invariant mass distribution for $\muh= 800\UGeV$ and the corresponding
uncertainty introduced by $\Gamma_{\tot}(\zeta)$ are shown in \refF{fig:HTO_14}.

The theoretical uncertainties coming from $\gh$ and $\Gamma_{\tot}$ can be combined
but a meaningful prediction requires that we cut the Higgs virtuality ($\zeta$) at
some upper value for which we have selected $\zeta_{\max} = (1.5\UTeV)^2$. Results are
shown in \refT{tab:HTO_THUC7} and in \refT{tab:HTO_THUC8}, otherwise the non-perturbative 
increasing of $\Gamma_{\tot}(\zeta)$ overcompensates the decreasing of $\sigma_{ij \to \PH}$ 
with a negligible smearing from the invariant mass distribution.
\begin{table}[t]
\begin{center}
\caption[]{\label{tab:HTO_THUC7}{Total theoretical uncertainty on the production 
cross-section at $7\UTeV$.
The total is obtained by considering the THU on $\gh$ and on $\Gamma_{\tot}$ with a cut
$\sqrt{\zeta} < 1.5\UTeV$.} }
\vspace{0.2cm}
\begin{tabular}{cc}
\hline 
$\muh$[GeV]                  & 
$\delta\sigma^{\myprod}[\%]$ \\
\hline
$600$  & ${-}5.5\;{}\;{+}5.9$  \\
$700$  & ${-}7.0\;{}\;{+}7.5$  \\
$800$  & ${-}7.7\;{}\;{+}8.8$  \\
$900$  & ${-}7.0\;{}\;{+}8.9$  \\
\hline
\end{tabular}
\end{center}
\end{table}
\begin{table}[t]
\begin{center}
\caption[]{\label{tab:HTO_THUC8}{Total theoretical uncertainty on the production 
cross-section at $8\UTeV$.
The total is obtained by considering the THU on $\gh$ and on $\Gamma_{\tot}$ with a cut
$\sqrt{\zeta} < 1.5\UTeV$.} }
\vspace{0.2cm}
\begin{tabular}{cc}
\hline 
$\muh$[GeV]                  & 
$\delta\sigma^{\myprod}[\%]$ \\
\hline
$600$  & $  {-}5.2\;{}\;{+}        5.8$ \\
$700$  & $  {-}6.3\;{}\;{+}        7.4$ \\
$800$  & $  {-}6.7\;{}\;{+}        8.4$ \\
$900$  & $  {-}6.0\;{}\;{+}        8.1$ \\
\hline
\end{tabular}
\end{center}
\end{table}
For completeness we present the leading (in the limit $M^2_{\PH} \to \infty$) $K\,$-factor 
for the $\PH \to \PV\PV$ amplitude ($\PV= \PW, \PZ$) in the OS-scheme and in the CPP-scheme:
\bqa
\hbox{OS} &\qquad & K \sim 1 + a_1\,\frac{G_{\ssF} M^2_{\PH}}{16 \sqrt{2} \pi^2} 
+ a_2\,\Bigl(\frac{G_{\ssF} M^2_{\PH}}{16 \sqrt{2} \pi^2}\Bigr)^2
\nl 
\hbox{CPP} &\qquad & K \sim 1 + a_1\,\frac{G_{\ssF} \cph}{16 \sqrt{2} \pi^2} 
+ \lpar a_2 + 3 i \pi\,a_1 \rpar\,\Bigl(\frac{G_{\ssF} \cph}{16 \sqrt{2} \pi^2}\Bigr)^2,
\eqa
where the coefficients are~\cite{Frink:1996sv}
\bq
a_1= 1.40 - 11.35\,i \qquad a_2 = - 34.41 - 21.00\,i
\eq
Above $1\UTeV$ the NNLO term dominates the $K\,$-factor.
\subsection{Interference signal/background}
The most important factor that has to be observed here is that our analysis of the
residual theoretical uncertainty refers to the signal only. In the current experimental
analysis there are additional sources of uncertainty, \eg background and Higgs interference 
effects. As a matter of fact, this interference is partly available and should not be included
as a theoretical uncertainty; for a discussion and results we refer to 
\Brefs{Campbell:2011cu,Kauer:2012ma}.
In particular, from \Brefs{Campbell:2011cu} we see that at $\muh= 600\UGeV$ (the highest value
reported) the effect is about $+40\%$ in the window $\zeta= 440{-}560\UGeV$, is
practically zero at the peak and reaches $-50\%$ after $\zeta= 680\UGeV$ (no cuts applied). 
For the total cross-section in $\Pg\Pg \to \Pl\PGn\Pl'\PGn'$ at $\muh = 600\UGeV$ the
effect of including the interference is already $+34\%$ and rapidly increasing with
$\muh$.

We stress that setting limits without including the effects of the interference induces 
large variations in rate and shape that will propagate through to all distributions. Therefore, 
any attempt to analyze kinematic distributions which are far from the SM shape may result 
in misleading limits.
\section{Conclusions}
In the last two decades many theoretical studies lead to improved estimates of the
SM Higgs boson total cross section at hadron 
colliders~\cite{Dawson:1991zj,Djouadi:1991tka,Spira:1995rr,Kramer:1996iq,Harlander:2001is,Anastasiou:2002yz,Ravindran:2002dc,Catani:2003zt} (including electroweak NLO 
corrections~\cite{Actis:2008ug,Actis:2008ts}) and of Higgs boson partial 
decay widths~\cite{Bredenstein:2007ec,Prophecy4f}. Less work has been devoted to studying 
the Higgs boson invariant mass distribution 
(Higgs--boson-lineshape)~\cite{Anastasiou:2011pi,Anastasiou:2012hx,Alioli:2008tz}.

In this work we made an attempt to resolve the problem by comparing different theoretical 
inputs to the off-shellness of the Higgs boson. There is no question at all that the 
zero-width approximation should be avoided, especially in the high-mass region where the 
on-shell width becomes of the same order as the on-shell mass, or higher.

The propagation of the off-shell $\PH$ is usually parametrized in terms of some Breit-Wigner 
distribution; several variants are reported in the literature, \eg \eqn{BWdef} and
\eqn{XBWdef}, see \Bref{Seymour:1995qg} for a discussion.
We have shown evidence that only the Dyson-resummed propagator should be used, 
leading to the introduction of the $\PH$ complex pole, a gauge-invariant property
of the $S\,$-matrix.

Finally, when one accepts the idea that the Higgs boson is an off-shell intermediate
state the question of the most appropriate choice of QCD scales arises. We have shown
the effect of including dynamical choices for QCD scales, $\muR, \muF$, instead of a 
static choice.

For many purposes the well-known and convenient machinery of the on-shell approach can
be employed but one should become aware of its limitations and potential pitfalls.
Although the basic comparison between different schemes was presented in this work, at
this time we are repeating that differences should not be taken as the source of
additional theoretical uncertainty; in \refS{Sect_gi} we have shown that the
CPP-scheme, described in \refS{schemes} (see \eqn{cpp}), is the only the only
theoretically consistent tool for describing the lineshape; it requires, however, 
the computation of Feynmann diagrams on the second Riemann-sheet. As a consequence, our 
simple pragmatic recommendation is to use the OFFP-scheme described in \refS{schemes} (see 
\eqn{offp}) with running QCD scales. 
The associated theoretical uncertainty has been discussed in \refS{Sect_THU} where we also
examine the limitations induced by an apparent breakdown of the perturbative expansion in
the ultra-heavy Higgs boson limit.

Finally, the most severe problem faced by a critical appraisal of the heavy Higgs boson lineshape 
is the missing inclusion of interference with the background, especially in the light of
gauge invariance issues. 
\Acknowledgments
We gratefully acknowledge several important discussions with C.~Anastasiou, M.~Cacciari, 
G.~Carrillo, A.~Denner, S.~Dittmaier, S.~Forte, S.~Frixione, M.~Grazzini, F.~Maltoni, 
C.~Mariotti, P.~Nason, C.~Oleari, S.~Pozzorini, B.~Quayle, D.~Rebuzzi, M.~Spira, C.~Sturm, 
R.~Tanaka and S.~Uccirati.
This work has been performed within the Higgs Cross-Section Working Group\\
{\tt https://twiki.cern.ch/twiki/bin/view/LHCPhysics/CrossSections}.


\clearpage

\clearpage

\clearpage

\clearpage
\appendix
\section{Appendix: Nielsen identities at work \label{appA}}
In this Appendix we consider one specific example of application of Nielsen identities (
see \Bref{Gambino:1999ai}) at one loop level, namely we explain of to get gauge-parameter 
independence in the process
\bq
\PH(P) \to \PZ(p_1,\mu) + \PZ(p_2,\nu).
\eq
We work in the $R_{\xi}\,$-gauge; for any quantity $f(\xi)$ we write a decomposition
\bq
f(\xi) = f(1) + \Delta f(\xi), \qquad \Delta f(1)= 0.
\eq
The most important quantity is the Higgs self-energy,
\bq
S_{\HH}(s) = S^{(1)}_{\HH} + \ord{g^4} =\frac{g^2}{16\,\pi^2}\,\Sigma_{\HH}(s) +
\ord{g^4}, \qquad s = - P^2.
\eq
Let $M_{\PH}$ be the renormalized Higgs mass, we obtain that the one-loop,
$\xi\,$-dependent, part can be given as
\bq
\Delta\Sigma^{(1)}_{\HH}(\xi,s,M^2_{\PH}) = \lpar s - M^2_{\PH} \rpar\,
\sigma^{(1)}_{\HH}(\xi,s,M^2_{\PH}).
\eq
The main equation is the one for the Higgs complex pole,
\bq
\cph - M^2_{\PH} + S^{(1)}_{\HH}\lpar \xi\,,\,\cph\,,\,M^2_{\PH} \rpar = 0,
\eq
from wich we derive 
\bq
M^2_{\PH} = \cph + \ord{g^2}, 
\qquad
\frac{\partial}{\partial\xi}\,S^{(1)}_{\HH}\lpar \xi\,,\,\cph\,,\,\cph \rpar = 0.
\eq
Next we consider the one-loop vertices contributing to $\PH \to \PZ\PZ$ and obtain
a one-loop amplitude
\bq
A^{(1)}_{\ssV}\lpar \PH \to \PZ\PZ\rpar =
\bigl( V^{(1)}_d\,\delta_{\mu\nu} + V^{(1)}_p\,p_{2\mu}\,p_{1\nu} \bigr)\,
e^{\mu}\lpar p_1,\lambda_1\rpar\,e^{\nu}\lpar p_2,\lambda_2\rpar.
\eq
Using the decomposition
\bq
V^{(1)}_{d,p}\lpar \xi\,,\,s\,,\,M^2_{\PH}\rpar = 
V^{(1)}_{d,p}\lpar 1\,,\,s\,,\,M^2_{\PH}\rpar  + 
\Delta V^{(1)}_{d,p}\lpar \xi\,,\,s\,,\,M^2_{\PH}\rpar 
\eq
we obtain the following results:
\begin{enumerate}
\item after reduction to scalar form-factors there are no scalar vertices remaining
in $\Delta V^{(1)}_d\lpar \xi\,,\,\cph\,,\,\cph\rpar$.
\item Furthermore, $\Delta V^{(1)}_p\lpar \xi\,,\,\cph\,,\,\cph\rpar = 0$.  
\end{enumerate}
Both results are expected since the corresponding terms could not cancel against anything
else. Furthermore we compute renormalization $\PZ\,$-factors for the external 
legs~\cite{Lehmann:1957zz}
\bqa  
s - M^2_{\PH} + S^{(1)}_{\HH}\lpar \xi\,,\,s\,,\,M^2_{\PH} \rpar &=& 
\lpar s - \cph \rpar\,
\Bigl[ 1 + 
\frac{S^{(1)}_{\HH}\lpar \xi\,,\,s\,,\,\cph \rpar -
      S^{(1)}_{\HH}\lpar \xi\,,\,\cph\,,\,\cph \rpar}{s - \cph} \Bigr]
\nl
{}&=& \lpar 1 + \PZ_{\PH} \rpar\,\lpar s - \cph \rpar + \ord{\lpar s - \cph\rpar^2}.
\eqa
It follows that
\bq
\Delta V^{(1)}_d \lpar \xi\,,\,\cph\,,\,\cph\rpar - 
\Bigl[ \frac{1}{2}\,\Delta \PZ_{\PH}(\xi) + \Delta \PZ_{\PZ}(\xi)\Bigr]\,A^{(0)} = 0,
\eq
where the lowest-order is defined by
\bq
A^{(0)} = -\,g\,\frac{M_{\PW}}{\cos^2\theta},
\qquad cos^2_{\theta} = \frac{M^2_{\PW}}{M^2_{\PZ}}, \quad 
M^2_{\PW,\PZ} = \cpwz.
\eq
At one loop it is enough to evaluate all functions at $s = M^2_{\PH}$ but only
$\cph$ is $\xi\,$-independent.
\section{Appendix: equivalence theorem for (off-shell) virtual bosons \label{appB}}
In this Appendix we discuss the extension of the equivalence theorem to virtual vector-bosons.
Formally, the theorem states, at the $S\,$-matrix level, that
\bq
S\bigl[ \PV_{\ssL} \bigr] = (i\,C)^n\,S\bigl[ \phi \bigr],
\label{ET}
\eq
where $n$ is the number of external, longitudinal, vector-bosons and $C = 1$ at tree level.
In \Bref{Bagger:1989fc} it is shown how one can get $C = 1$ at higher orders through
a choice of the $\phi$ renormalization constant.
We consider the amplitude for the process
\bq
\PH(P) \to \PAf(q_1) + \Pf(k_1) + \PAf(q_2) + \Pf(k_2),
\eq
where all fermion are massless. The amplitude can be split according to the number of
resonant vector-bosons,
\bq
A = A_{2\ssV} + A_{1\ssV} + A_{0\ssV},
\eq
where $A_{2\ssV}$ starts from LO in perturbation theory while $A_{1\ssV}, A_{0\,\ssV}$
start from NLO. Furthermore we have
\bq
A_{2\ssV} = \bigl[ F_d\,\delta^{\mu\nu} + F_p\,\frac{p^{\mu}_1\,p^{\nu}_2}{M^2_{\PH}} \bigr]\,
\Delta^{\ssV}_{\mu\alpha}(p_1)\,\Delta^{\ssV}_{\nu\beta}(p_2)\,
J^{\alpha}(q_1,k_1)\,J^{\beta}(q_2,k_2),
\label{dres}
\eq
where $\Delta_{\ssV}$ is the vector-boson propagator and $J^{\alpha,\beta}$ are
(conserved) fermion currents. In the $\xi = 1$ gauge there are no scalar propagators
involved and each $\PV$-propagator is proportional to $\delta_{\mu\alpha}$ \etc

We are interested in the leading term of the amplitude in the limit $M_{\PH} \to \infty$.
This term derives from a balance between $\PH\,\phi\,$-couplings ($\phi$ stands for a 
generic Goldstone-boson) and $\PH\,$-propagators; at one-loop (or higher) $A_{1\ssV}$ is given 
by boxes and $A_{1\ssV}$ by pentagons. Since external fermions are massless we can easily see 
that there are not enough internal scalars to contribute to the leading term. 

We will use the formalism of \Brefs{Passarino:1986bw,Passarino:1983zs,Passarino:1983bg} 
and will write
\bq
\delta_{\mu\nu} - \frac{p_{\mu} p_{\nu}}{p^2} = e_{\ssL\,\mu}(p)\,e^*_{\ssL\,\nu}(p) +
\sum_{\lambda=-1,+1}\,e_{\perp\,\mu}(p,\lambda)\,e^*_{\perp\,\nu}(p,\lambda).
\eq
Since the fermion currents are conserved we can replace the numerator in the $\PV$-propagators
with the sum over polarizations. A convenient choice for the polarizations is the following:
\bq
e_{\ssL\,\mu}(p_1) = - N^1_{\ssL}\,\bigl( \spro{p_1}{p_2}\,p_{1\mu} + s_1\,p_{2\,\mu} \bigr),
\qquad
e_{\ssL\,\mu}(p_2) = - N^2_{\ssL}\,\bigl( \spro{p_1}{p_2}\,p_{2\mu} + s_2\,p_{1\,\mu} \bigr),
\eq
where $N_{1,2}$ are the normalizations, $p^2_i = - s_i$ and
\bq
e_{\perp\,\mu}(p_i,\lambda) = \frac{1}{\sqrt{2}}\,
\bigl[ n_{\mu}(p_i) + i\,\lambda\,N_{\mu}(p_i) \bigr],
\quad
N_{\mu}(p_i) = (s_i)^{-1/2}\,\ep_{\mu\alpha\beta\rho}\,
n^{\alpha}(p_i)\,e^{\beta}_{\ssL}(p_i)\,p^{\rho}_i,
\eq
\bq
n_{\mu}(p_1) = i\,N^1_{\perp}\,\ep_{\mu\alpha\beta\rho}\,
k^{\alpha}_1\,p^{\beta}_1\,p^{\rho}_2,
\qquad
n_{\mu}(p_2) = i\,N^2_{\perp}\,\ep_{\mu\alpha\beta\rho}\,
k^{\alpha}_2\,p^{\beta}_2\,p^{\rho}_1.
\eq
With this choice one obtains
\bq
\sum_{\lambda=-1,+1}\,e_{\perp\,\mu}(p,\lambda)\,e^*_{\perp\,\nu}(p,\lambda)=
\delta_{\mu\nu} - \frac{p_{\mu} p_{\nu}}{p^2} - e_{\ssL\,\mu}(p)\,e^*_{\ssL\,\nu}(p).
\eq
The first question to answer is about $L\,$-polarization dominance in Higgs decay for 
$M_{\PH} \to \infty$\footnote{We gratefully acknowledge A.~Denner and S.~Dittmaier for an 
important discussion on this point.}. For the $F_d, F_p$ parts in \eqn{dres} we have to evaluate
\bqa
{}&{}& \sum_{\lambda_1,\lambda_2}\,\spro{e(p_1\,\lambda_1)}{e(p_2\,\lambda_2)}\,
e^*_{\alpha}(p_1,\lambda_1)\,e^*_{\beta}(p_2,\lambda_2),
\nl
{}&{}& \sum_{\lambda_1,\lambda_2}\,\spro{e(p_1\,\lambda_1)}{p_2}\,
\spro{e(p_2\,\lambda_2)}{p_1}\,e^*_{\alpha}(p_1,\lambda_1)\,e^*_{\beta}(p_2,\lambda_2).
\eqa
First we introduce invariants for the complete process
\bqa
\spro{q_1}{k_1} &=& - \frac{1}{2}\,s_1,
\quad
\spro{q_2}{k_2} = - \frac{1}{2}\,s_2,
\quad
\spro{q_1}{q_2} = - \frac{1}{2}\,s_3,
\nl
\spro{q_1}{k_2} &=& - \frac{1}{2}\,s_4,
\quad
\spro{k_1}{q_2} = - \frac{1}{2}\,s_5,
\quad
\spro{k_1}{k_2} = - \frac{1}{2}\,s_6,
\eqa
where the invariants satisfy $\sum_i\,s_i = M^2_{\PH}$. Therefore, not all scales in the problem 
can be small as compared to $M^2_{\PH}$. For the $\PV\,$ squared-propagators we have
\bq
\frac{1}{(s - M^2_{\PV})^2 + \Gamma^2_{\PV}\,M^2_{\PV}} =
\frac{\pi}{M_{\PV}\,\Gamma_{\PV}}\,\delta\lpar s - M^2_{\PV}\rpar +
\mbox{PV}\,\Bigl[\frac{1}{(s - M^2_{\PV})^2}\Bigr] + \dots 
\eq
Therefore, always in the limit $M_{\PH} \to \infty$, we can assume that 
$s_{1,2} \approx M^2_{\PV} \to 0$. In the rest-frame of the Higgs boson and using massless fermions
in the final state, the condition $s_{1,2} = 0$ requires $s_i = M^2_{\PH}/4$ for
$i=3,\dots,6$. Next we introduce
\bq
D_{\ssL\,;\,\ssL} = \bmid \spro{e_{\ssL}(p_1)}{e_{\ssL}(p_2)} \bmid^2,
\quad
D_{\perp -\,;\,\perp -} = \bmid \spro{e_{\perp}(p_1,-1)}{e_{\perp}(p_2,-1)} \bmid^2,
\eq
\etc in the limit $s_{1,2} \to 0$ and $M_{\PH} \to \infty$. We obtain
\bq
D_{\ssL\,;\,\ssL} = \frac{1}{4}\,\frac{M^4_{\PH}}{s_1 s_2} -
\frac{1}{2}\,\frac{s_1 + s_2}{s_1 s_2}\,M^2_{\PH} + \ord{1},
\label{Dres}
\eq
while all other combinations are finite in the limit. We also introduce non-diagonal elements,
\bq
N_{\lambda\sigma\rho\tau} = \spro{e(p_1,\lambda)}{e(p_2,\sigma)}\,
\bigl[ \spro{e(p_1,\rho)}{e(p_2,\tau)} \bigr]^*,
\eq
where $e(p,0) = e_{\ssL}(p)$ and $e(p,\pm 1) = e_{\perp}(p,\pm 1)$. Also these elements
are finite or zero in the limit $s_1, s_2 \to 0$.
Furthermore, we introduce other quantities
\bq
P_{\ssL\,;\,\ssL} = \bmid \spro{e_{\ssL}(p_1)}{p_2}\,\spro{e_{\ssL}(p_2)}{p_1} \bmid^2,
\eq
\etc (including non-diagonal terms) and find
\bq
P_{\ssL\,;\,\ssL} = \frac{1}{16}\,\frac{M^8_{\PH}}{s_1 s_2} -
\frac{1}{4}\,\frac{s_1 + s_2}{s_1 s_2}\,M^6_{\PH} + \ord{M^4_{\PH}},
\label{Pres}
\eq
while all other combinations are finite in the limit. Finally we define the fermion part of the
amplitude, squared and summed over spins:
\bq
\Gamma_{\ssL} = \sum_{\rm spins}\,\bmid \spro{J(q,k)}{e_{\ssL}(p)} \bmid^2,
\qquad
\Gamma_{\perp\,;\,\lambda} = \sum_{\rm spins}\,\bmid \spro{J(q,k)}{e_{\perp}(p,\lambda)} \bmid^2,
\eq
where $p = q + k$ and
\bq
J^{\mu}(q,k) = {\overline{u}}(k)\,\gamma^{\mu}\,\lpar V_+\,\gamma_+ + V_-\,\gamma_-\rpar\,v(q),
\qquad \gamma_{\pm} = 1 \pm \gamma^5.
\eq
We obtain
\bq
\Gamma_{\ssL} = 4\,\lpar V^2_+ + V^2_- \rpar\,s + \ord{s^2}\,
\qquad
\Gamma_{\perp\,;\,\lambda} = 4\,\lpar V^2_+ + V^2_- \rpar\,s + \ord{s^2},
\eq
where $s = - p^2$. We also compute non-diagonal elements 
\bq
\Gamma_{\lambda,\sigma} = \sum_{\rm spins}\, 
\spro{J(q,k)}{e(p,\lambda)}\,\bigl[ \spro{J(q,k)}{e(p,\sigma)} \bigr]^{\dagger},
\eq
and find
\bq
\Gamma_{0\,,\, \pm 1} =  2\,\sqrt{2}\,i \lpar V^2_+ - V^2_- \rpar\,s + \ord{s^2},
\qquad
\Gamma_{-1\,,\, +1} =  2 \lpar V^2_+ + V^2_- \rpar\,s + \ord{s^2}.
\eq
Therefore, all $\Gamma$ factors are suppressed in the low-$\,s$ region and,
according to \eqns{Dres}{Pres}, only longitudinal vector bosons contribute to the leading term 
for $M_{\PH} \to \infty$.
In conclusion, for the double-resonant part of the amplitude, only the longitudinal polarizations 
matter. One-loop corrections are computed exactly in \Bref{Prophecy4f} which also includes 
two-loop leading corrections; they are extracted from the Higgs-Goldstone model and 
{\sc Prophecy4f}{} multiplies the complete LO width with the NNLO correction factor. 
For the contributions of the transverse vector-bosons it is not correct but irrelevant in the
limit $M_{\PH} \to \infty$, as we have explicitly shown. 

\begin{figure}
\vspace{-3.5cm}
\centering
\begin{minipage}{.9\textwidth}
\centerline{\includegraphics[bb=  0 0 595 842,width=14.cm]{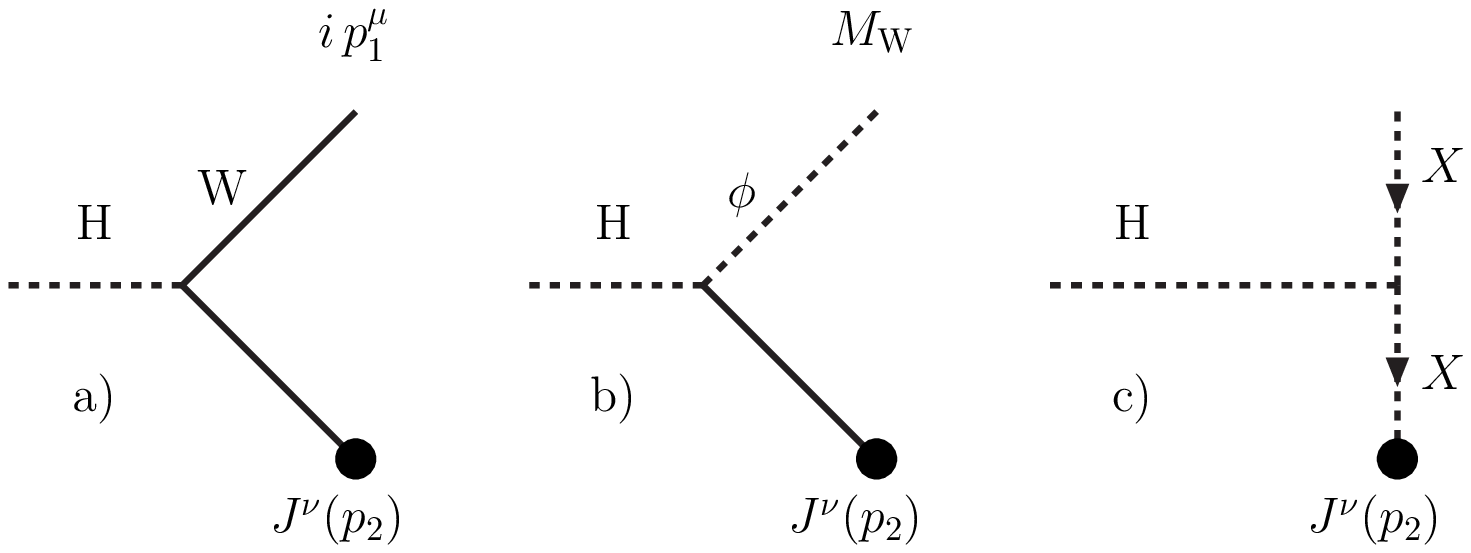}}
\vspace{-11.4cm}
\caption[]{Example of a LO simply-contracted WSTI where the dash-line represents the 
Higgs boson, solid line represent vector-bosons and dash-arrow--lines represent 
Goldstone-bosons or Faddeev-Popov ghosts. If the source $J$ is physical then
$a + b = c = 0$, otherwise $a + b + c = 0$}
\label{HLS:LOWI}
\end{minipage}
\end{figure}

\begin{figure}
\vspace{-4.5cm}
\centering
\begin{minipage}{.9\textwidth}
\centerline{\hspace{3.5cm}\includegraphics[bb=  0 0 595 842,width=14.cm]{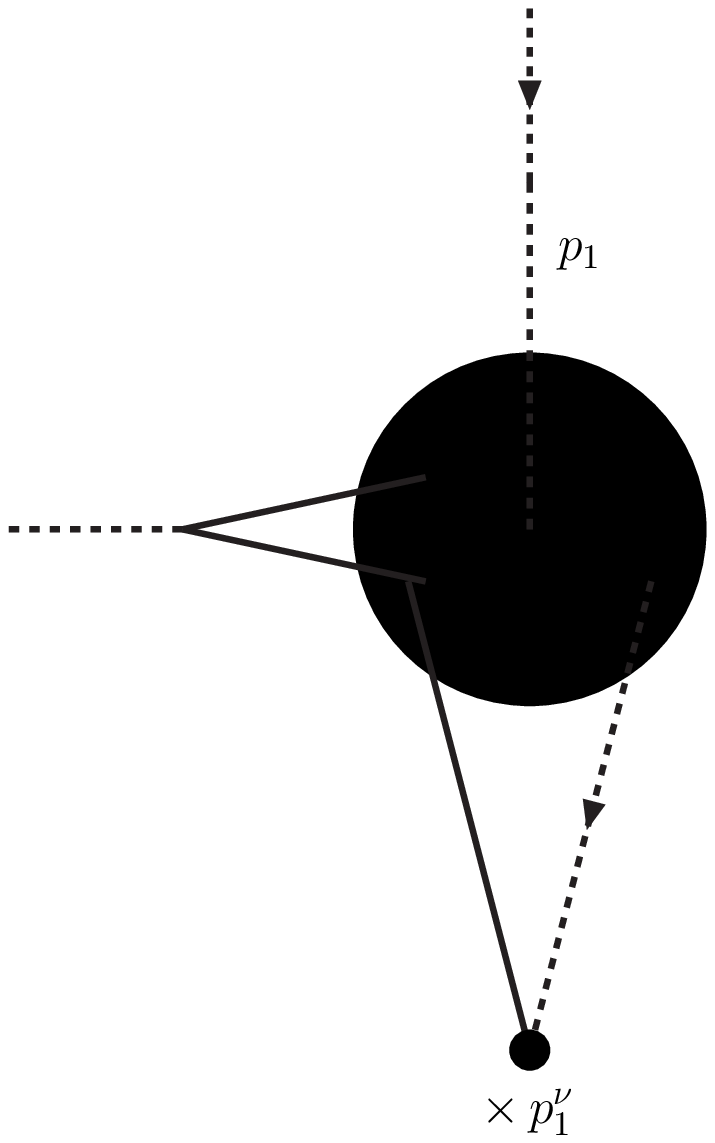}}
\vspace{-8.4cm}
\caption[]{Example of a remainder for a NLO simply-contracted WSTI where the dash-line 
represents the Higgs boson, solid line represent vector-bosons and dash-arrow--lines 
represent Faddeev-Popov ghosts.}
\label{HLS:ETWI}
\end{minipage}
\end{figure}

Before concluding our argument we need to explain how the equivalence theorem (see 
\Bref{Bagger:1989fc}) works in practice for our off-shell vector bosons.
The theorem is based on a Ward-Slavnov-Taylor identity which relates the amplitude for a 
$\PV\,$-boson contracted with its momentum to the corresponding amplitude where the 
$\PV\,$-boson is replaced by a Goldstone-boson $\phi$. However, in our case we have
\bq
e_{\ssL\,\mu}(p_1) = - N^1_{\ssL}\,\bigl( \spro{p_1}{p_2}\,p_{1\mu} + s_1\,p_{2\,\mu}\bigr),
\eq
and the longitudinal polarization is not simply proportional to the momentum. To see how it works
in the $M_{\PH} \to \infty$ limit, we consider the LO $\PH\PWp\PWm$ vertex contracted with
longitudinal polarizations
\bq
V_{\ssL\,;\,\ssL} = - g\,M_{\PW}\,\spro{e_{\ssL}(p_1)}{e_{\ssL}(p_2)},
\qquad
e_{\ssL\,\mu}(p_1) = - N^1_{\ssL}\,\bigl( L\,\spro{p_1}{p_2}\,p_{1\mu} + 
R\,s_1\,p_{2\,\mu} \bigr),
\eq
where we have artificially introduced the $L, R$ coefficients. It is easy to show that
only the $L^2$ part of $V_{\ssL\,;\,\ssL}$ survives in the limit, giving
\bq
V_{\ssL\,;\,\ssL} \sim \frac{1}{2}\,g\,\frac{M_{\PW}}{\sqrt{s_1 s_2}}\,M^2_{\PH},
\eq
which, after identification of $s_i$ with $M^2_{\PW}$ is exactly minus the $\PH \phi^+ \phi^-$
vertex, as dictated by the equivalence theorem (but only in the limit $M_{\PH} \to \infty$). 
As expected, the WSTI relating $\partial_{\mu} \PV^{\mu}$ to $M_{\PV} \phi$, is valid only for 
on-shell particles. The same holds for the combination
\bq
- g\,M_{\PW}\,\spro{e_{\ssL}(p_1)}{p_2}\,\spro{e_{\ssL}(p_2)}{p_1},
\eq
showing that, in the limit $M^2_{\PH} \to \infty$, we can extract the leading behavior from
the Higgs-Goldstone model and we can multiply the LO width with the NNLO correction factor,
channel-by-channel. The only delicate point concerns the WSTI which substitutes two 
(contracted) $\PV\,$-lines with $\phi\,$lines. When only one of the two is contracted and the other
is multiplied by the wrong momentum there is an additional term in the WSTI.
We illustrate this singly-contracted WSTI at LO in \refF{HLS:LOWI}: if the $J\,$-source is
physical ($\spro{p^{\mu}_2}{J_{\mu}(p_2)} = 0$) then the sum of the first two diagrams is zero,
otherwise there is a remainder given by diagram c) of \refF{HLS:LOWI} where Faddeev-Popov ghosts 
are involved.
A similar situation happens also at higher orders, as shown in \refF{HLS:ETWI}, where the 
dash-arrow line represents Faddeev-Popov ghosts and the off-shell $\PV\,$-source contracted 
with the wrong momentum can absorb a $\PV\,$-ghost pair. One can argue, on the basis
of the explicit calculation, that this term is never leading at one-loop but we have not
found a convincing argument that this is the case to all orders.

The above arguments, strictly speaking, apply to the decay of an on-shell Higgs boson. If
this decay is part of a complete process, \eg $\Pg\Pg \to 4\,\Pf$ (where we have 
Higgs-resonant diagrams) then the following happens: let $g$ be the $SU(2)$ coupling
constant and $\lambda= g^2\,M^2_{\PH}/M^2_{\PW}$; the work of \Bref{Bagger:1989fc}
shows that, for $g^2/\lambda \to 0$, the equivalence theorem is true to lowest nonzero order
in $g$ an to all orders in $\lambda$. However, there is no reason to expect the theorem to hold 
when using a propagator with a Breit-Wigner width, which is equivalent to summing a subset of 
diagrams of higher order in $\lambda$.

We conclude the Appendix with another comment: technically speaking \eqn{ET} requires that
$S\bigl[ \phi \bigr]$ is computed in the full SM, not only in the Higgs-Goldstone model. We
have verified in simple examples that the difference is subleading when $M_{\PH} \to \infty$.
\section{Appendix: complex poles, real kinematics and pseudo-observables \label{appC}}
In this Appendix we discuss how to implement complex poles in the amplitude for
$\PH \to \PF$ within the CPP-scheme. For sake of simplicity we consider the process $\PH(P) \to
\PZ(p_1) + \PZ(p_2)$ where both $\PZ$s are on-shell while the Higgs boson is
off-shell, \ie $P^2 = - s$. At LO the $S\,$-matrix element is
\bq
S_{\LO} = - g\,\frac{M_{\PZ}}{\cos\theta}\,\spro{e(p_1)}{e(p_2)}.
\eq
Squaring the amplitude and summing over polarization introduces a $P^2$ which is a
kinematic factor and should be replaced with $s$
\bq
\sum_{\rm spins}\,\bmid S_{\LO}\bmid^2 = \sqrt{2}\,G_{\ssF}\,s^2\,\bigl(
1 - 4\,x_{\PZ} + 12\,x^2_{\PZ} \bigr),
\qquad
x_{\PZ} = \frac{M^2_{\PZ}}{s}.
\eq
At LO there is never a problem with gauge parameter dependence. At NLO there is an
amplitude 
\bq
A^{\mu\nu}_{\NLO}\lpar s\,,\,M^2_{\PH} \rpar =
s\,F_d\lpar s\,,\,M^2_{\PH} \rpar\,\delta^{\mu\nu} + 
F_p\lpar s\,,\,M^2_{\PH} \rpar\,p^{\nu}_1\,p^{\mu}_2,
\eq
where the $M^2_{\PH}$ arises from vertices and propagators, always with $P^2 = - s$.
Renormalization in the complex-mass scheme requires the replacement $M^2_{\PH} \to
\cph$; taking the residue at the complex pole also requires $s \to \cph$ and guarantees
gauge parameter independence. When we go to the $S\,$matrix element and compute the NLO
correction we obtain
\bq
2\,\Re\,\sum_{\rm spins}\,S^{\dagger}_{\LO}\,S_{\NLO} =  
- 144\,\bigl( \sqrt{2} G_{\ssF} \bigr)^{3/2}\,M_{\PW} s^2\bigl[
x_{\PZ}\,\frac{\Re \cph F_d }{s} + \frac{1}{12}\,
\bigl( 1 - \frac{1}{6} \frac{1}{x_{\PZ}} + \frac{4}{3}\,x_{\PZ} \bigr)\,\Re F_p\bigr],
\eq
where the form factors are computed at $s = M^2_{\PH} = \cph$. When there are more particles
in the final state the analytical continuation of the form factors should be done
according to \eqns{mACone}{mACtwo} and generalizations, followed by the integration over the 
phase space with real angles and energies.
\subsection{Analytic continuation \label{appCa}}
We continue this Appendix emphasizing, yet once more, that complex kinematics is not needed
since we need to perform analytic continuation in the $S\,$-matrix element (because of
gauge invariance) while keeping the phase-space real; this is the approach introduced in 
\Bref{Denner:2000bj}, the so-called leading-pole-approximation (LPA).
In Appendix A of \Bref{Weldon:1975gu} it is shown that, in order to have a pole on the second 
Riemann sheet independently of the other momentum variables, analytic continuation of the 
scattering amplitude in the variable $k^0$ is required ($k$ being the momentum of the unstable 
particle). Therefore the $S\,$-matrix has to be analytically continued in $k^0$ around the 
positive energy cut to the second sheet and the state vector for the corresponding unstable 
particle must not depend on $k^0$; QFT provides an explicit construction for this state as 
described in details in \Bref{Weldon:1975gu}.

Consider the process where an off-shell Higgs boson of virtuality $\sqrt{\zeta}$ decays into
$n$ massless particles of momentum $q_i$; then the quantity of interest is
\bqa
\Gamma_{\PH \to n}(\zeta) &=&
\frac{1}{2\,\sqrt{\zeta}}\,\int\,d\Phi_{1 \to n}\,\sum_{s,c} 
\bmid A_{\PH \to n}\lpar \{\spro{q_i}{q_j}\} \rpar\bmid^2
\nl
{}&=& \frac{1}{2\,\sqrt{\zeta}}\,\int\,d\PI_{3\,n - 7}\,\sum_{s,c} 
\bmid A_{\PH \to n}\lpar \{I\} \rpar\bmid^2,
\label{ACID}
\eqa
where $\{I\}$ is the set of $3\,n-7$ invariants, linearly independent, needed to describe 
the process.
Another quantity of interest is the production cross-section for an off-shell Higgs
boson plus jets, \eg $\Pg(p_1) + \Pg(p_2) \to \Pg(q_1) + \PH(q_2)$, with $q^2_2= \zeta$,
\bqa
\sigma_{\Pg\Pg \to \PH + n\,Pg}(\zeta) &=&
\frac{1}{2\,s}\,\int\,d\Phi_{2 \to n + 1}\,\sum_{s,c} 
\bmid A_{\Pg\Pg \to \PH + n\,\Pg}\lpar \{\spro{q_i}{q_j}\} \rpar\bmid^2
\nl
{}&=& \frac{1}{2\,s}\,\int\,d\PI_{3\,n - 1}\,\sum_{s,c} 
\bmid A_{\Pg\Pg \to \PH + n\,\Pg}\lpar \{I\} \rpar\bmid^2,
\label{ACIP}
\eqa
An extension to complex kinematics is possible and the relevant work can be found 
in \Bref{Weldon:1975gu}; here we repeat the construction for completeness.
We consider distorsions of the original real contour that project one-to-one onto the 
original contour under the projection that sends complex momentum onto its real part.
The contour of integration may be distorted in any bounded region provided the
integrand is analytic through the region of distorsion.

Suppose that, in a given process, $k_c$ is the momentum of an unstable )incoming or outgoing)
Higgs particle and $k$ is its projection back to the real axis; the general form of $k_c$ will 
be (here we work in the Bjorken-Drell metric)
\bqa
k_c^{\mu} &=& p^{\mu} + i\,q^{\mu},
\qquad
k_c^2= \muhs - i\,\muh \gh = {\hat{\mu}_{\PH}}^2,
\nl
{\hat \mu}_{\PH} &=& \frac{1}{\sqrt{2}}\,\muh\lpar 1 + \frac{\mBh}{\mh}\rpar^{1/2} -
\frac{i}{\sqrt{2}}\,\gh\lpar 1 + \frac{\mBh}{\mh}\rpar^{-1/2}.
\label{ACone}
\eqa
The vectors $p, q$ must satisfy
\bq
p^2 - q^2 = \muhs,
\qquad
\spro{p}{q} = - \frac{1}{2}\,\muh \gh.
\label{Ccon}
\eq
To have a resonance $k$ must be a timelike momentum and we can go to a frame
$k = \lpar \sqrt{\zeta}\,,\,{\vec{0}}\rpar$ where $\zeta$ is the Higgs virtuality.
We have chosen the following analytic continuation, yielding the complex pole: 
\bq
\mbox{Step}\;1) \quad k_{\zeta} = L\lpar k_{\zeta}\,,\,k\rpar\,k, \qquad
k_{\zeta} = \lpar \sqrt{\zeta}\,,\,{\vec{0}}\rpar,
\label{OSCone}
\eq
where $L$ is a Lorentz boost;
\bq
\mbox{Step}\;2) \quad k_{\zeta} \to k_{\ssM} = \lpar \muh\,,\,{\vec{0}}\rpar, 
\label{OSCtwo}
\eq
where the analytic continuations is defined by
\bq
k_{\lambda} = \frac{1}{2\,\sqrt{\zeta}}\,
\Bigl( (2 - \lambda)\,\zeta + \lambda\,\muhs\,,\,0\,,\,0\,,\,2 \lambda (\muhs - \zeta)\Bigr),
\quad 0 \le \lambda \le 1.
\eq
Finally, we boost back in the ${\vec k}$ direction and continue into the complex plane
as follows:
\bq
\mbox{Step}\;3) \quad \to k_c = p + i\,q,
\quad
p = \lpar \Re\,k_c^0\,,\,{\vec k}\rpar,
\quad 
q = \lpar Im\,k_c^0\,,\,{\vec 0}\rpar.
\label{ACtwo}
\eq
The continuation in \eqns{OSCone}{OSCtwo} is the real, on-shell, projection in LPA.
Since $q^2 > 0$ from \eqn{Ccon} it follows that $p^2 > 0$. Any such state may be
transformed by a real Lorentz transformation into one with standard momentum
${\overline k}$ defined by
\bq
{\overline p} = \lpar \frac{\spro{p}{q}}{\sqrt{q^2}}\,,0\,,\,0\,,\,
\sqrt{\frac{G(p,q)}{q^2}}\rpar,
\qquad
{\overline q} = \lpar \sqrt{q^2}\,,{\vec{0}}\rpar,
\label{standm}
\eq
where $G(p,q)= p^2 q^2 - (\spro{p}{q})^2$ is the Gram determinant of $p, q$.

Once complex boosts are allowed states with momentum $k_c$ satisfying $k_c^2 = {\hat{\mu}}^2_{\PH}$
can be constructed from \eqn{standm}. However, states appearing in scattering processes have a
timelike momentum. Therefore, the only unstable particle states that appear in
physical processes have a momentum $k_c$ that becomes timelike when continued back to
the real axis, \ie $p^2 >0$ is needed to produce resonances while the associated $q$ is not
constrained. In the additional two cases we have standard momenta (\ie in each case
$k_c$ is related by a real Lorentz transformation to the standard momentum ${\overline k}$
\bqa
q^2 &=& 0,
\quad 
{\overline p} = \lpar \frac{\lambda^2 p^2 + (\spro{p}{q})^2}{2 \lambda\,\spro{p}{q}}\,,0\,,0\,,
\frac{\lambda^2 p^2 - (\spro{p}{q})^2}{2 \lambda\,\spro{p}{q}}\rpar,
\quad
{\overline q} = \lpar \lambda\,,0\,,0\,,\lambda\rpar,
\nl
q^2 &<& 0,
\quad 
{\overline p} = \lpar \sqrt{\frac{G(p,q)}{q^2}}\,,\,0\,,0\,\frac{\spro{p}{q}}{\sqrt{- q^2}}\rpar
\quad
{\overline q} = \lpar {\vec 0}\,,\,\sqrt{- q^2}\rpar.
\label{standmm}
\eqa
The possibility of having two parametrizations connected by a complex boost (an element of
$\ssL(\ssC)$ the complex, homogeneous Lorentz group) may seem like 
posing a question of uniqueness, but it does not since scalar products are defined without
complex conjugation of one of the complex four-vectors.

Since we perform our analytic continuation on the invariants (see \eqns{ACID}{ACIP}) we have to 
define the mapping. Consider again the process $\Pg(p_1) + \Pg(p_2) \to \Pg(k_1) + \PH(k_2)$
and define $s = (p_1 + p_2)^2$, $t = (p_1 - k_2)^2$ and
\bq
t= -\,\frac{s}{2}\,\lpar 1 - \frac{\zeta}{s}\rpar\,\lpar 1 - \cos\theta\rpar
\quad \to \quad 
t_c= -\,\frac{s}{2}\,\lpar 1 - \frac{\cph}{s}\rpar\,\lpar 1 - \cos\theta\rpar,
\eq
where $s$ has been kept real to preserve $s= 4\,E^2$, ($E$ being the c.m.s. energy) and $\theta$
is the scattering angle, also real. We add one condition to the mapping $\{I\} \to \{I_c\}$,
that after continuation all the invariants are analytic functions of $\cph$. It is
easily seen that the condition is fulfilled by 
\bq
{\overline k}^c_2 = {\overline p} + i\,{\overline q},
\qquad
{\overline p} = \frac{1}{2\,\sqrt{s}}\,\lpar s + \muhs\,,\,0\,,\,0\,,\, - s + \muhs\rpar,
\quad
{\overline q} = -\,\frac{\muh \gh}{2\,\sqrt{s}}\,\lpar 1\,,\,0\,,\,0\,,\,1\rpar,
\eq
\ie $q^2= 0$, corresponding to
\bq
{\overline k}^c_2 = \frac{1}{2\,\sqrt{s}}\,\lpar \cph + s\,,\,0\,,\,0\,,\, \cph - s\rpar.
\eq
In principle there is no need to discuss crossed channels separately. Indeed, any
function $A(\{p\})$, after going through complex momenta, may be analytically continued 
back to the real axis but with negative energies for some of the incoming or outgoing particles. 
The result of this analytical continuation is the amplitude for a different physical process
(\eg $\PH \to \Pg\Pg\Pg$). One should only pay attention to the fact that
\bq
A\lpar\{I\}\rpar =\, < a | S | b >
\quad
{\overline A}\lpar\{I\}\rpar =\, < b | S | a >
\qquad
{\overline A}_c\lpar\{I_c\}\rpar \not= A^{\dagger}_c\lpar\{I_c\}\rpar,
\eq
for complex invariants. Let us consider the decay of a Higgs boson of virtuality $P^2= \zeta$
into thre massless particles of momenta $k_1 \dots k_3$. We define invariants
$s_{ij} = (k_i + k_j)^2$ and angles $\theta_{ij}$. Select one invariant, say $s_{12}$, to
remain real and keep $\theta_{12}$ also real. The analytic continuation is defined by 
\bq
s^c_{23} = \frac{1}{2}\,\Bigl[ \zeta - s_{12} - \Bigl( (\zeta - s_{12})^2 - 
4\,c\,\zeta s_{12}\Bigr)^{1/2}\Bigr],
\quad  
s^c_{13} = \frac{1}{2}\,\Bigl[ \zeta - s_{12} + \Bigl( (\zeta - s_{12})^2 - 
4\,c\,\zeta s_{12}\Bigr)^{1/2}\Bigr],
\eq
with $c= \cot^2(\theta_{12}/2)$. Of course, we can keep any of the $s_{ij}$ real and define
analytic continuations of the same function, $\Gamma_{\PH \to 3}$, in different portions of 
the $3\,$-dimensional, complex, space $s_{12}, s_{13}$ and $s_{23}$: in the sense of analytic
continuation there is no ambiguity. The choice of which invariant remains real is equivalent
to the choice of which momentum is fixed by conservation. To follow the argument we write
\bqa
\Gamma_{\PH \to 3}(\zeta) &=&
\frac{1}{2\,\sqrt{\zeta}}\,\int\,d\PI_2
\,\sum_{s,c} \bmid A_{\PH \to 3}\lpar \{I\} \rpar\bmid^2 = 
\frac{1}{2\,\sqrt{\zeta}}\,\int_0^{\zeta}\,d s_{12}\,\int_0^{\zeta-s_{12}}\,d s_{13}
\,\sum_{s,c} \bmid A_{\PH \to 3}\lpar \zeta\,,\,s_{12}\,,\,s_{13} \rpar\bmid^2 
\nl
{}&=& \frac{1}{2\,\sqrt{\zeta}}\,\int_0^{\zeta}\,d s_{12}\,\int_{-1}^{+1}\,d\cos\theta_{12}
\,J\lpar \cos\theta_{12}\,,\,s_{12}\,,\,\zeta\rpar\,\Theta\,
\sum_{s,c} \bmid A_{\PH \to 3}\lpar \zeta\,,\,s_{12}\,\,\cos\theta_{12} \rpar\bmid^2,
\eqa
where $\Theta$ is the condition $(\zeta - s_{12})^2 \ge 4\,c\,\zeta s_{12}$, 
\bq
\Gamma_{\PH \to 3}(\zeta) =
\frac{1}{2\,\sqrt{\zeta}}\,\int_{-1}^{+1}\,d\cos\theta_{12}\,\int_0^{\zeta_-}\,d s_{12}\,
\,J\lpar \cos\theta_{12}\,,\,s_{12}\,,\,\zeta\rpar\,
\sum_{s,c} \bmid A_{\PH \to 3}\lpar \zeta\,,\,s_{12}\,\,\cos\theta_{12} \rpar\bmid^2,
\eq
\bq
\zeta_- = \lpar \cot\frac{\theta_{12}}{2} - \csc\frac{\theta_{12}}{2}\rpar^2\,\zeta.
\eq
A more democratic approach in dealing with $\PH$ decay into $n$ massless states is 
defined by the following choice:
\bq
p_i \to p_i(\lambda\,,\,\kappa) = \lpar \lambda + i\,\kappa\rpar\,p_i
    \to p^c_i = \lpar \alpha + i\,\beta\rpar\,p_i,
\qquad i=1\,\dots\,n+1
\eq
where $p^2_i = 0$ for $i > 1$ and
\bq
\alpha^2 - \beta^2 = \frac{\muhs}{p^2_1}, 
\qquad
\alpha\,\beta= -\,\frac{1}{2}\,\frac{\muh \gh}{p^2_1}.
\eq
This continuation respects momentum conservation and mass-shell condition for (massless) stable 
particles and works as long as the continuation passes through analytic regions of the integrand 
as $\lambda$ varies from $1$ to $\alpha$ and $\kappa$ varies from $0$ to $\beta$.

There is never any problem with fully inclusive quantities but a careful discussion is needed if
we want to consider differential distributions with cuts: in this case one would like
to integrate a function, whose arguments have been analytically continued into the complex plane, 
over a real (fixed) phase-space. In this case $\Gamma_{\PH \to 3}(\zeta)$ becomes 
\bq
\frac{1}{2\,\sqrt{\zeta}}\,\int_{-1}^{+1}\,d\cos\theta_{12}\,\int_0^{\zeta_-}\,d s_{12}\,
\,J\lpar \cos\theta_{12}\,,\,s_{12}\,,\,\zeta\rpar\,
\sum_{s,c} \bmid A_{\PH \to 3}\lpar \cph\,,\,s_{12}\,\,\cos\theta_{12} \rpar\bmid^2
\eq
To close the discussion we should say that the definition of signal is always conventional
since differences (that are non-resonant) will always be included into background plus
interference. Furthermore, when main emphasis is on differential distributions in 
$\Pp\Pp \to 4\,\Pf$ there is little meaning in splitting the whole process into components;
vice versa, when the problem is with extracting pseudo-observables, \eg $\Gamma_{\PH \to 4\,\Pf}$,
analytic continuation is performed only after integration over all variables but the Higgs 
virtuality. Nevertheless, if one wants to introduce cuts on differential distributions we have
two alternatives. We can define\footnote{We gratefully acknowledge S.~Pozzorini for an important 
suggestion on this point.} 
\bq 
\Gamma_{\PH \to 3}(\cph) = \Re\,\Gamma^c_{\PH \to 3}(\cph),
\eq
\bq
\Gamma^c_{\PH \to 3}(\cph) =
\frac{1}{2\,\sqrt{\cph}}\,\int_{-1}^{+1}\,d\cos\theta_{12}\,\int_{\gamma}\,d s_{12}\,
\,J\lpar \cos\theta_{12}\,,\,s_{12}\,,\,\zeta\rpar\,
\sum_{s,c} \bmid A_{\PH \to 3}\lpar \cph\,,\,s_{12}\,\,\cos\theta_{12} \rpar\bmid^2
\eq
where $\gamma$ is a curve in the complex $s_{12}\,$-plane connecting the origin with the
pont $s_{12} = \cph^-$ ($\cph^-$ being the continuation of $\zeta_-$ from $\zeta$ to $\cph$)
provided the integrand is anlaytic through $\gamma$.
Finally, we project back into the real $s_{12}\,$-axis with a change of variable
\bq
\int_{\gamma}\,d s_{12} = \int_0^{\zeta_-}\,d s_{12}\,J_c\lpar s_{12}\,,\,\zeta\,,\,\cph\rpar,
\eq
where $J_c$ is the Jacobian of the transformation. 

Alternatively the NLO corrections to a process with $n+1$ external legs can be written as a sum 
over $D$, the set of scalar one-loop, $n+1\,$-leg diagrams
\bq
\sum_{s,c}\,\bmid A_{\PH \to n}\lpar \{I\} \rpar\bmid^2 =
\Re\,\sum_{d \in D}\,Q_d\lpar \zeta\,,\,\{I\}\rpar\,
\int_{\ssS_d}\,[dx_d]\,V^{-\ssN_d-\ep}_d\lpar \{x_d\}\,,\,\zeta\,,\,\{I\}\rpar,
\label{Ifirst}
\eq
where $Q_d$ are functions of the invariants, $\{x\}$ are the Feynman parameters that
parametrize the loop integral for diagram $d$, $S_d$ is a simplex in the $x\,$-space, $V_d$ is
a quadratic(linear) form in the Feynman parameters(invariants) and $N_d$ is a positive
integer ($4 - \ep$ is the space-time dimensionality). We can introduce cuts on the real kinematics,
\bq
\Gamma^{\kcut}_{\PH \to n}(\zeta) =
\frac{1}{2\,\sqrt{\zeta}}\,\int\,d\PI^{\kcut}_{3\,n - 7}\,
\sum_{s,c}\,\bmid A_{\PH \to n}\lpar \{I\} \rpar\bmid^2,
\eq
perform first the integration over the phase space with cuts, perform analytic continuation in
one variable ($\zeta \to \cph$) while doing the integration over Feynman parameters
(which requires continuation into the second Riemann sheet). 
The functions $Q_d$ may contain square roots, depending on the masses of the final states and
on their number. For massless final states and $n \le 3$ each $Q_d$ is a rational
function of the invariants. For higher values of $n$ it is well-known that phase-space integrals
may lead to elliptic functions.

To give an example we consider $\PH(p) \to \Pg(k_1) + \Pg(k_2) + \Pg(k_3)$ with
$s= (p_1 - k_1)^2$ and $t= (p - k_3)^2$. For instance, there will be a contribution from the 
direct box with internal top quarks; the fully extrapolated contribution corresponding 
to $Q= 1$ from \eqn{Ifirst} can be cast into the following (manifestly finite) form:
\bq
\intfxy{x}{y}\,\Bigl[
\li{2}{\frac{r_b - r_a}{1 - r_a}} + \li{2}{r_a} - \li{2}{r_b} \Bigr]\,
\frac{1}{(1 - x)\,y\,(x - y)},
\eq
\bq
r_a = \frac{\zeta}{m^2_{\PQt}}\,y\,(1 - x),
\qquad
r_b = \frac{\zeta}{m^2_{\PQt}}\,y\,(1 - y).
\eq
If we impose a cut, \eg  $(1-z_1)\,\zeta \le s \le (1-z_2)\,\zeta$, the result is equally
compact,
\bqa
\intfxy{x}{y}\,\Bigl[
{}&{}& \li{2}{z_1\,\frac{r_b - r_a}{1 - r_a}} + 
       \li{2}{z_1\,\frac{r_a}{1 - z_1\,r_a}}  -
       \li{2}{z_1\,\frac{r_b}{1 - z_1\,r_a}}  
\nl
{}&-& \li{2}{z_2\,\frac{r_b - r_a}{1 - r_a}} - 
      \li{2}{z_2\,\frac{r_a}{1 - z_2\,r_a}}  +
      \li{2}{z_2\,\frac{r_b}{1 - z_2\,r_a}}  
\Bigr]\,
\frac{1}{(1 - x)\,y\,(x - y)}.
\eqa
The technique we have described is easily extendible to an arbitrary number of legs.
Consider a process $N \to 0$ with all incoming momenta; a convenient way of 
describing the boundaries of the phase space, the relations between vectors
and invariants and the non-linear constraints that arise when $N \ge 6$
id the following. If $p_1$ is the momentum associated to the Higgs boson
we redefine n-dimensional momenta and introduce components
\bq
p_1 = \lpar \muh\,,\,0_{n-1}\rpar,
\eq
where $0_k$ is a string of $k$ zero components. Then we introduce 
\bq
\lpar h^{\ssN} \rpar_{ij} = -\,\spro{p_i}{p_j}, \qquad i,j= 1,\,\cdots\,,N,
\eq
\bqa
h^{\ssN}_{kl} &=& h^{\ssN} \qquad \mbox{with row}\;k\;\mbox{and column}\;l\;\mbox{removed},
\nl
h^{\ssN}_{kl\,;\,rs} &=& h^{\ssN}_{kl} \qquad 
\mbox{with row}\;r\;\mbox{and column}\;s\;\mbox{removed},
\eqa
\etc The vector $p_2$ is defined by
\bq
p_2 = \lpar \frac{h^2_{12}}{\muh}\,,\,0_{n-2}\,,\,
\sqrt{\frac{{\rm det}\,h^2}{{\rm det}\,h^1}} \rpar.
\eq
Next n-dimensional vectors will be
\bqa
p_3 &=& 
\lpar \frac{h^3_{13}}{\muh}\,,\,0_{n-3}\,,\,
-\,\frac{
         {\rm det}\,h^{3}_{32}
        }
        {
         \sqrt{
               {\rm det}\,h^1\,{\rm det}\,h^2
              }
        }\,,\,
\sqrt{\frac{{\rm det}\,h^3}{{\rm det}\,h^2}} \rpar,
\nl
p_4 &=& 
\lpar \frac{h^4_{14}}{\muh}\,,\,0_{n-4}\,,\,
-\,
\frac{
      {\rm det}\,h^{4}_{43\,;\,32}
     }
     {
      \sqrt{
            {\rm det}\,h^1\,{\rm det}\,h^2
           }
     }\,,\,
\frac{{\rm det}\,h^{4\,;\,43}}{\sqrt{{\rm det}\,h^2\,{\rm det}\,h^3}}\,,\,
\sqrt{\frac{{\rm det}\,h^4}{{\rm det}\,h^3}}\rpar,
\eqa
etc. If $N = 5$ then momentum conservation gives $p_5= -\sum\,p_i$. If, 
instead $N= 6$, $p_6$ follows from momentum conservation whereas $p_5$ 
requires a fifth component; if $n = 4$ the vanishing of the fifth component 
requires ${\rm det}\,h^5 = 0$. The boundary of the phase space is determined 
by the equations
\bq
{\rm det}\,h^2 < 0, \,\dots\,,{\rm det}\,h^{\ssN-1} < 0.
\eq
As we have mentioned several times the analytic continuation requires knowledge of the branch 
cuts of the integrand. When the Higgs boson is the only massive external line a box can be 
expressed in terms of $R\,$-functions (see Appendix B of \Bref{'tHooft:1978xw}):
\bq
R\lpar y_0\,,\,y_1\rpar = \int_0^1 dy \frac{1}{y - y_0}\,\Bigl[ \ln(y - y_1) -
\ln(y_0 - y_1)\Bigr],
\eq
where $y_{0,1}$ depend on the invariants of the problem. The function $R$ has the following
branch cuts in the complex $y_0, y_1\,$-space:
\[
0 \le x \le 1 
\left\{
\begin{array}{ll}
1) & y_1 = (1 - x)\,y_0 \\
2) & y_1 = (1 - x)\,y_0 + x \\
3) & y_0 = x\,y_1 \\
4) & y_0 = x\,y_1 + 1 - x
\end{array}
\right.
\]
Pentagons etc. are always linear combination of boxes.
\subsection{Pseudo-observables}
We consider now the problem of extracting pseudo-observables from matrix elements. First we 
introduce the two-body phase-space,
\bqa
\Phi_2\lpar s,s_1,s_2\rpar &=& \frac{1}{8\,\pi^2\,\sqrt{s}}\,\int \prod_{i=1,2}\,d^4 p_i\,
\delta^+ \lpar p^2_i + s_i\rpar\,\delta^4\lpar P - p_1-p_2\rpar 
\nl
{}&=& \frac{1}{16\,\pi\sqrt{s}}\,\frac{\lambda^{1/2}\lpar s,s_1,s_2\rpar}{s}\,
\theta\lpar s - (\sqrt{s_1} + \sqrt{s_2})^2 \rpar\,
\stackrel{\sim}{s \to \infty} \frac{1}{16\,\pi\,\sqrt{s}},
\eqa
with $P^2= -s$. Consider again the process $\PH(P) \to \PZ(p_1) + \PZ(p_2)$ with an amplitude
given by
\bq
A^0_{\mu\nu} =  - g\frac{M_{\PW}}{\cos^2\theta}\,\delta_{\mu\nu},
\qquad
A^1_{\mu\nu}=  \frac{g^3}{M_{\PW}}\,\Bigl[
\cph F_d\,\delta_{\mu\nu} + F_p\,p_{1\nu}\,p_{2\mu} \Bigr].
\eq
The form factors are function of the external invariants and of the internal masses but
they are computed at the Higgs complex pole, i.e.
\bq
F_{d,p} = F_{d,p} \lpar \cph,\cph\,,\,M^2_{\PZ},M^2_{\PZ} \rpar,
\eq
so that they are gauge parameter independent. We introduce the following quantity:
\bq
D_2 = \frac{1}{\mid \Delta_{\PH}(s)\mid^2}\,\Phi_2 \lpar s,M^2_{\PZ},M^2_{\PZ} \rpar\,
      \bmid \bigl( A^0_{\mu\nu} + A^1_{\mu\nu}\bigr)\,
      e^{\mu}_{\ssL}(p_1)\,e^{\nu}_{\ssL}(p_2)\bmid^2,
\eq
which is the product of a Higgs propagator times the $S\,$-matrix element for the Higgs
boson into a pair of longitudinal $\PZ$ bosons. We obtain
\bqa
D_2 &=& \frac{1}{\mid \Delta_{\PH}(s)\mid^2}\,\Gamma_{\PH \to \PZ_{\ssL}\PZ_{\ssL}}(s),
\nl
\Gamma_{\PH \to \PZ_{\ssL}\PZ_{\ssL}}(s) &=& \frac{g^2}{16\,\pi\,\sqrt{s}}\,
\frac{\lambda^{1/2}\lpar s,M^2_{\PZ},M^2_{\PZ}\rpar}{s}\,
\bigl( \Gamma_0 + \frac{g^2}{\cos^2\theta}\,Re \Gamma_1 \bigr),
\eqa
\bq
\Gamma_0 = \frac{M^2_{\PZ}}{\cos^2\theta} +
\frac{1}{4}\,\frac{\lambda(s,M^2_{\PZ},M^2_{\PZ})}{M^2_{\PZ}\,\cos^2\theta},
\quad
\Gamma_1 = \frac{1}{4}\,\frac{s - M^2_{\PZ}}{M^4_{\PZ}}\,\lambda(s,M^2_{\PZ},M^2_{\PZ})\,F_p
- \frac{\cph}{2}\,\Bigl[ 4 + \frac{\lambda(s,M^2_{\PZ},M^2_{\PZ})}{M^4_{\PZ}} \Bigr]\,F_d.
\label{offon}
\eq
There is no problem with gauge parameter dependence in \eqn{offon}; nevertheless, it is useful
to separate the residue of the Higgs complex pole and also to introduce $\PZ$ complex
poles obtaining a pseudo-observable 
\bq
\Gamma_{\PH^c \to \PZ^c_{\ssL}\PZ^c_{\ssL}} = \frac{g^2}{16\,\pi\,\sqrt{\mid \cph\mid}}\,
\Re\,\bigl( {\overline\Gamma}_0 + \frac{g^2}{\cos^2\theta}\,{\overline\Gamma}_1 \bigr),
\qquad \mid \cph\mid \gg \mid \cpz\mid,
\eq
where ${\overline\Gamma}$ is obtained with the substitution $s \to \cph$ and
$M^2_{\PZ} \to \cpz$ in $\Gamma$. 
As a second step we consider $\PH \to \PZ\PZ \to \Pep\Pem\PGmp\PGmm$ and define
\bq
D_4 = \frac{1}{\mid \Delta_{\PH}(s)\mid^2}\,\sum_{\rm spins}\,\Phi_4\,
      \bmid \bigl( A^0_{\mu\nu} + A^1_{\mu\nu}\bigr)\,
      J^{\mu}\,J^{\nu}\bmid^2\,
      \frac{1}{\mid \Delta_{\PZ}(s_1)\mid^2\,\mid \Delta_{\PZ}(s_2)\mid^2},
\eq
where $J$ is the (conserved) leptonic current and
\bq
\Phi_4 = \frac{1}{(2\,\pi)^8}\,\frac{1}{2\,\sqrt{s}}\,
           \int \prod_{i=1,2} d^4 q_i\,d^4 k_i\,\delta^+(q^2_i)\,\delta^+(k^2_i)\,
           \delta^4\lpar P - \sum_i (q_i+k_i) \rpar,
\eq
is the phase-space for $\PH \to\,$leptons. We write it as
\bqa
\Phi_4 &=& \frac{1}{(2\,\pi)^8}\,\frac{1}{2\,\sqrt{s}}\,
           \int d^4 p_1\,d^4 p_2\,\delta^4\lpar P - p_1 - p_2 \rpar
           \int \prod_{i=1,2} d^4 q_i\,d^4 k_i\,\delta^+(q^2_i)\,\delta^+(k^2_i)\,
\nl
{}&\times& \delta^4\lpar p_1 - q_1 - k_1) \rpar\,
           \delta^4\lpar p_2 - q_2 - k_2) \rpar
\nl
{}&=& \frac{\sqrt{s_1 s_2}}{\pi^2}\,\int ds_1\,ds_2\,
      \Phi_2\lpar s,s_1,s_2\rpar\,\Phi_2\lpar s_1,0,0\rpar\,\Phi_2\lpar s_2,0,0\rpar.
\eqa
The form factors $F_{d,p}$ are computed with $s = \cph$ and $s_{1,2} = \cpz$ to
guarantee gauge parameter independence. Based on longitudinal polarization dominance
(see \refA{appB}) for $s \to \infty$ we introduce longitudinal polarization vectors in the 
numerator of the $\PZ\,$-propagators and obtain
\bq
\sum_{\rm spins}\,\bmid \bigl( A^0_{\mu\nu} + A^1_{\mu\nu}\bigr)\,J^{\mu}\,J^{\nu}\bmid^2 =
\sum_{\rm spins}\,\bmid \spro{J}{e_{\ssL}(p_1)} \bmid^2\,\bmid \spro{J}{e_{\ssL}(p_2)} \bmid^2
g^2\,\Re\,{\cal A},
\eq
\bqa
{\cal A} &=& \frac{\cpz}{\cos^2\theta} + \frac{g^2}{\cos^2\theta}\,\Bigl\{
\frac{1}{4}\,\frac{s - s_1 - s_2}{s_1 s_2}\,\lambda\lpar s,s_1,s_2\rpar\,F_p -
\frac{1}{2}\,\frac{\cph}{s_1 s_2}\,\Bigl[ \lambda\lpar s,s_1,s_2\rpar + 4\,s_1 s_2\Bigr]\,F_d
\Bigr\}.
\eqa
Therefore, taking the residue of all complex poles, in the limit $\mid \cph \to \infty$ we obtain
\bq
\frac{d^2 D_4}{ds_1 ds_2} = \frac{\mid \cpz\mid}{\pi^2}\,
\frac{1}{\mid \Delta_{\PH}(s)\mid^2}\,
\frac{\Gamma_{\PZ^c \to \Pep\Pem}}{\mid \Delta_{\PZ}(s_1)\mid^2}\,
\frac{\Gamma_{\PZ^c \to \PGmp\PGmm}}{\mid \Delta_{\PZ}(s_2)\mid^2}\,
\Gamma_{\PH^c \to \PZ^c_{\ssL}\,\PZ^c_{\ssL}}\,
 + \mbox{non-resonant terms}.
\eq
For an Heavy Higgs boson is straightforward to extract the pseudo-observable defining
$\PH \to \PZ_{\ssL} \PZ_{\ssL}$ from the decay $\PH$ into two pairs of leptons of
different flavor. Interference effects, as well as transverse polarizations, will
be analyzed in a subsequent paper.
\clearpage
\bibliographystyle{atlasnote}
\bibliography{HLS}{}

\end{document}